\documentclass[fleqn,usenatbib]{mnras}

\usepackage{newtxtext,newtxmath}

\usepackage[T1]{fontenc}

\DeclareRobustCommand{\VAN}[3]{#2}
\let\VANthebibliography\thebibliography
\def\thebibliography{\DeclareRobustCommand{\VAN}[3]{##3}\VANthebibliography}

\usepackage{graphicx}	
\usepackage{amsmath}	
\usepackage[dvipsnames]{xcolor}

\usepackage{isotope}
\usepackage{setspace}
\usepackage{multirow}
\usepackage[tight]{units}
\usepackage[acronym,nomain,nonumberlist,toc,shortcuts]{glossaries}

\newcommand{\de}{\partial}
\newcommand{\Msun}{M_{\odot}}
\newcommand{\Rsun}{R_{\odot}}
\newcommand{\Ni}{\isotope[56]{Ni}}
\newcommand{\X}{\isotope[56]{X}}
\newcommand{\Co}{\isotope[56]{Co}}
\newcommand{\Fe}{\isotope[56]{Fe}}
\newcommand{\Si}{\isotope[28]{Si}}
\newcommand{\C}{\isotope[12]{C}}
\newcommand{\Ti}{\isotope[44]{Ti}}
\newcommand{\Mns}{M_\mathrm{PNS}}
\newcommand{\Rns}{R_\mathrm{PNS}}
\newcommand{\Vns}{v_\mathrm{PNS}}
\newcommand{\VnsVector}{\textbf{v}_\mathrm{PNS}}
\newcommand{\Jns}{J_\mathrm{PNS}}
\newcommand{\JnsVector}{\textbf{J}_\mathrm{PNS}}
\newcommand{\Rpro}{R_\mathrm{prog}}
\newcommand{\Mpro}{M_\mathrm{prog}}
\newcommand{\Mzams}{M_\mathrm{ZAMS}}

\newacronym[longplural={core-collapse supernovae}, shortplural={CCSNe}]{ccsn}{CCSN}{core-collapse supernova}
\newacronym{snr}{SNR}{supernova remnant}
\newacronym{zams}{ZAMS}{zero-age main-sequence}
\newacronym{rsg}{RSG}{red supergiant}
\newacronym{bsg}{BSG}{blue supergiant}
\newacronym[longplural={pre-supernovae}, shortplural={preSNe}]{presn}{preSN}{pre-supernova}
\newacronym[longplural={Rayleigh-Taylor instabilities}]{rti}{RTI}{Rayleigh-Taylor instability}
\newacronym[longplural={supernovae}, shortplural={SNe}]{sn}{SN}{supernova}
\newacronym{lmc}{LMC}{Large Magellanic Cloud}
\newacronym{sasi}{SASI}{standing accretion shock instability}
\newacronym{pns}{PNS}{proto-neutron star}
\newacronym[longplural={equations of state}]{eos}{EOS}{equation of state}
\newacronym{nse}{NSE}{nuclear statistical equilibrium}
\newacronym[description={Consistent multifluid advection (scheme) by \cite{plewa1999}}]{cma}{CMA}{consistent multifluid advection}
\newacronym[description={piecewise parabolic method by \cite{colella1984}}]{ppm}{PPM}{piecewise parabolic method}
\newacronym[description={improved advection upstream splitting method Riemann solver by \cite{liou1996}}]{ausmp}{AUSM+}{improved advection upstream splitting method}
\newacronym[description={Harten-Lax-van Leer-Einfeldt Riemann solver by \cite{einfeldt1988}}]{hlle}{HLLE}{Harten-Lax-van Leer-Einfeldt}
\newacronym{cfl}{CFL}{Courant-Friedrich-Levy}
\newacronym{ns}{NS}{neutron star}
\newacronym{jwst}{JWST}{\emph{James Webb Space Telescope}}
\newacronym{csm}{CSM}{circumstellar medium}

\title[Instabilities in CCSNe from RSG stars]{Hydrodynamic instabilities in long-term three-dimensional simulations of neutrino-driven supernovae of 13 red supergiant progenitors}

\author[B. Giudici, M. Gabler and H.-T. Janka]{
Beatrice Giudici,$^{1}$\thanks{E-mail: beatrice.giudici@uv.es}
Michael Gabler$^{1}$
and Hans-Thomas Janka$^{2}$
\\
$^{1}$Departamento de Astronomía y Astrofísica, Universitat de València, E-46100 Burjassot (València), Spain\\
$^{2}$Max Planck Institute for Astrophysics, Karl-Schwarzschild-Str.~1, 85748 Garching, Germany\\
}

\date{Accepted XXX. Received YYY; in original form ZZZ}

\pubyear{2025}

\begin{document}
\label{firstpage}
\pagerange{\pageref{firstpage}--\pageref{lastpage}}
\maketitle

\begin{abstract}
We present long-term three-dimensional (3D) simulations of Type-IIP supernovae (SNe) for 13 non-rotating, single-star, red-supergiant (RSG) progenitors with zero-age-main-sequence masses between 12.5\,$\Msun$ and 27.3\,$\Msun$. The explosions were modelled with a parametric treatment of neutrino heating to obtain defined energies, $\Ni$ yields, and neutron-star properties in agreement with previous results. Our 3D SN models were evolved from core bounce until 10~days to study how the large-scale mixing of chemical elements depends on the progenitor structure. Rayleigh-Taylor instabilities (RTIs), which grow at the (C+O)/He and He/H interfaces and interact with the reverse shock forming after the SN shock has passed the He/H interface, play a crucial role in the outward mixing of $\Ni$ into the hydrogen envelope. We find most extreme $\Ni$ mixing and the highest maximum $\Ni$ velocities in lower-mass (LM) explosions despite lower explosion energies, and the weakest $\Ni$ mixing in the 3D explosions of the most massive RSGs. The efficiency of radial $\Ni$ mixing anti-correlates linearly with the helium-core mass and correlates positively with the magnitude of a local maximum of $\rho r^3$ in the helium shell. This maximum causes shock deceleration and therefore facilitates high growth factors of RTI at the (C+O)/He interface in the LM explosions. Therefore fast-moving $\Ni$ created by the asymmetric neutrino-heating mechanism is carried into the ubiquitous RT-unstable region near the He/H interface and ultimately far into the envelopes of the exploding RSGs. Our correlations may aid improving mixing prescriptions in 1D SN models and deducing progenitor structures from observed SN properties.
\end{abstract}



\section{Introduction}

The detection of Supernova SN~1987A in the \ac*{lmc} on February 23, 1987 marked
a groundbreaking event in the history of astronomy. Until now, this \ac*{ccsn} and its remnant are among the most thoroughly studied objects in modern astrophysics, which reveal the evolution of a massive blue supergiant star from its pre-collapse state through the \ac*{sn} explosion into the remnant stage of the gaseous ejecta in extraordinary detail
(see e.g., \citealt{arnett1989a, bethe1990, mccray2017, podsiadlowski2017}).
Meanwhile, a growing number of different kinds of observations clearly show the genuinely three-dimensional (3D) character of this event \citep{Sinnott2013,utrobin2015,larsson2016,Abellan2017,Alp2019,Cigan2019}.
A new breakthrough was marked by the advent of the \ac*{jwst}, which allows observations with unprecedented resolution. \cite{jones2023} were able to spatially resolve the ejecta, equatorial ring, and outer rings of SN~1987A in the mid-infrared, and \cite{larsson2023} constructed the emissivity maps of the [Fe I] $\unit[1.443]{\mu m}$ and [He I] $\unit[1.083]{\mu m}$ lines. In particular, the [Fe I] line revealed a highly asymmetric morphology of the ejecta and the interaction of Fe-rich material with the reverse shock. Recently, \citet{Larsson2025} connected the 3D structure of SN~1987A’s innermost ejecta with the possible location of the compact remnant. The remnant, which is most likely a \ac*{ns} \citep{Page+2020,Greco+2021,Greco+2022}, seems to move towards the observer and to the north-east of the putative explosion centre according to the 3D [Fe I] 1.443\,$\mu$m line map of \citet{larsson2023}, but to the south-east according to the position of the JWST [Ar VI] source discussed by \citet{Fransson+2024}. The north-east direction would be consistent with conclusions drawn from the 679~GHz emission of a warm dust `blob' observed with ALMA by \citet{Cigan2019}.

Another well studied, young \ac*{snr} is Cassiopeia~A (Cas~A), whose \ac*{sn} light reached the earth about 350 years ago \citep{fesen2006}. This remnant is characterized by a reverse-shock heated shell enclosing the unshocked inner volume, a flattened ejecta distribution seen as a tilted thick disk, and two opposing wide-angle, jet-like ejecta structures \citep{delaney2010}. See also \cite{isensee2010} for \emph{Spitzer Space Telescope} infrared observations,
\cite{lazendic2006} for \emph{Chandra} X-ray observations, and \cite{fesen1996, fesen2001} for optical observations, as well as the more recent, highly detailed infrared survey by \ac*{jwst} \citep{milisavljevic2024}. Studies of such kind can reveal the geometrical distributions of chemical elements in the SN ejecta and have unfolded the morphologies not only of SN~1987A and Cas~A, but also of many other remnants of CCSNe \citep[e.g.,][]{aschenbach+1995,vink2012,vink2017,holland-ashford+2017,katsuda+2018,larsson2021,mayer+2022,mayer+2023}.

Time sequences of spectra during the light-curve evolution of SNe are also powerful observational probes of radial mixing, in particular of heavy elements from the innermost into the outermost layers of the exploding stars. The mixing of $\Ni$ has important implications for the shapes of the light curves, for example the duration of the plateau \citep[e.g.][]{nakar+2016,goldberg+2019} and the shape of the light-curve maximum~\citep[e.g.,][]{utrobin2019,utrobin2021}. A fast rise of the observed light curve is related to rapidly expanding $\Ni$-rich structures, while the width of the emission peak can be related to the total amount of $\Ni$ present and when the matter becomes transparent. The shape of the light curve is thus directly related to the distribution of $\Ni$.

After the SN shock has been launched, it propagates through the outer shells of the progenitor, which consist of progressively lighter elements. As shown for a spherically symmetric blast-wave solution \citep{sedov1959}, the acceleration of the shock wave depends on variations of the density gradient. The shock accelerates where the density gradient is steeper than $r^\mathrm{-3}$ and decelerates otherwise. Strong acceleration (deceleration) usually occurs when the shock approaches (moves away from) composition interfaces, e.g. the (C+O)/He and He/H interfaces. This combination can lead to a (locally) positive pressure gradient, while the density gradient remains negative. Consequently, these layers are prone to the growth of \ac*{rti} \citep{chevalier1976}. This instability is seeded by asymmetries generated in the innermost part of the ejecta \citep{kifonidis2003} 
and it can cause large-scale radial mixing of heavy elements into the outer layers of the star. In addition, the lighter elements (mostly hydrogen) can be mixed inwards \citep{fryxell1991, muller1991}.

First long-term 2D and 3D simulations investigating the role of \acp*{rti} relied on an artificial initiation of the explosion either by a point-like thermal explosion 
\citep{arnett1989b, hachisu1990, hachisu1992, hachisu1994, yamada1990, fryxell1991, muller1991, herant1991, herant1992, herant1994, iwamoto1997, nagataki1998, kane2000}, by piston-driven explosions \citep{hungerford2003, hungerford2005, joggerst2009, joggerst2010a, joggerst2010b}, or by injection of thermal and kinetic energy into a given location \citep{couch2009, couch2011, ono2013}. As a next step, \citet{ellinger2012, ellinger2013} performed asymmetric explosions with artificially imposed, global, low-mode asymmetries.

On the theoretical-phenomenological side of SNR modelling, there has been remarkable success in reproducing observed morphological features of SN~1987A \citep{ono2020, ono2024, orlando2020, nakamura2022} and of Cas~A \citep{orlando2016} by relying on specifically defined initial conditions (partially even at shock breakout) with imposed explosion asphericities. Though matching some observations well, this kind of simulations cannot answer the question how the asymmetries are seeded and how they are linked to the \ac*{sn} mechanism. In particular, the growth of hydrodynamic instabilities in the phases preceding the shock breakout is not captured self-consistently.

In a parallel theoretical development, this shortcoming was overcome by long-term simulations where the SN explosion was initiated and powered by neutrino energy deposition in the innermost ejecta, based on the paradigm of the neutrino-driven explosion mechanism, where the stalled \ac*{sn} shock is revived by the interaction of neutrinos radiated by the hot \ac*{pns} with the still infalling matter of the collapsing stellar core \citep{colgate+1966,arnett1966,bethe+1985}. Inspired by the radial mixing of core material into the stellar envelope observed in SN~1987A, it was shown by multi-dimensional simulations that hydrodynamic instabilities develop in the neutrino-heated gain layer behind the stalled shock. Depending on the shock radius, strong convection
\citep{bethe1990, herant1992, herant1994, burrows1995, janka1995, janka1996} and global mass motions by the \ac*{sasi} \citep{foglizzo2002, blondin2003, blondin2006, ohnishi2006, foglizzo2007, scheck2008, fernandez2010, buellet2023} can occur. Both push shock expansion and therefore increase the volume and mass in the gain layer, thus facilitating easier explosions because more matter in the post-shock region absorbs more energy from neutrinos. Moreover, the joint effect of these instabilities leads to non-radial flows and turbulent mass motions in the neutrino-heated gas and imposes large-scale asymmetries on the \ac*{sn} ejecta already at the onset of the explosion.

Based on the paradigm of the neutrino-driven mechanism, long-term simulations have achieved to connect the final ejecta asymmetries of \acp*{ccsn} with these early ejecta asymmetries that emerged from the mentioned hydrodynamic instabilities during the neutrino-heating phase. The first such simulations that followed the \ac*{sn} evolution from the onset of the explosion until the breakout of the \ac*{sn} shock from the stellar surface and beyond were performed in 2D by \citet{kifonidis2003, kifonidis2006}, who initiated the neutrino-driven blast by a parametric prescription (``light bulb'' or neutrino ``engine'') of the neutrino luminosities that accomplished the energy deposition in the gain layer. Meanwhile, a larger number of 3D simulations of this kind has been performed, including explosions of \ac*{rsg} and \ac*{bsg} progenitors of Type-IIP \acp*{sn} and of stripped progenitors of Type-IIb and Ib/c \acp*{sn} \citep{hammer2010, wongwathanarat2013, wongwathanarat2015, wongwathanarat2017,utrobin2015, utrobin2017,utrobin2019,utrobin2021,vanbaal+2023,vanbaal+2024}. Some of these simulations have even been continued into the early evolution of the \acp*{snr} for hundreds of days up to hundreds of years \citep{gabler2021,orlando+2021,orlando2022}.

Although the shock revival phase by neutrino heating can now be simulated fully self-consistently in 3D \citep[e.g.,][]{kuroda2012, Takiwaki2014, Melson2015,melson+2015, lentz+2015, mueller+2017, muller2019, burrows2019, powell2019, powell2020, bollig2021, vartanyan2022, burrows+2024, janka2024,nakamura+2025}, these simulations involve (more or less) elaborate and time consuming treatments of the neutrino transport. Therefore they usually cover periods of at most the first seconds after core bounce with few exceptions where the evolution was followed until shock breakout for ultrastripped progenitors \citep{muller2018} and a small sample of \ac*{rsg} progenitors \citep{stockinger2020,sandoval2021,vartanyan+2025,vartanyan2025b}. Simulations with a rigorous treatment of the neutrino physics over such long periods of time for larger sets of models are still prohibitive because of their enormous demands of computing resources. Only recently, in a study parallel to ours, \citet{vartanyan2025b} presented a small set of such 3D simulations that were continued into the long-time evolution beyond shock breakout for \acp*{sn} originating from \ac*{rsg} progenitors.

In our project we will proceed along the lines of the 3D \ac*{ccsn} modelling with neutrino engine-driven explosions mentioned above. 
In order to reduce the computing time requirements for the complex neutrino physics and transport in the newly formed \ac*{pns}, the high-density core of the \ac*{pns} with a neutrino optical depth of more than 10--100 (the exact value depends on time, model, and neutrino species) is replaced by a moving inner grid boundary, which mimics the contraction of the cooling compact remnant. At this inner boundary the neutrino luminosities radiated by the \ac*{pns} are prescribed parametrically such that the explosion energy can be tuned to a chosen value.

In the near-surface layers of the \ac*{pns} and the neutrino-heating layer behind the \ac*{sn} shock the neutrino-transport equations and neutrino-matter interactions are solved with a grey ray-by-ray approximation \citep{scheck2006,wongwathanarat2013}. This approach allows us to perform a larger suite of 3D simulations that capture the fundamental aspects of neutrino-driven explosions and that follow long evolution times with acceptable computational costs. Thus we can investigate a set of 13 \ac*{rsg} progenitors and study how the hydrodynamic instabilities of the shock revival phase act later as seeds of the \acp*{rti} at the composition-shell interfaces after the passage of the outward propagating \ac*{sn} shock. This interplay of early and subsequent instabilities is essential to determine the final morphology of the explosion and the large-scale radial mixing of the ejecta, which shape many observational properties of \acp*{sn} and young \acp*{snr}. 

The manuscript is organized as follows: In Section~\ref{sec:numerics}, we briefly describe the \ac*{presn} models, the numerical methods, and the input physics employed in our simulations.
In Section~\ref{sec:early_times} the results for the early phase of the \ac*{sn} explosions up to 2.5\,s after core bounce are presented with a focus on the explosion energies, nickel yields, and \ac*{pns} properties. 
In Section~\ref{sec:long_times}, we discuss the results of our 3D simulations for the evolution until 10 days after shock revival, considering a linear stability analysis for \acp*{rti} in 1D models, the radial mixing of the ejecta in 3D, the changes of the 3D morphologies of our models until 10\,d, and the asymmetric shock breakout in our 3D \ac*{rsg} explosions. We also generalize a framework developed by \citet{utrobin2019, utrobin2021} to quantify and characterize the radial $\Ni$ mixing in a larger set of \ac*{bsg} explosions to our \ac*{rsg} explosions, and we correlate the degree of mixing with characteristic parameters of the explosion and information obtained from the 1D stability analysis. In Section~\ref{sec:dep_prog}, we go one step further and establish correlations between the extent of the radial $\Ni$ mixing and the structure of the progenitor star. Our findings are summarized in Section~\ref{sec:conclusion}.

\section{Model overview and numerical setup}\label{sec:numerics}

We investigate the explosion of 13 \ac*{rsg} progenitor models that were generated with the stellar evolution code \textsc{Kepler} \citep{weaver1978}. These models have been discussed in detail previously \citep{sukhbold2014, sukhbold2016}. Therefore, we will summarize their main properties briefly in Section~\ref{subsec:presn}. In Section~\ref{subsec:num}, we describe the \textsc{Prometheus-HotB} code, which we employ to
evolve these models after spherical collapse from the bounce in full 3D. Further details of the setup of the simulations are then given in Section~\ref{subsec:physics}

\subsection{Progenitor models}\label{subsec:presn}

All stellar progenitors are non-rotating stars of solar metallicity and were evolved from the main sequence until the onset of the Fe-core collapse 
\citep{sukhbold2014, woosley2015}. When referring to our simulations, we use SW to indicate models based on the first reference \citep{sukhbold2014} and WH for the second \citep{woosley2015}. The model names are completed by three digits indicating the ZAMS mass of the respective progenitor.

The calculations included the entire stars with their hydrogen envelopes, using a mass loss prescription \citep{nieuwenhuijzen1990, wellstein1999, woosley2002}
and considering a rate of $\isotope[12]{C}(\alpha,\gamma)\isotope[16]{O}$ equal to 1.2 times that of \citet{buchmann1996}, $S$(300keV) = 146 keVb.

A simple parameter, introduced to characterize the 1D core structure of the pre-collapse star, is given by the compactness \citep{oconnor2011}:
\begin{equation}
    \xi_{M} = \left.\frac{M / \Msun}{R(M) / \unit[1000]{km}} \right|_{t_\mathrm{bounce}}\,.
    \label{eq:compactness}
\end{equation}
These authors found that with a higher compactness, the star is less likely to explode. However, this simple criterion turned out not to be perfectly conclusive \citep[e.g.,][]{ugliano2012, mueller+2016, ertl2016, sukhbold2016, ertl2020}, and it is in tension with some of the most recent self-consistent multi-dimensional \ac*{ccsn} simulations as well as 1D explosion modelling that includes post-shock turbulence in a parametrized mixing-length-like treatment \citep[e.g.][]{wang2022,boccioli+2025}.\footnote{The dependence of explodability on the compactness is still controversial. It is sensitive to the strength of the neutrino heating and its dependence on the neutrino emission of the hot PNS (e.g., the growth of the mean energy of the radiated neutrinos with NS mass), in competition with the explosion-damping influence of a high mass-infall rate in high-compactness progenitors. In particular, recent 2D and 3D simulations that produce explosions relatively readily \citep{burrows2020, vartanyan2023, chen2026} obtained successful explosions for high-compactness stars due to very high neutrino heating, for example also in black-hole forming core-collapse events \citep{burrows2023,sykes2025, eggenberger2025}. Such results were taken by \citet{boccioli+2025} as a guidance for the choice of parameter values in a method that treats SN turbulence in reduced-dimensionality \citep[STIR;][]{couch2020}. The outcomes of those simulations, however, are in conflict with simulations by the Garching group using different codes, as recently discussed by \citet{janka2025}. However, this still unsettled question of stellar explodability as a function of compactness or pre-collapse mass of the progenitor is not of direct relevance for our work studying the asymmetries and mixing in long-term, 3D CCSN models.}

There are substantial differences in the stellar structure between models with masses below $\unit[20]{\Msun}$ and those above that value. The compactness parameter of equation~\eqref{eq:compactness}, which is an indication of the density structure of the core, increases rapidly in the range of $\unit[12-15]{\Msun}$ and then it remains roughly constant until $\sim\unit[20]{\Msun}$.
In stellar models with masses of up to about $\unit[20]{\Msun}$, carbon ignites exoergically and powers convection, creating highly variable \ac*{presn} core structures \citep{sukhbold2014}.\footnote{We note that the behavior of the compactness as a function of the progenitor mass varies with the stellar evolution code used by different groups. It differs between the \textsc{Kepler} models considered in our study and the \textsc{Franec} models of \citet{chieffi2020} and \textsc{MESA} models of \citet{laplace2021}, mostly depending on the $^{12}$C$(\alpha,\gamma)$ rate and the adopted wind mass-loss prescription.}

In stars with masses between 20 and $\unit[30]{\Msun}$, on the contrary, 
carbon burning releases less energy than is emitted by neutrinos. This efficient neutrino loss prevents the stellar cores from being convective.
Thus more massive stars develop extended and dense shells that undergo rapid and intense shell burning. As a result of their more evolved and massive cores, these stars tend to have a higher compactness than lower-mass models (consistent with the thermal/entropy corrections to the Chandrasekhar mass shown in equation~6 of \citealt{timmes1996}).

\begin{table}
    \centering
    
    \begin{tabular}{c c c c c c c}
        \hline
        Model& $\xi_{2.5}$ & $\Rpro$& $\Mpro$ & $M^\mathrm{core}_\mathrm{He}$ & $M^\mathrm{core}_\mathrm{C+O}$ \\
        \multicolumn{2}{c}{} & [$\unit{\Rsun}$] &  {[$\unit{\Msun}$]}  & {[$\unit{\Msun}$]}  & {[$\unit{\Msun}$]}\\
        \hline
        WH12.5 & 0.039 &  683.60 & 11.12 & 3.36 & 2.28 \\
        SW13.1 & 0.071 &  734.85 & 11.47 & 3.67 & 2.53 \\
        SW14.2 & 0.130 &  788.03 & 12.32 & 4.04 & 2.84 \\
        SW16.3 & 0.161 &  875.86 & 13.86 & 4.82 & 3.57 \\
        SW18.2 & 0.186 &  989.20 & 15.02 & 5.57 & 4.27 \\
        SW19.8 & 0.128 & 1058.01 & 15.85 & 6.24 & 4.88 \\
        SW20.8 & 0.161 & 1107.33 & 16.02 & 6.68 & 5.30 \\
        SW21.0 & 0.139 & 1124.55 & 16.12 & 6.76 & 5.37 \\
        SW25.5 & 0.250 & 1413.83 & 15.45 & 8.58 & 7.21 \\
        SW25.6 & 0.217 & 1428.27 & 15.60 & 8.64 & 7.28 \\
        SW26.2 & 0.243 & 1458.76 & 15.40 & 8.88 & 7.51 \\
        SW27.0 & 0.244 & 1508.21 & 15.40 & 9.26 & 7.92 \\
        SW27.3 & 0.249 & 1511.77 & 15.07 & 9.35 & 7.99 \\
        \hline
    \end{tabular}
    \caption{Properties of the \ac*{presn} models.
    The first column contains the name of the models, where the first two letters indicate the source of the model, WH for \citet{woosley2015} and SW for \citet{sukhbold2014}, and the number defines the \ac*{zams} mass in solar masses.
    The second column shows the compactness $\xi_{2.5}$ as defined in \citet{oconnor2011} and equation \eqref{eq:compactness}. Columns~3 and~4 contain the radius and the mass of the \ac*{presn} star, respectively.
    In columns~5 and~6 we list the He and C+O core masses, measured at the locations where the hydrogen and helium mass fractions, respectively, drop below 50\% of their maximum values.}
    \label{tab:presn}
\end{table}

The \ac*{zams} masses of the \ac*{presn} models analysed in this paper are in the range between $\unit[12]{\Msun}$ and $\unit[30]{\Msun}$.
The selected models and their basic properties are listed in Table~\ref{tab:presn}. The locations of the (C+O)/He and He/H composition interfaces are defined as the positions at the bottom of the helium and hydrogen layers of the star where the He and H mass fractions, respectively, drop below half of the maximum value in the respective cell \citep{wongwathanarat2015}. Note, that this is different from the $20\%$ threshold taken in \cite{sukhbold2016}. The difference is minor and affects the C+O core masses in Table~\ref{tab:presn} only in the second digit and the He core masses at most on the level of $\sim$0.1\,M$_\odot$.

\begin{figure*}
    \centering
    \includegraphics[width=\linewidth]{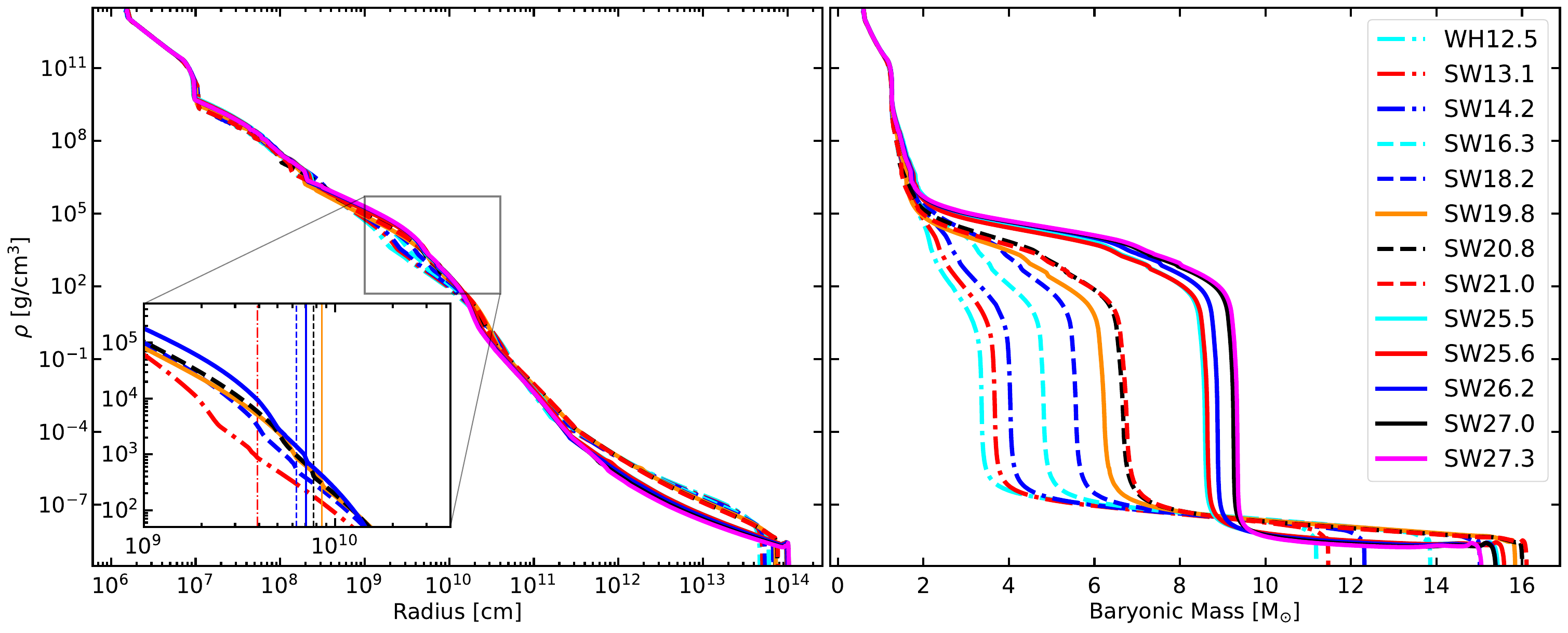}
    \caption{Density profiles as functions of radius (left) and enclosed baryonic mass (right) for all of the considered \ac*{sn} models roughly some 10\,ms after core bounce when the shock has reached a radius of $\sim$100\,km and an enclosed mass of $\sim$1.3\,M$_\odot$. The vertical lines in the inset depict the location of the (C+O)/He composition interface for a selected set of models.
    }
    \label{fig:densityProfile}
\end{figure*}

In Fig.~\ref{fig:densityProfile}, we display the density profiles of all the progenitor models analysed in this paper just after core bounce as functions of radius (left) and enclosed mass (right). 

In order to perform a larger number of 3D simulations that follow the explosions from core bounce until well after shock breakout with reasonable computational resources, we rely in our treatment of neutrino effects on the approximate ray-by-ray neutrino transport scheme of \citet{scheck2006}, supplemented by the long-term simulation setup introduced by \citet{wongwathanarat2013} and \cite{wongwathanarat2015}, as detailed in the next section. 
Within this approach, we have the freedom to tune the explosion energy and the timescale for significant neutrino energy deposition by parameters. For our 3D models we chose parameter values such that the explosion energies and nickel yields are comparable to those obtained by \citet{sukhbold2016} and \citet{ertl2020} in their large sets of 1D models. Those models were exploded with a similar, parametric neutrino-engine treatment including a 1D version of the ray-by-ray description of neutrino transport applied in the present work. Due to remaining differences between the two approaches and fundamental 1D vs.\ 3D differences, we were not able to exactly recover the explosion parameters obtained by \citet{sukhbold2016} and \citet{ertl2020}.

\subsection{Numerical setup}\label{subsec:num}

We performed the simulations with the code \textsc{Prometheus} \citep{fryxell1991, muller1991} in its version \textsc{Prometheus-HotB} \citep{janka1996, kifonidis2003, kifonidis2006, scheck2006, arcones2007, wongwathanarat2013, wongwathanarat2015, wongwathanarat2017, gessner2018, stockinger2020, gabler2021}.
The code is a 3D, explicit finite-volume hydrodynamics scheme that includes a simplified neutrino radiation module, an \ac*{eos} covering high and low densities including \ac*{nse}, and a description of nuclear burning in the form of an $\alpha$ network when NSE does not apply.
The equations of hydrodynamics are solved with the \ac*{ppm} \citep{colella1984} employing the exact Riemann solver for real gases of \citet{colella1985}. Different species are evolved via the \ac*{cma} scheme \citep{plewa1999}. Inside the grid cells where strong shocks are present, either the \ac*{hlle} \citep{einfeldt1988} or the \ac*{ausmp} \citep{liou1996} solver are employed in order to suppress numerical artifacts that can arise due to odd-even decoupling. Until shock breakout, we usually use the \ac*{ausmp} solver with a static grid. Only later, when following the expanding ejecta with a quasi-Lagrangian computational grid \citep{gabler2021}, the more diffusive \ac*{hlle} solver is used to avoid problems with numerical stability.

The computational domain is discretized by means of an axis-free overlapping Yin-Yang grid technique \citep{kageyama2004} implemented into \textsc{Prometheus-HotB} by \citet{wongwathanarat2010}.
The Yin-Yang approach avoids numerical artifacts that can arise near the polar axis in a standard spherical grid. It also alleviates the time-step constraints imposed by the \ac*{cfl} condition, which is very restrictive for a spherical polar grid due to the small azimuthal grid step size near the axis. Thus, the Yin-Yang grid allows for larger time steps in the simulations and, hence, for computationally less expensive long-time simulations.

Since we are only interested in the matter that is ejected during the \ac*{sn} explosion but not in the detailed evolution of the \ac*{pns}, where the \ac*{cfl} limit is particularly small, the computational efficiency of our simulations is further increased by excluding a time-dependent spherical volume around the coordinate centre from the computational grid. This central region is replaced by an inner grid boundary. For the first 2.5\,s after core bounce the excised domain contains the high-density core of the \ac*{pns} and at later times a larger central volume (details will be provided later in this section). 

Newtonian self-gravity is taken into account by solving the integral form of Poisson's equation with a multipole expansion method optimized for parallel computations \citep{muller1995, wongwathanarat2019}.
The central mass that is not included on our computational grid is taken into account as a point mass in the gravity solver. We also apply general relativistic corrections to the monopole of the potential following \citet{scheck2006} and \citet{arcones2007}, with a treatment based on Case~A of \citet{marek2006}.

During the first 2.5\,s we handle neutrino effects by an approximate, grey transport treatment based on an analytic solution of the Boltzmann equations for neutrino number density and energy density with a ray-by-ray approach for multi-dimensional simulations. For the basic equations, numerical treatment and input concerning the neutrino rates, see \cite{scheck2006}. During this early phase the high-density core of the \ac*{pns} (up to $\sim \unit[1.1]{\Msun}$), where the mean optical depth for neutrinos typically exceeds 10--100 (depending on time and neutrino species), is excised from the computational domain and replaced by a closed inner grid boundary at radius $R_\mathrm{ib}$. This $R_\mathrm{ib}$ is contracted to mimic the shrinking of the new-born \ac*{pns} due to cooling and deleptonization by neutrino emission.
At this inner grid boundary, the neutrino luminosities from the high-density core of the PNS are imposed with a functional behaviour in time and parameter values given in Section~\ref{subsec:physics}. The spectral properties of the inflowing neutrinos at the inner boundary are chosen according to local thermal equilibrium with the stellar plasma in the innermost radial grid zones, i.e., following \citet{ugliano2012} and \citet{wongwathanarat2013,wongwathanarat2015}, their spectra are assumed to have Fermi–Dirac shape with mean energies determined by the temperature in the innermost radial cells of our computational grid, consistent with the fact that neutrinos in highly optically thick regions thermalize quickly with the stellar medium.

The use of our ray-by-ray transport approximation below an average neutrino optical depth of about 10 (or more) allows us to account for luminosity changes connected to the neutrino emission from matter accreted by the PNS. Therefore the luminosities and mean energies of the radiated neutrinos are time-dependent, direction-dependent, and different between the different neutrino species ($\nu_e$, $\bar\nu_e$, and heavy-lepton neutrinos $\nu_x$). This also implies that the neutrino emission of the PNS and the post-shock neutrino heating in our simulations are not spherically symmetric.

Initially, the inner boundary is placed at a radius of $R_\mathrm{ib} \gtrsim 60$\,km. The first 25 zones in radial direction are equispaced with a $\Delta r/r \le 0.01$ (typically $\Delta r \sim 6\times10^4$\,cm), and farther out the distance between radial grid points increases logarithmically. We use a total number of 400 radial grid points.
The outer boundary of the radial grid at $R_\mathrm{ob}=4\times 10^5$\,km is treated as an outflow boundary. 
The final radius of the inner grid boundary after $\unit[2.5]{s}$ is $\approx\unit[20]{km}$.
The angular part of the computational grid covers the entire $4\pi$ solid angle with the constant resolution of $\Delta \theta = \Delta \phi = 2^{\circ}$ in both angular directions. In each Yin and Yang patch, we thus have 45 zones in $\theta$-direction and 135 zones in $\phi$-direction.

We follow the shock-revival phase and shock propagation on this grid until $t_\mathrm{map}$ = 2.5\,s after bounce. At this point the simulation is remapped to a larger radial domain in order to track the propagation of the explosion shock through the outer stellar layers.
The new boundaries are chosen at $R_\mathrm{ib} = 500$\,km and twice the progenitor radius $R_\mathrm{ob} \approx  2 \times R_\mathrm{prog}$. The radial grid now contains 1200 points initially and is again logarithmically spaced with the innermost cell having a size of $\Delta r = 2.5$ km. The angular grid remains unchanged. At this time we describe the inner boundary as an inflow boundary where a neutrino-driven wind is imposed with a functional behaviour and parameters provided in Section~\ref{subsec:physics}.

During the subsequent long-term simulations, we remove the innermost radial zone whenever $R_\mathrm{ib}$ becomes smaller than 1$\%$ of the minimum radius of the aspherical \ac*{sn} shock, or when the material in every cell of the innermost layer is falling inward with velocities greater than the speed of sound in the respective cell. This progressively reduces the number of radial grid zones, thus speeding up the simulations. The mass of the excised zones is added to the central point mass.

The flow through the inner boundary after the neutrino-driven wind ceases and, to a lesser extent, the successive cutting out of the innermost cells can cause some of the matter to leave our numerical grid. In the most extreme cases (the most massive progenitors) this may add up to the order of one $\Msun$ which is removed from our simulations. We cannot strictly determine what the fate of this mass will be. It may be accreted to the central \ac*{ns}, possibly causing the formation of a black hole in our simulations, or it may get re-accelerated outwards. However, this ``fallback'' could be a feature of our parametrized explosion set up and more realistic engine simulations may provide a more robust description of the slowest ejecta. Furthermore, we are not interested in what happens to the innermost ejecta, since these are not expected to be prone to RTI.

Once the shock leaves the surface of the progenitor, the grid is moved radially as the \ac*{sn} ejecta expands. This moving mesh relaxes the \ac*{cfl} condition more and more and reduces numerical diffusion through cell boundaries. The grid velocity is set to be linearly proportional to the radius. Depending on the model, the outermost grid cell moves with at most 150\% of the maximum fluid velocity, whereas the velocity of the innermost radial grid cell is always set to zero. Between these maximal radial extents, the mesh velocity increases linearly. In some cases, the shock may still expand faster than the such determined maximum grid velocity. To avoid the shock leaving the computational domain during the simulation, we add a new cell to the outer boundary whenever the shock gets closer than 20 cells to the outer boundary of the computational grid \citep{gabler2021}. At the same time we remove the innermost radial cell to keep the total number of cells constant. 

By the end of the simulations, at 10 days after the explosion, the radial grid contains between 400 and 500 zones depending on the model. The ratio of $\Delta r / R $ is usually of the order of 0.01, and always smaller than 0.05.

\subsection{Input physics}\label{subsec:physics}

We performed our simulations with the tabulated \ac*{eos} LS220 for high densities, $\rho>\unit[10^{11}]{g \cdot cm^{-3}}$ \citep{lattimer1991}, and a tabulated \ac*{eos} by \citet{timmes2000} based on the Helmholtz free energy for intermediate densities. 
The latter considers arbitrarily degenerate and relativistic Fermi gases for electrons and positrons, photons, and ideal Boltzmann gases for the baryonic matter. We adopt an alpha network with 19 nuclear species: protons, the 13 $\alpha$-nuclei from \isotope[4]{He} to $\Ni$, the beta decay products of $\Ni$ and \isotope[44]{Ti} ($\Co$, $\Fe$, \isotope[44]{Sc}, \isotope[44]{Ca}), and a tracer element $\X$ to represent the production of neutron-rich nuclear species when the electron fraction is $Y_\mathrm{e} < 0.49$.
These nuclei are treated as a mixture of ideal Boltzmann gases that are advected with the flow and can evolve due to nuclear reactions described by an $\alpha$-chain reaction network \citep{kifonidis2003, wongwathanarat2013} at temperatures $T < 8 \cdot 10^9$\,K, where we assume NSE not to hold. The nuclear composition is consistently coupled with the EOS and thus with the hydrodynamic medium. As we will show in Section~\ref{sec:earlyexplosion}, our results for the explosive production of $^{56}$Ni with our small $\alpha$-network do not sensitively depend on the chosen value of the NSE threshold temperature.

The use of the small $\alpha$-network allows us to track the evolution of more than a dozen nuclear species in the $\alpha$-chain from the progenitors into the explosions and to determine their abundance changes, radial mixing, and anisotropic distributions along with the hydrodynamic 3D supernova simulations. Of course, quantitatively more informative calculations also of the abundances of odd-$Z$ elements, isotopes, intermediate-mass species, and nuclei of the iron group (including $^{56}$Ni) and beyond require the post-processing of our 3D explosion models with tracer particles and big nucleosynthesis networks. This enterprise is beyond the scope of the present work.

When the density (temperature) in a zone drops below $\unit[10^{-10}]{g \cdot cm^{-3}}$ ($\unit[10^4]{K}$), which is the lowest value given in the \ac*{eos} table of \citet{timmes2000}, we switch to a simpler \ac*{eos} taking into account only a set of ideal Boltzmann gases and blackbody radiation.
The pressure $p$ and the internal energy $e$ per unit mass are given by

\begin{equation}
    p = \frac{1}{3} a T^4 + \frac{k_B}{\mu m_H} \rho T\text{,}
\end{equation}\label{eq:Boltzp}

\begin{equation}
    e = \frac{a T^4}{\rho} + \frac{3}{2} \frac{k_B}{\mu m_H} T\text{,}
\end{equation}\label{eq:Boltze}
where $a$ is the radiation constant, $\rho$ is the density, $T$ is the temperature, $k_\mathrm{B}$ is Boltzmann's constant, $\mu$ is the mean molecular weight, and $m_\mathrm{H}$ is the atomic mass unit.

For the early stages of the explosion, $t \le \unit[2.5]{s}$, when neutrinos are treated by the grey ray-by-ray approximation of \citet{scheck2006}, we follow \citet{wongwathanarat2013, wongwathanarat2015} in the parametrization of the neutrino luminosities and mean energies imposed at the inner grid boundary. The luminosities $L_0^\mathrm{tot}$ are set to remain constant during $\unit[1]{s}$ and then to decay exponentially with a timescale of 1\,s. Their initial values are specified such that our simulations reproduce roughly the 1D explosion energies reported by \citet{sukhbold2016} and are given in Table~\ref{tab:3Dexplosion}.

\begin{table*}
    \centering
    \setlength{\tabcolsep}{5pt}
    \begin{tabular}{c c c c c c c c c c c c c c c c c c}
    \hline
        Model & Properties & Reference & Model &$L^\mathrm{tot}_0$ & $E^\mathrm{1D}_\mathrm{exp}$ & $E^\mathrm{3D}_\mathrm{exp}$ & $t_\mathrm{exp}$ & $M_\mathrm{NiCoFe0.5X}$ & $\Mns$  & $\Vns$  & $\Jns^{46}$ & $\alpha_\mathrm{sk}$ & $T_\mathrm{spin}$ & $A_1$ & $A_2$ \\
        Class && Model & Name &[B s$^{-1}$] & [B] & [B] & [ms] & [$\unit{\Msun}$] & [$\unit{\Msun}$] & [\mbox{km s$^{-1}$}] &   & [${}^{\circ}$] & [s] & & \\
        \hline
        \multirow{3}{*}{LM}& Low & \multirow{3}{*}{SW13.1} & WH12.5 & 77 & 0.74 & 0.70 & 346 & 0.061 & 1.40 & 287.54 & 4.27  & 86.5  & 0.23 & 0.30 & 0.24 \\
        & $\Mzams$ & & SW13.1 & 77 & 0.68 & 0.69 & 358 & 0.059 & 1.43 & 258.87 & 6.24  & 127.5 & 0.16 & 0.32 & 0.27 \\
        & & & SW14.2 & 77 & 0.78 & 0.76 & 407 & 0.066 & 1.52 & 632.57 & 6.41  & 37.7  & 0.17 & 0.61 & 0.25 \\
        \hline
        \multirow{4}{*}{HM-LE}& Low & \multirow{4}{*}{SW18.2} & SW16.3 & 117 & 1.15 & 1.26 & 215 & 0.097 & 1.35 & 311.67 & 4.54  & 126.7 & 0.21 & 0.18 & 0.15 \\
        &  $E^\mathrm{3D}_\mathrm{exp}$  & & SW18.2 & 111 & 1.24 & 1.25 & 344 & 0.107 & 1.59 & 499.02 & 10.30 & 93.3 & 0.11  & 0.35 & 0.22 \\
        & High  & & SW20.8 & 77 & 0.83 & 0.83 & 382 & 0.076 & 1.47 & 680.60 & 9.09  &  49.7 & 0.12 & 0.52 & 0.36 \\
        & $\Mzams$  & & SW21.0 & 95 & 1.13 & 1.01 & 201 & 0.076 & 1.35 & 109.13 & 5.19  & 113.7 & 0.19 & 0.27 & 0.27 \\
        \hline
       \multirow{6}{*}{HM-HE}&  & \multirow{6}{*}{SW26.2} & SW19.8 & 131 & 1.67 & 1.73 & 209 & 0.077 & 1.36 & 58.25  & 7.34  & 103.9 & 0.13 & 0.04 & 0.06 \\
       & High & & SW25.5 & 131 & 1.30 & 1.45 & 274 & 0.107 & 1.60 & 84.64  & 9.29  & 74.3 & 1.05 & 0.13 & 0.27 \\
       &  $E^\mathrm{3D}_\mathrm{exp}$ & & SW25.6 & 119 & 1.28 & 1.31 & 270 & 0.097 & 1.56 & 395.55 & 0.32  & 78.4  & 3.50 & 0.26 & 0.34 \\
       &  & & SW26.2 & 107 & 1.28 & 1.23 & 245 & 0.100 & 1.53 & 216.34 & 2.63  & 45.0  & 0.42 & 0.26 & 0.19 \\
       & High& & SW27.0 & 125 & 1.26 & 1.42 & 233 & 0.110 & 1.46 & 348.20 & 7.92  & 129.7 & 0.13 & 0.20 & 0.18 \\
       &  $\Mzams$ & & SW27.3 & 119 & 1.44 & 1.41 & 240 & 0.109 & 1.48 & 618.51 & 14.42 & 45.2  & 0.07 & 0.41 & 0.14 \\
        \hline
    \end{tabular}
    \caption{Properties of the explosions and of the \acp*{pns} in the 3D models at $t_\mathrm{map} = \unit[2.5]{s}$ as defined in Section~\ref{sec:early_times}. $L_0^\mathrm{tot}$ is the assumed total initial neutrino luminosity at the inner grid boundary, distributed as 25\% for $\nu_\mathrm{e}$, 15\% for $\bar{\nu}_\mathrm{e}$, and 60\% for $\nu_\mathrm{x}$, where $\nu_\mathrm{x}$ represents the combined contributions from all heavy-lepton neutrino species.
    $E^\mathrm{1D}_\mathrm{exp}$ is the explosion energy in the corresponding 1D models.
    $E^\mathrm{3D}_\mathrm{exp}$ and $t_\mathrm{exp}$ are the explosion energy and time in 3D.
    $M_\mathrm{NiCoFe0.5X}$ is our best estimate for the total yield of $\Ni$ computed as the sum of the mass produced by the alpha network as $\Ni$ plus half of the mass created as tracer species $\X$. $\Mns$ is the baryonic \ac*{pns} mass in $\unit{\Msun}$. $\Vns$ is the hydrodynamic kick velocity of the \ac*{pns}. $\Jns^{46}$ is the angular momentum of the \ac*{pns} in units of $10^{46}$\,g\,cm$^2$s$^{-1}$.
    $\alpha_\mathrm{sk}$ is the angle obtained between the direction of $\Vns$ and $\Jns$.
    $T_\mathrm{spin}$ is the spin period of the final \ac*{ns}, calculated by considering the \ac*{ns} as a homogeneous sphere of mass $\Mns$ and radius $\unit[12]{km}$ with angular momentum $\Jns$. 
    $A_1$ and $A_2$ are the dipole and quadrupole amplitudes of the distribution of the radial momentum of the entire NiCoFe0.5X mass at $t_\mathrm{map}$, normalized by the corresponding monopole. 
    The grouping of our models reflects the final efficiency of the radial element mixing in our long-term 3D simulations. Since the mixing is determined and discussed only later in Section~\ref{sec:long_times}, we sort our models into categories that are named according to progenitor and explosion properties, which are determined earlier. Thus we introduce a low-ZAMS-mass category LM, which includes low-mass progenitors with typically low explosion energies and the strongest radial mixing. Our second group consists of higher-ZAMS-mass, low-energy (HM-LE) cases, which includes higher-mass representatives of our investigated models with lower explosion energies and less extreme element mixing than the LM models. A third category, named HM-HE, contains the higher-energy explosions of high-mass progenitors with the least extreme radial mixing.
}
    \label{tab:3Dexplosion}
\end{table*}

After remapping our simulations at $t_\mathrm{map}=\unit[2.5]{s}$, we account for the ongoing neutrino energy deposition into the ejecta in terms of a neutrino-driven wind. 
The latter is described as a spherically symmetric baryonic mass flow injected through the inner grid boundary with a constant velocity $v_\mathrm{w}$ and a power-law decay of density, specific internal energy, and pressure \citep{wongwathanarat2015}. 
The time-dependent density $\rho_\mathrm{w}(t)$, specific internal energy $e_\mathrm{w}(t)$, and pressure $p_\mathrm{w}(t)$ are given by
\begin{equation}
    \rho_\mathrm{w} (t) = \rho_\mathrm{w} (t_\mathrm{map}) \left(\frac{t}{t_\mathrm{map}}\right)^{-7/2}\text{,}
    \label{eq:rho_wind}
\end{equation}
\begin{equation}
    e_\mathrm{w} (t) = e_\mathrm{w} (t_\mathrm{map}) \left(\frac{t}{t_\mathrm{map}}\right)^{-7/6}\text{,}
    \label{eq:e_wind}
\end{equation}
and
\begin{equation}
    p_\mathrm{w} (t) = \frac{1}{3} e_\mathrm{w} (t) \rho_\mathrm{w} (t)\text{.}
    \label{eq:p_wind}
\end{equation}

The choice of the power-law indices in equations\,\eqref{eq:rho_wind} and \eqref{eq:e_wind} is motivated by extrapolation of the density and internal energy evolution at $r=\unit[500]{km}$. The values of $\rho_\mathrm{w}(t_\mathrm{map})$ and $e_\mathrm{w}(t_\mathrm{map})$ at mapping time $t_\mathrm{map}$ are model-dependent. This neutrino-wind boundary condition is applied until fallback sets in and the matter in the innermost radial zone moves inward in all cells. At this point, the inner boundary treatment is switched to a free outflow condition. The usage of the neutrino-driven-wind boundary condition replaces the computationally expensive tracking of the PNS evolution at $t > t_\mathrm{map} = 2.5$\,s in our 3D simulations. It is certainly less accurate than the most-modern, fully self-consistent, long-term 3D simulations that follow the neutrino transport in the PNS over periods of several seconds \citep[e.g.,][]{bollig2021,burrows+2024,burrows+2024a,janka2024,janka2025,vartanyan+2025,vartanyan2025b}. This difference, however, should only affect the innermost ejecta, the onset of the fallback, and the final PNS mass, but it should not have any substantial influence on our discussion of nonradial mixing by RTI at the composition-shell interfaces and the corresponding asymmetric evolution of SN explosions over long periods of time beyond the shock breakout from the stellar surface.

Outside of the progenitor at $r > R_\mathrm{prog}$  
we consider a spherically symmetric, \ac*{csm} given by
\begin{align}
    \rho_\mathrm{e}(r) = \max \Bigg\{& 
        \unit[10^{-12}]{g\,cm^{-3}} \left( \frac{\Rpro}{r} \right)^4,\, \rho_0 \left( \frac{1000\,\Rsun}{r} \right)^2, \notag \\
        & \unit[10^{-25}]{g\,cm^{-3}} 
    \Bigg\}\,,
    \label{eq:rho_e}
\end{align}
where $\rho_0 = 1.46 \times \unit[10^{-13}]{g \cdot cm^{-3}}$ for model WH12.5, and $\rho_0 = 1.14 \times \unit[10^{-15}]{g \cdot cm^{-3}}$ for the other models. The same values were used in \cite{sukhbold2016}. We have introduced the first term with the $r^{-4}$ dependence to avoid numerical problems at too steep density gradients at the stellar surface. For typical wind velocities of $20\,\mathrm{km\,s}^{-1}$, these values correspond to mass-loss rates of $2.82\times\unit[10^{-4}]{\Msun \mathrm{yr}^{-1}}$ and $2.19\times\unit[10^{-6}]{\Msun \mathrm{yr}^{-1}}$, respectively. In the simulations we neglect the small velocities of the stellar wind in the boundary conditions and set the velocity of the CSM to zero.

The temperature of the surrounding medium is chosen to have a constant temperature of $T_\mathrm{e} = \unit[1.0]{K}$ for model WH12.5, whereas it is given by the relation
\begin{equation}
    T_\mathrm{e} (r) = T_0 \left(\frac{\Rpro}{r}\right)^3\text{,}
    \label{eq:T_e}
\end{equation}
with $T_0 = \unit[275]{K}$ for the other models. These conditions are not realistic for stellar winds of \acp*{rsg}, and progenitors of different masses are expected to have different and possibly time-dependent and non-spherical winds and environments \citep[e.g.,][]{Bronner+2025, Ma+2025, Laplace+2026}. Since we are not interested in radiative phenomena around SN shock breakout, but only in the ejecta asymmetries and mixing instabilities inside the exploding stars, the detailed conditions in the CSM do not matter for our simulations. The gas is placed around the stars mainly for numerical reasons to permit us extending our simulations beyond the point of shock breakout from the stellar surface. Furthermore, as long as the CSM density remains sufficiently low, its specific structure does not significantly affect the simulation results on timescales of some years (the evolution periods we focus on here are roughly 10~days), which are much shorter than the transition time from the ejecta-dominated stage to the Sedov-Taylor phase of \acp*{snr} \citep{McKee+1995}.
In contrast, higher CSM densities can lead to stronger deceleration of the ejecta that catch up with the forward shock, potentially resulting in more pronounced Rayleigh-Taylor mushroom-like structures. To minimize the impact of the chosen CSM configuration on our results and to focus on the effects of the stellar structure on the RTIs, we adopt a low-density environment. 

As is \citet{stockinger2020} and \citet{gabler2021}, we allow for radioactive $\beta$ decay of $\Ni$ through $\Co$ to $\Fe$. The released decay energy is taken into account as described in \citet{gabler2021} and causes additional heating of the $\Ni$-rich ejecta. 
In addition to the decay chain of $\Ni$, the radioactive decay of $\Ti$ to $\isotope[44]{Sc}$ and then to $\isotope[44]{Ca}$ is implemented into \textsc{Prometheus-HotB}, but it is irrelevant on the timescales studied in this paper.

\section{Early-time evolution: explosion and PNS properties}
\label{sec:early_times}

In our study of the 3D mixing during the long-term evolution of CCSN explosions, it turned out that the 13 investigated models can be grouped into three different classes (see Table~\ref{tab:3Dexplosion}), depending on the strength of the radial mixing of chemical elements and the stellar and explosion properties that determine the mixing. Since we will analyse the mixing behaviour and the underlying processes of one reference model for each class in detail only in Section~\ref{sec:long_times}, we sort our models into categories that are named according to progenitor and explosion properties that are determined earlier. In the present section, these three categories of models and their representative cases are introduced, and our results for the early phase of the explosions until $t\leq t_\mathrm{map} = \unit[2.5]{s}$ after bounce are discussed.

\subsection{Model categories}
\label{sec:modelgroups}

The first class is formed by models with the lowest investigated \ac*{zams} masses (Low $M_\mathrm{ZAMS}$, LM) including WH12.5, SW13.1, and SW14.2. As we will show later, these models are characterized by strong mixing and the presence of two phases of \ac*{rti}. The corresponding growth factors of the instabilities are highest at the locations of the (C+O)/He and He/H composition interfaces. Our reference model of this group is model SW13.1. 

All the other models have higher \ac*{zams} masses, $M_\mathrm{ZAMS} > \unit[16]{\Msun}$ (High $M_\mathrm{ZAMS}$, HM) and only one dominant phase of \ac*{rti}. In contrast to the LM models, the growth factors of the instability at the (C+O)/He interface are small or vanishing, and high growth factors are present only at the location of the He/H composition interface. We split this HM group further into models with lower explosion energies (HM-LE), for which the lighter elements like H and He obtain very high final velocities, and models with higher explosion energies (HM-HE), which develop tails of lower velocities for H and He.
Models SW16.3, SW18.2, SW20.8, and SW21.0 belong to the HM-LE class, for which model SW18.2 is chosen as reference model.
The third group, HM-HE, consists of models SW19.8, SW25.5, SW25.6, SW26.2, SW27.0, and SW27.3, with SW26.2 as reference model. 

The grouping into the three classes and the reference model for each class are implemented in Table~\ref{tab:3Dexplosion}. Note that within these classes, model SW20.8 is a slight outlier with respect to its relatively low explosion energy of $E_\mathrm{exp}=\unit[0.83]{B}$, which is closer to the LM models than to the typical value in the group of HM-LE models. However, since the radial structure of the star and the corresponding mixing behaviour are more similar to those of the HM-LE models, we count model SW20.8 to the HM-LE group.

\subsection{The first 2.5\,s of the explosions}
\label{sec:earlyexplosion}

\begin{figure*}
    \centering
    \includegraphics[width=\linewidth]{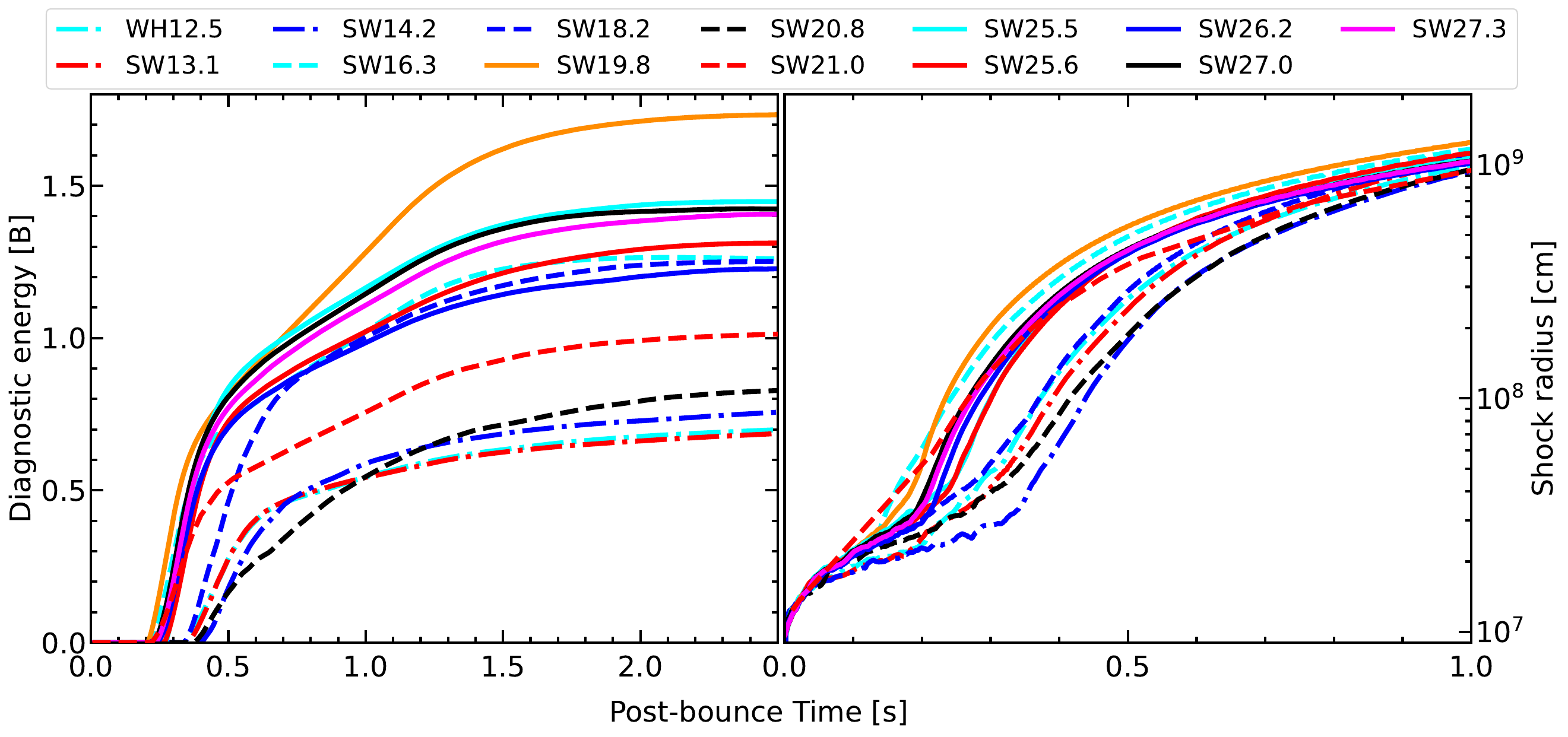}
    \caption{Evolution of the diagnostic (explosion) energies $E_\mathrm{exp}$ (left; as defined in the main text) and of the angle-averaged shock radii (right) for all computed 3D \ac*{sn} models versus post-bounce time. The evolution in the left panel is followed until the remapping time $t_\mathrm{map} = \unit[2.5]{s}$, after which the diagnostic energies have effectively reached their terminal values. In the right panel, the shock expansion is displayed only until 1\,s after bounce for better visibility.}
    \label{fig:exp}
\end{figure*}

In the left panel of Fig.\,\ref{fig:exp}, the time evolution of the diagnostic or ``explosion'' energies $E_\mathrm{exp}$ of all models is depicted. The energy
$E_\mathrm{exp}$ is defined as the sum of internal plus kinetic plus gravitational energy of all grid cells behind the shock where this total energy is positive, but it does not include the negative binding energy of the progenitor star ahead of the shock. At 2.5\,s after bounce the latter is very small and $E_\mathrm{exp}$ has effectively reached its terminal value as measurable for an observed \ac*{sn}.
Once this energy reaches $E_\mathrm{exp}\geq \unit[0.005]{B}$ ($\unit[1]{B} = 1\,\mathrm{bethe} \equiv \unit[10^{51}]{erg}$), the model is considered to explode and the corresponding instant defines the explosion time $t_\mathrm{exp}$.

Overall, all models show similar trends: once the explosion sets in, the energy grows very quickly to reach a plateau well before our mapping time $t_\mathrm{map}$.
For our less energetic models (WH12.5, SW13.1, SW14.2, and SW20.8) the initial phase of slow shock expansion lasts longer than for our more energetic models (Fig.\,\ref{fig:exp}, right panel). In these models the neutrino heating is weaker and it takes more time to build up positive energy behind the shock and to launch the explosion. 
This extended delay of the explosion creates the perfect conditions for the development of early-time hydrodynamic instabilities, such as Rayleigh-Taylor overturn, convection, and \ac*{sasi} \citep{foglizzo2006, scheck2008}.
The values of $t_\mathrm{exp}$ and of $E_\mathrm{exp}$ at 2.5\,s are given in Table~\ref{tab:3Dexplosion}. The explosion energies at this remapping time lie in the range between $\unit[0.69]{B}$ and $\unit[1.73]{B}$.

\begin{figure*}
    \centering
    \includegraphics[width=\linewidth]{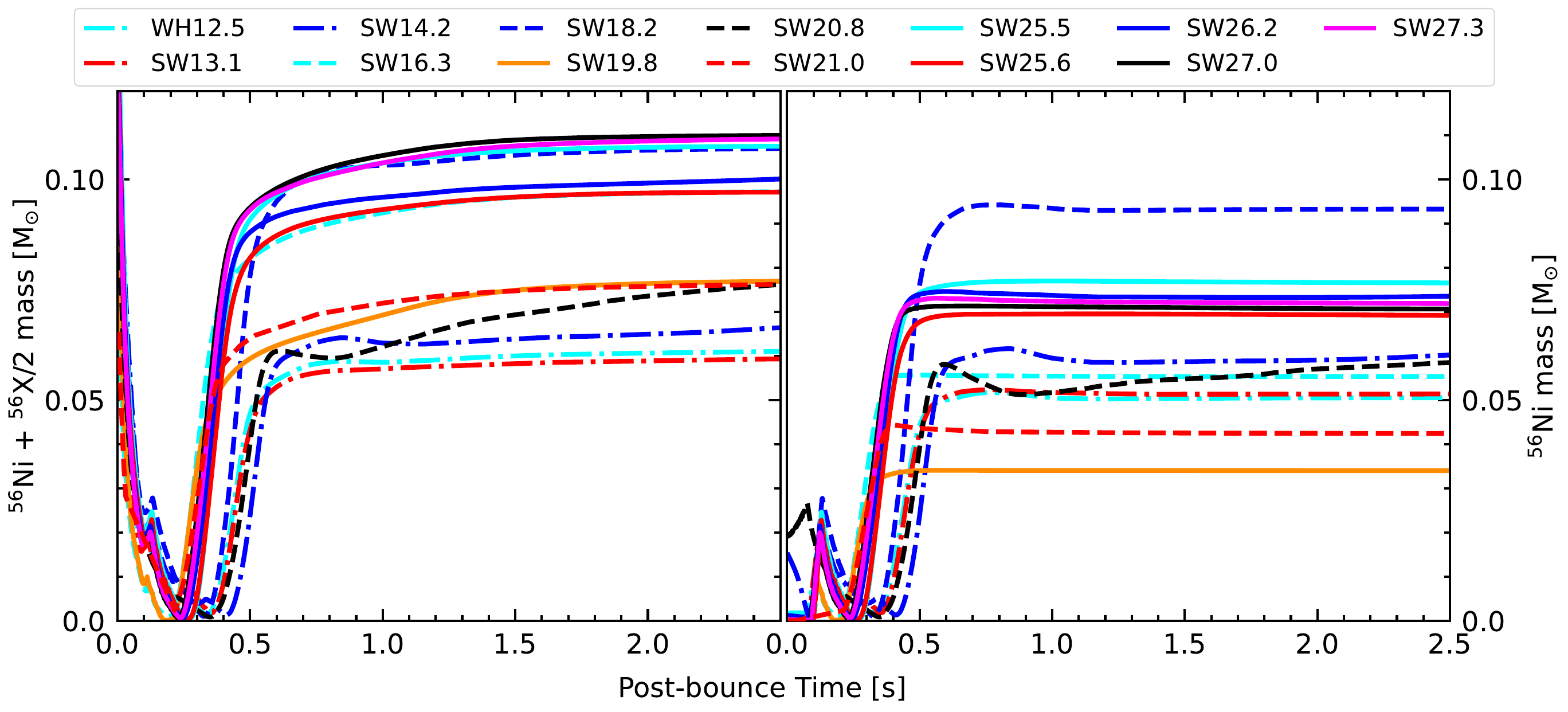}
    \caption{Evolution of the mass of $\Ni$ plus 50\% of tracer $\X$ (left) and only $\Ni$ (right) in our 3D \ac*{sn} models. While $\Ni$ is predominantly synthesized during the first 0.5 s and plateaus after that (right panel), minor amounts of $\X$ are produced for at least one second longer.}
    \label{fig:nimass}
\end{figure*}

\begin{figure*}
    \centering
    \includegraphics[width=\linewidth]{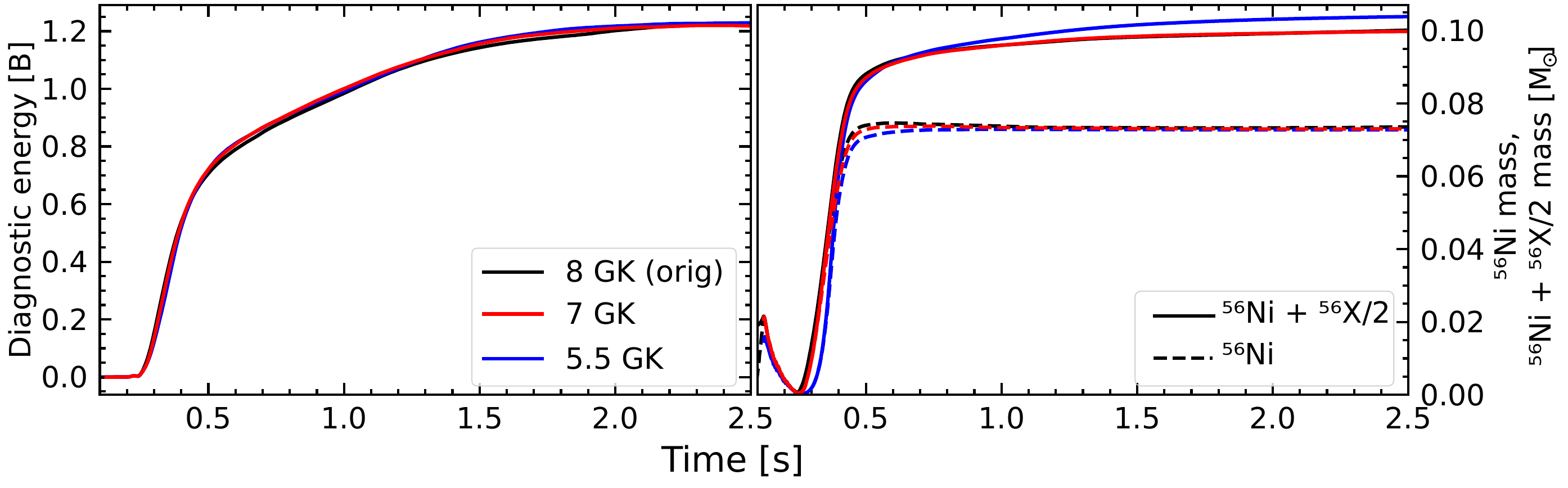}
    \caption{Diagnostic explosion energy (left) and masses of $^{56}$Ni as well as $^{56}$Ni plus $50\%$ of tracer $^{56}$X (right) as functions of time in model SW26.2 for different threshold temperatures where the NSE description for the nuclear composition is replaced by the use of our small $\alpha$-network, which is applied at temperatures below 8\,GK in our standard treatment. The final yields do not sensitively depend on the choice of the threshold temperature between 5.5\,GK and 8\,GK.}
    \label{fig:burning_thresh}
\end{figure*}
\begin{figure*}
    \centering
    \includegraphics[width=\linewidth]{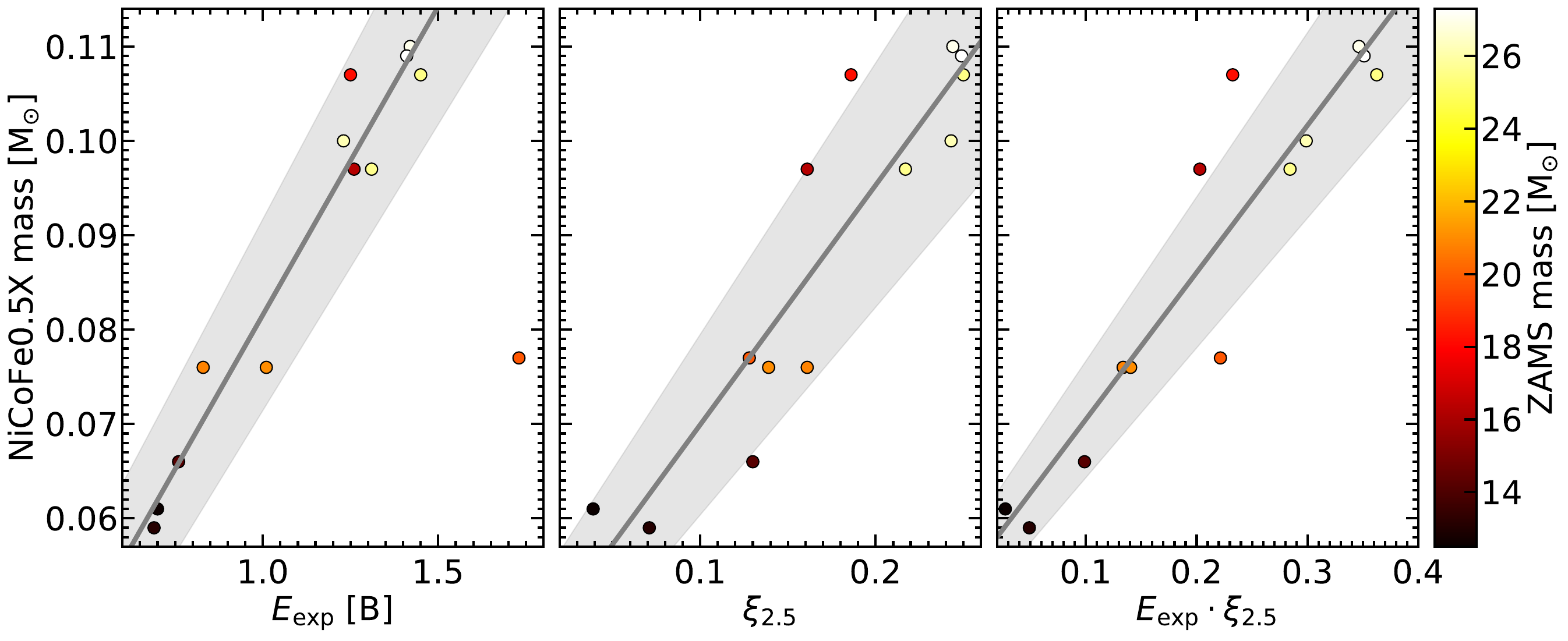}
    \caption{Correlations of the amount of produced NiCoFe0.5X at $t_\mathrm{map} = \unit[2.5]{s}$ with the (diagnostic) explosion energy $E_\mathrm{exp}$ (left panel), compactness $\xi_\mathrm{2.5}$ (middle panel), and the product of both (right panel). The grey lines denote linear fits of the points, and the grey-shaded bands are the 1$\sigma$ scatter regions. Note that the fit in the left panel does not include the outlier model SW19.8.}
    \label{fig:NiEexp}
\end{figure*}

As discussed in \cite{wongwathanarat2013} and \cite{gabler2021}, our simplified neutrino treatment leads to uncertainties in the electron fraction $Y_\mathrm{e}$ of the neutrino-heated ejecta and some of the matter is expelled with slightly neutron-rich conditions ($Y_\mathrm{e} \sim 0.46$--0.49). Therefore, a certain fraction of the matter that is assigned to the tracer nuclei $\X$ could be, in fact, $\Ni$. Depending on the model, values between $30\%$ and $70\%$ are expected to be reasonable. For simplicity and to be in line with previous literature \citep{sukhbold2016,utrobin2019}, we will assume a representative value of $50\%$. Therefore, when remapping our simulations at $t=t_\mathrm{map}$, we add half of the tracer material to $\Ni$. Consequently, 
from here on we will always refer to $\Ni$ plus $50\%$ of the tracer $\X$ as NiCoFe0.5X, which also includes the decay products $\Co$ and $\Fe$ .
The corresponding yields $M_\mathrm{NiCoFe0.5X}$ produced during the explosion are listed in Table~\ref{tab:3Dexplosion}. 
The different time evolutions of the pure $\Ni$ mass (right panel) and that of NiCoFe0.5X (left) are given in Fig.~\ref{fig:nimass}. Figure~\ref{fig:burning_thresh}, by using the example of model SW26.2, demonstrates that the final yields of these nuclei change by less than 5\% (mostly through the amount of tracer material $^{56}$X) when the NSE threshold temperature is varied between $8$\,GK and $5.5$\,GK. Such a change is negligible in the context of the present work.

Since the yields depend on how much mass can be heated to high temperatures either by neutrinos or by the outward moving \ac*{sn} shock, it is expected that the total mass of NiCoFe0.5X depends on the explosion energy and on the compactness of the models \citep[e.g.,][]{nakamura+2015,sukhbold2016,burrows+2024}. We plot the dependence of the NiCoFe0.5X mass on $E_\mathrm{exp}$ in the left panel of Fig.\,\ref{fig:NiEexp}. All models, except model SW19.8, follow a roughly linear correlation.\footnote{We point out that model SW19.8, similar to the 19.7, 20.2, 20.4, 20.5, 25.7, and 25.8 M$_\odot$ explosion models of \citet{sukhbold2016}, has a progenitor that experienced a carbon-oxygen shell merger \citep[see discussion in][]{bocciolo2026}, which reduced the density and lifted the entropy over a large radial domain. Therefore, all these progenitors possess relatively low compactness values and show earlier explosions and thus higher explosion energies than other models in the same ZAMS-mass range, yet relatively small $^{56}$Ni masses (in the KEPLER calculations) in \citet{sukhbold2016}. This trend is also present for SW19.8 in our 3D simulations.} When excluding this outlier model, the fit has a coefficient of determination $R_\mathrm{fit}^2 = 0.95$. When plotting the NiCoFe0.5X yield as function of compactness (central panel), we do not find any strong outlier, but the scatter increases ($R_\mathrm{fit}^2 = 0.84$). The combination of the two parameters into a product, $E_\mathrm{exp}\cdot \xi_\mathrm{2.5}$, as shown in the right panel of Fig.\,\ref{fig:NiEexp}, leads to a slightly lower scatter compared to the linear fit with $\xi_\mathrm{2.5}$ alone $(R_\mathrm{fit}^2 = 0.87)$. 

Based on the evolution of the central region of our simulations until $t_\mathrm{map}$, we can estimate the properties of the \ac*{pns} at this time. 
The \ac*{pns} mass $\Mns$, defined by the baryonic mass above a density of $\unit[10^{11}]{g \cdot cm^{-3}}$ (which we consider as the location of the PNS surface), is listed in column~10 of Table~\ref{tab:3Dexplosion}.
{In contrast to our parametrically triggered neutrino-driven explosions, current fully self-consistent multi-dimensional simulations yield long-lasting accretion downflows to the PNS \citep[see, e.g., ][]{bollig2021,burrows+2024,janka2025}. On the one hand, this accretion increases the PNS mass over longer timescales compared to our models.
On the other hand, a significant fraction of this infalling matter is re-ejected after having absorbed energy from neutrinos and thus contributes to a long-term gradual growth of the explosion energy and synthesis of selected elements.
Such a behavior is not seen in our models because of strong initial neutrino heating and the use of a spherical neutrino-wind boundary after $t_\mathrm{map}$ (equations~\ref{eq:rho_wind}--\ref{eq:p_wind}). This difference in the flow dynamics, however, affects only the final PNS masses as a function of the progenitor mass and compactness (causing a tendency to slightly underestimated PNS masses) as well as a smaller fraction of the innermost ejecta. This is of secondary relevance for our discussion of 3D explosion asymmetries and the associated outward mixing of NiCoFe0.5X during the explosion.}

In order to compute the kinematic quantities of the \ac*{pns}, we follow \citet{scheck2006} and assume conservation of the linear and angular momenta. Since the progenitor star was at rest at the beginning of the simulation, we can estimate the linear momentum and the velocity of the \ac*{pns} from the interaction with the surrounding matter. While the \ac*{pns} is forced to stay at the origin, because the centre of the gravitational potential cannot move out of the centre of the numerical grid, the ejecta are allowed to evolve freely. Therefore, the kick velocity of the \ac*{pns} due to the asymmetric \ac*{sn} explosion can be computed from the negative of the total linear momentum of the ejecta as
\begin{equation}
    \VnsVector(t) = - \textbf{P}_\mathrm{gas}(t) / \Mns(t)\text{,}
    \label{eq:NSvelocity}
\end{equation}
where
\begin{equation}
    \textbf{P}_\mathrm{gas}(t) = \int_{\Rns}^{R_\mathrm{ob}} dV \rho \textbf{v}
    \label{eq:gasmomentum}
\end{equation}
is the integral of the gas momentum between the \ac*{pns} radius, $\Rns$, and the outer grid boundary, $R_\mathrm{ob}$. 
We obtain magnitudes of $\VnsVector$ between several tens of km\,s$^{-1}$ and almost $700\, \unit{km\,s^{-1}}$ (see Table~\ref{tab:3Dexplosion}, column~11).
These hydrodynamic kick velocities of the \acp*{pns} are compatible with those obtained in previous simulations \citep{wongwathanarat2013, muller2019, bollig2021, coleman2022, janka2022, janka2024, burrows+2024a}, and the PNS velocity directions point opposite to the main NiCoFe0.5X-rich ejecta as previously discussed by \citet{wongwathanarat2013,wongwathanarat2017,bollig2021} and more recently also reported by \citet{vartanyan2025b}. Because of our approximate treatment of the neutrino transport with spherical neutrino luminosities imposed at the inner grid boundary,\footnote{Note that our inner grid boundary is usually at $\nu_e$ optically depths of 10 and higher, and grey ray-by-ray neutrino transport is applied at lower optical depths in the outer PNS layers. This makes neutrino (accretion) emission and neutrino heating aspherical. However, our treatment does not fully include effects that originate from the convective layer in the deeper PNS interior and that also affect the neutrino-emission asymmetry, for example the lepton-number emission self-sustained asymmetry \citep[LESA;][]{Tamborra+2014}. This phenomenon can contribute to a neutrino-induced PNS kick, as discussed by \citet{janka2024}.} we do not evaluate our models for the \ac*{pns} kicks associated with anisotropic neutrino emission.

\begin{figure*}
    \centering
    \includegraphics[width=0.32\linewidth]{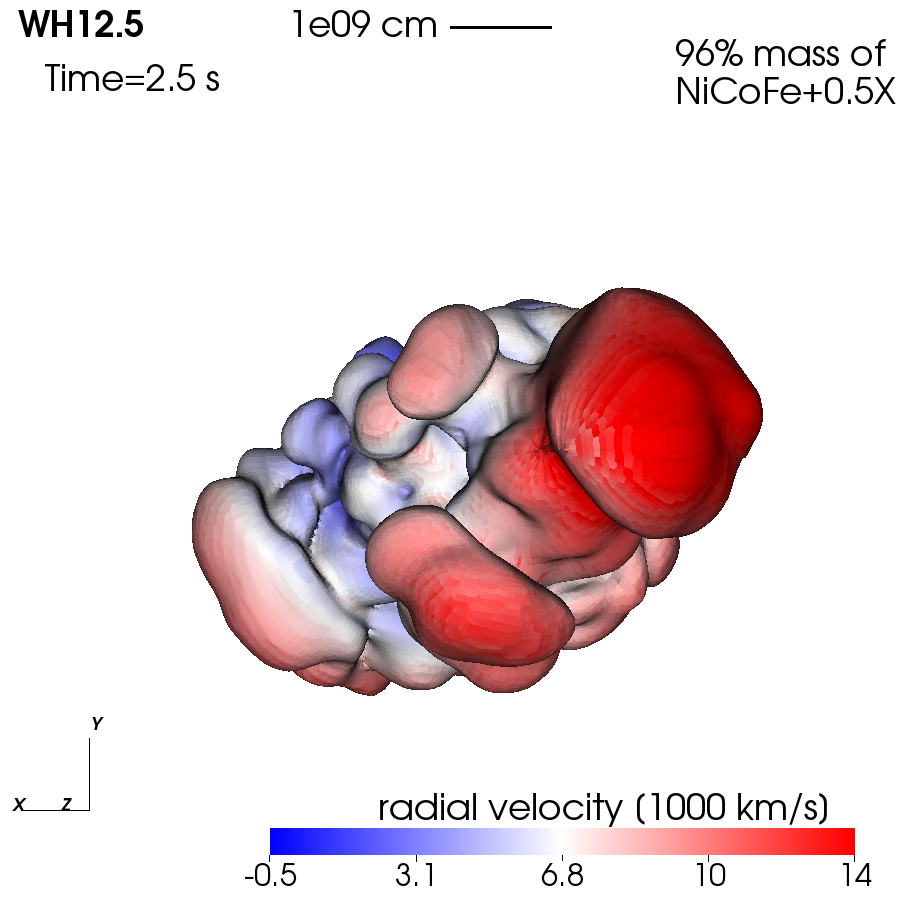}
    \includegraphics[width=0.32\linewidth]{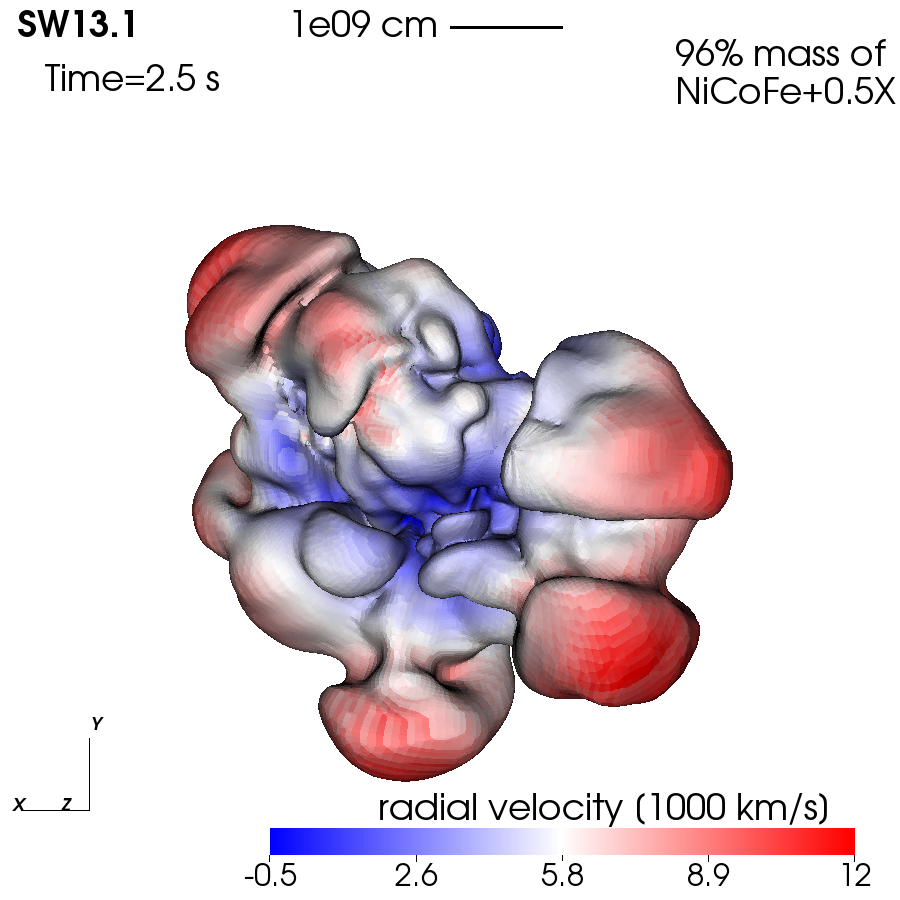}
    \includegraphics[width=0.32\linewidth]{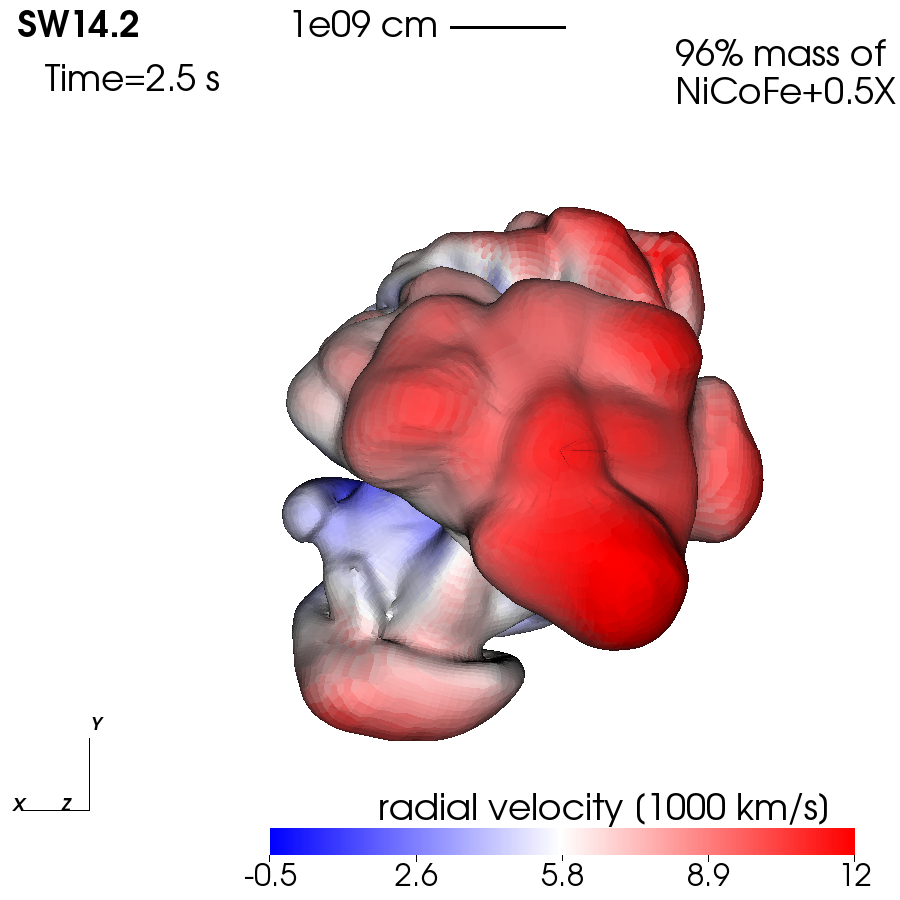}
    \caption{Isosurfaces of the NiCoFe0.5X mass fraction containing $96\%$ of the total mass of NiCoFe0.5X at $t = t_\mathrm{map} = \unit[2.5]{s}$, colour-coded with the radial velocity, for our three models in the LM category with its reference case SW13.1 (middle panel). The mass fraction is typically $X_\mathrm{NiCoFe0.5X} \approx 0.11$ for all the models.}
    \label{fig:ni-plumes_cat1}
\end{figure*}

\begin{figure*}
    \centering
    \includegraphics[width=0.40\linewidth]{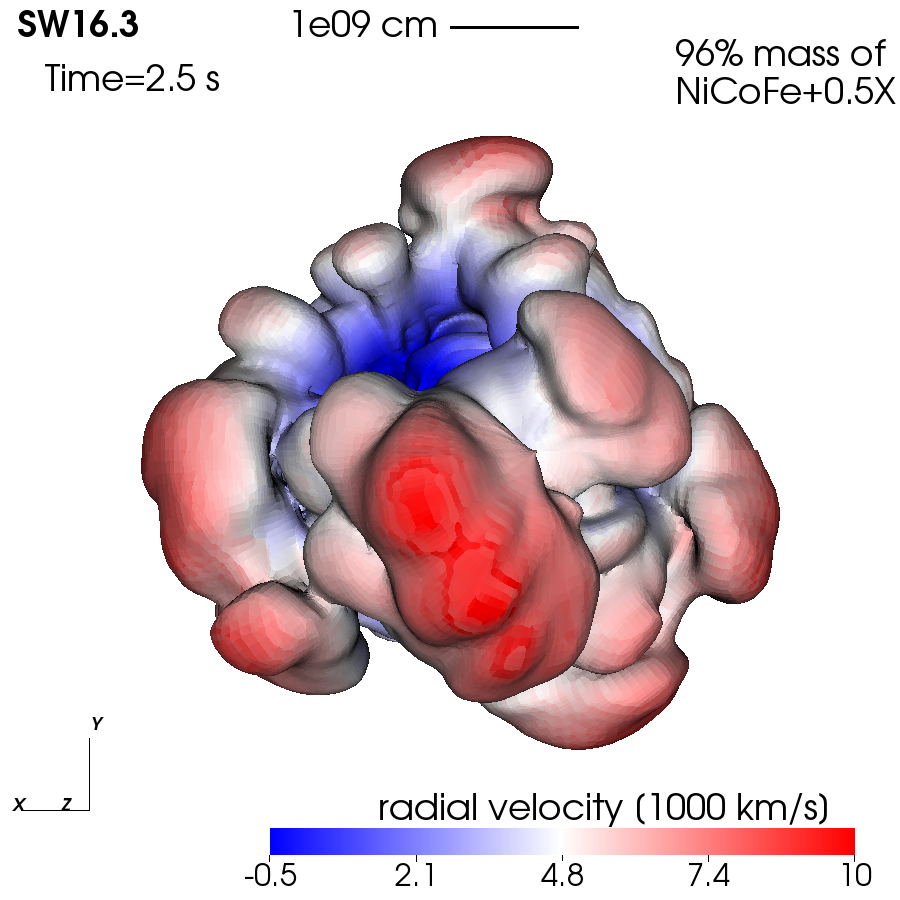}
    \includegraphics[width=0.40\linewidth]{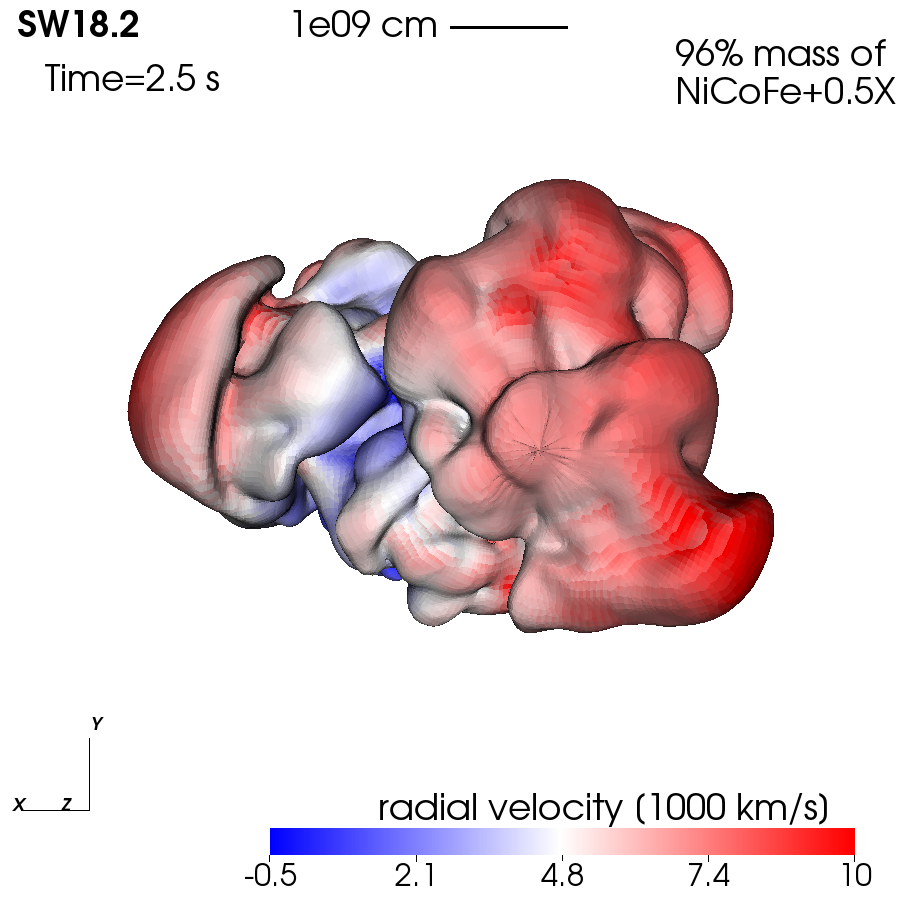}\\
    \includegraphics[width=0.40\linewidth]{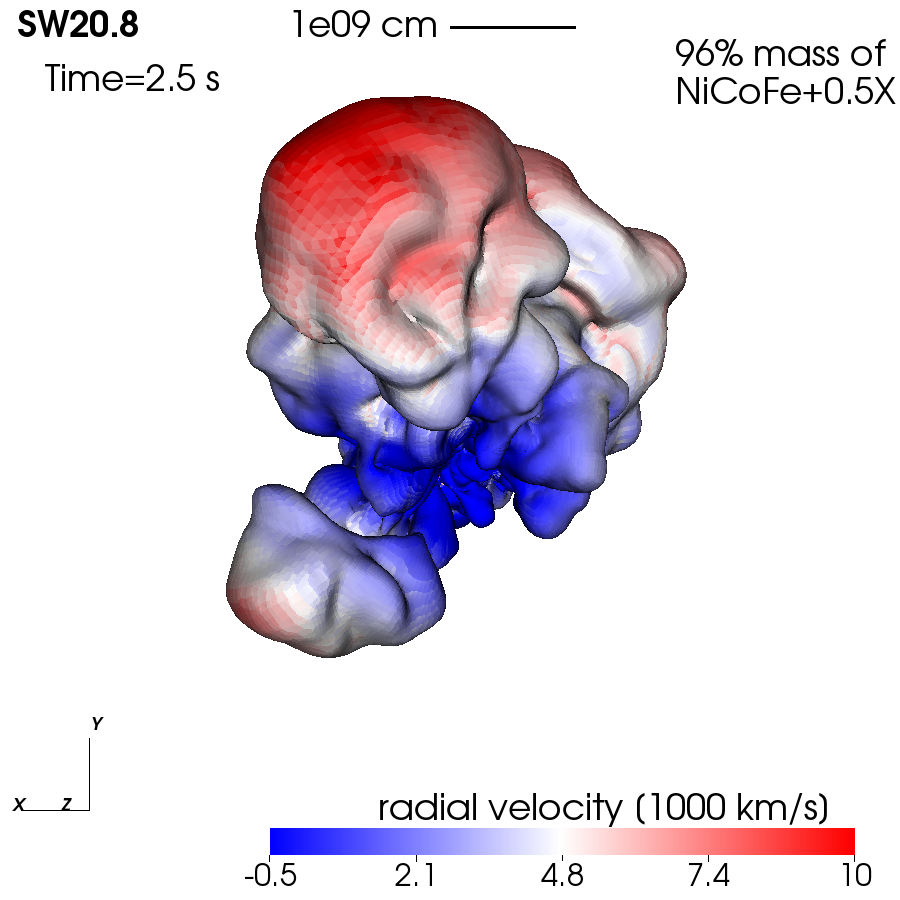}
    \includegraphics[width=0.40\linewidth]{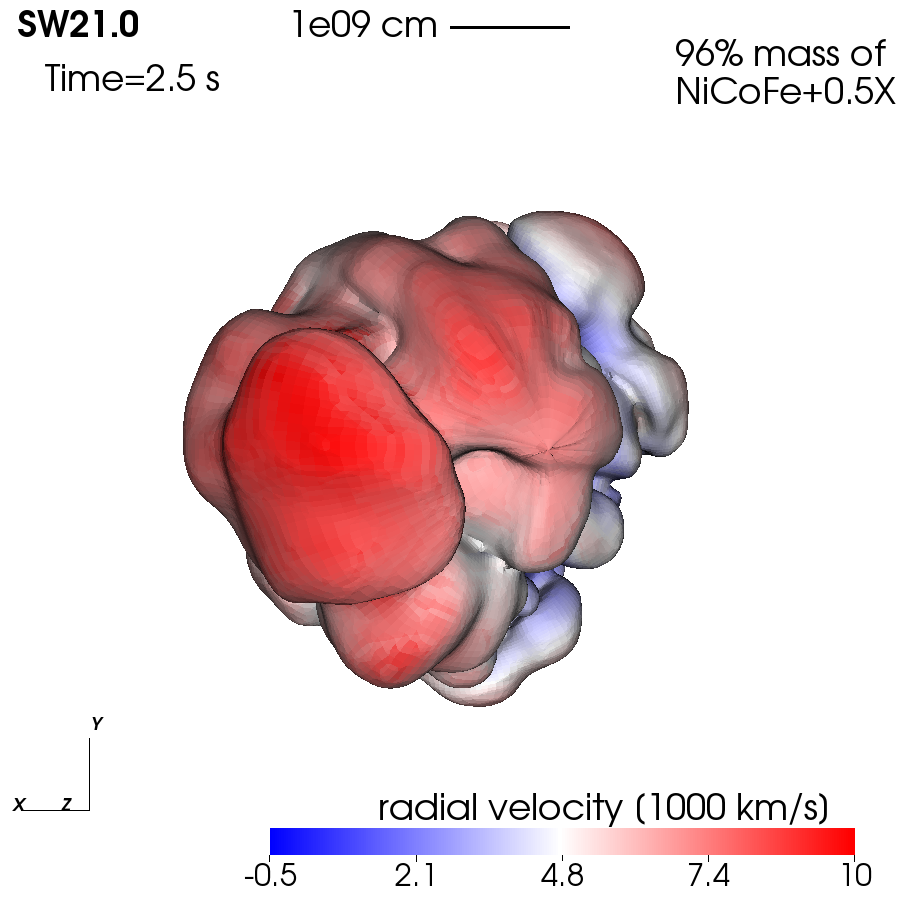}   
    \caption{Isosurfaces of the NiCoFe0.5X mass fraction containing $96\%$ of the total mass of NiCoFe0.5X at $t = t_\mathrm{map} = \unit[2.5]{s}$, colour-coded with the radial velocity, for our four models in the HM-LE category with its reference case SW18.2 (upper right panel). The mass fraction is typically $X_\mathrm{NiCoFe0.5X} \approx 0.11$ for all the models. 
    }
    \label{fig:ni-plumes_cat2}
\end{figure*}

\begin{figure*}
    \centering
    \includegraphics[width=0.32\linewidth]{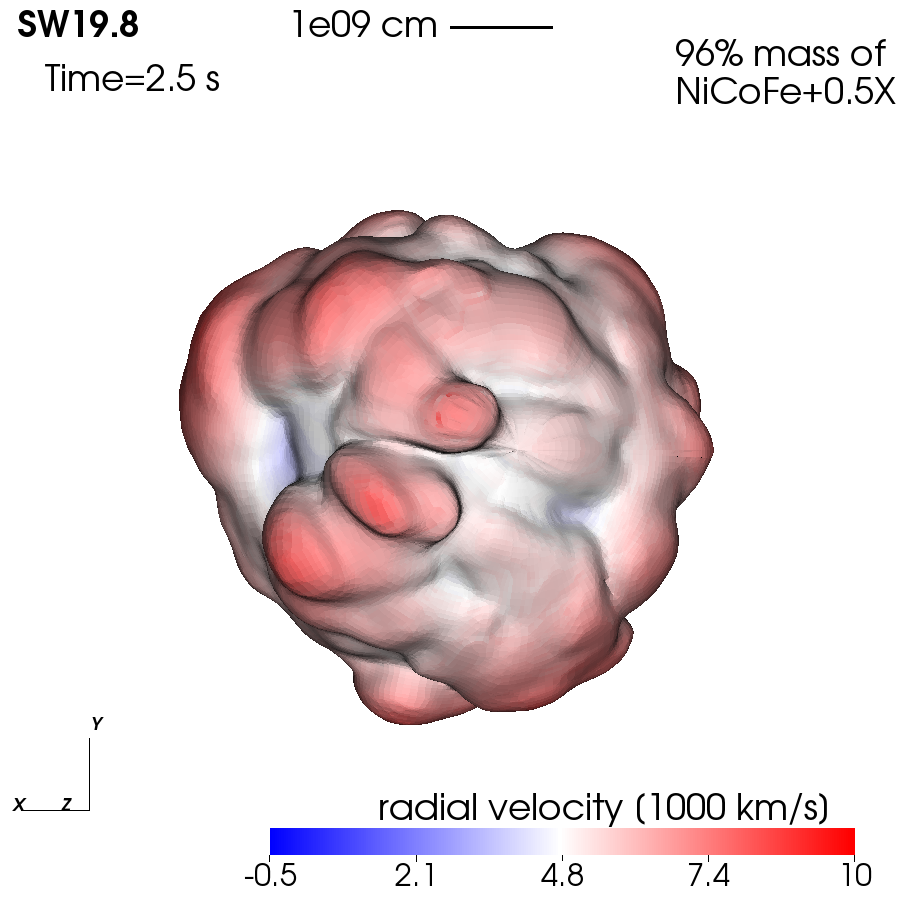}
    \includegraphics[width=0.32\linewidth]{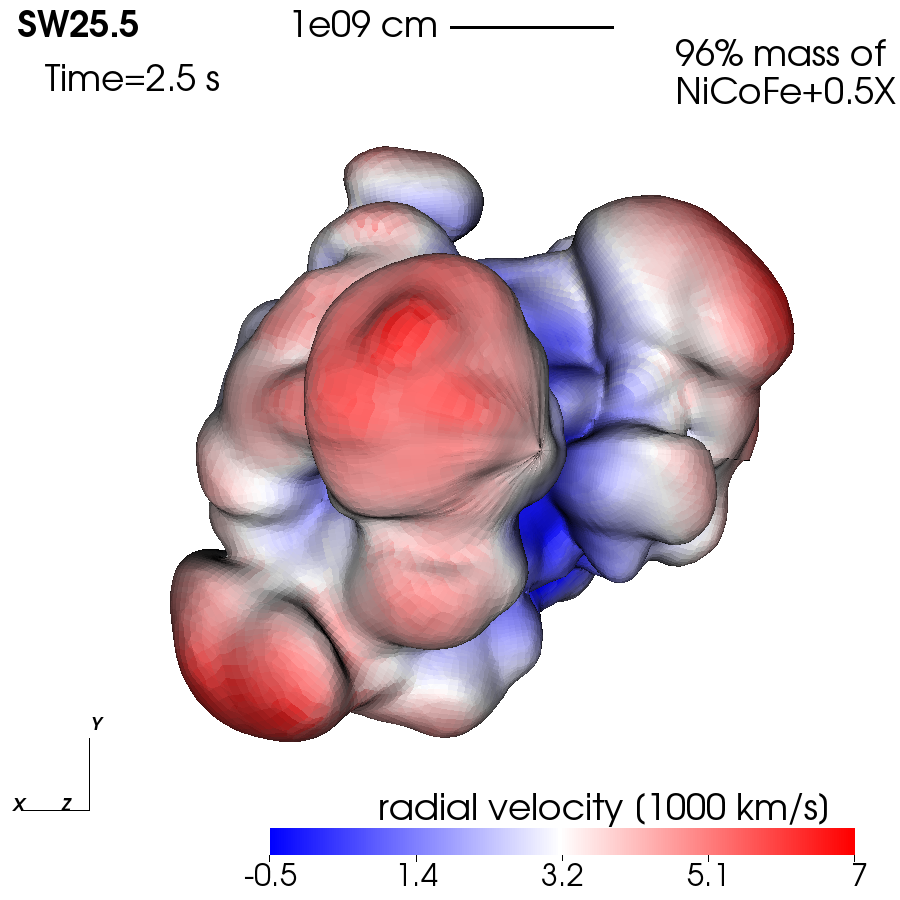}
    \includegraphics[width=0.32\linewidth]{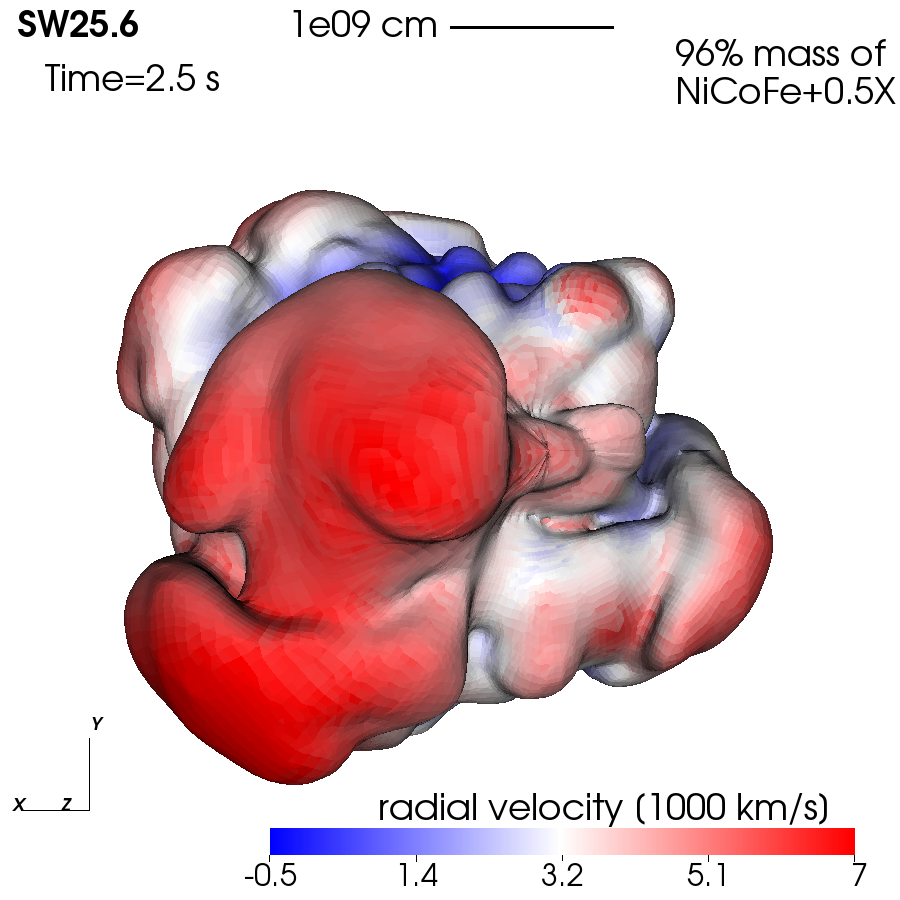}\\
    \includegraphics[width=0.32\linewidth]{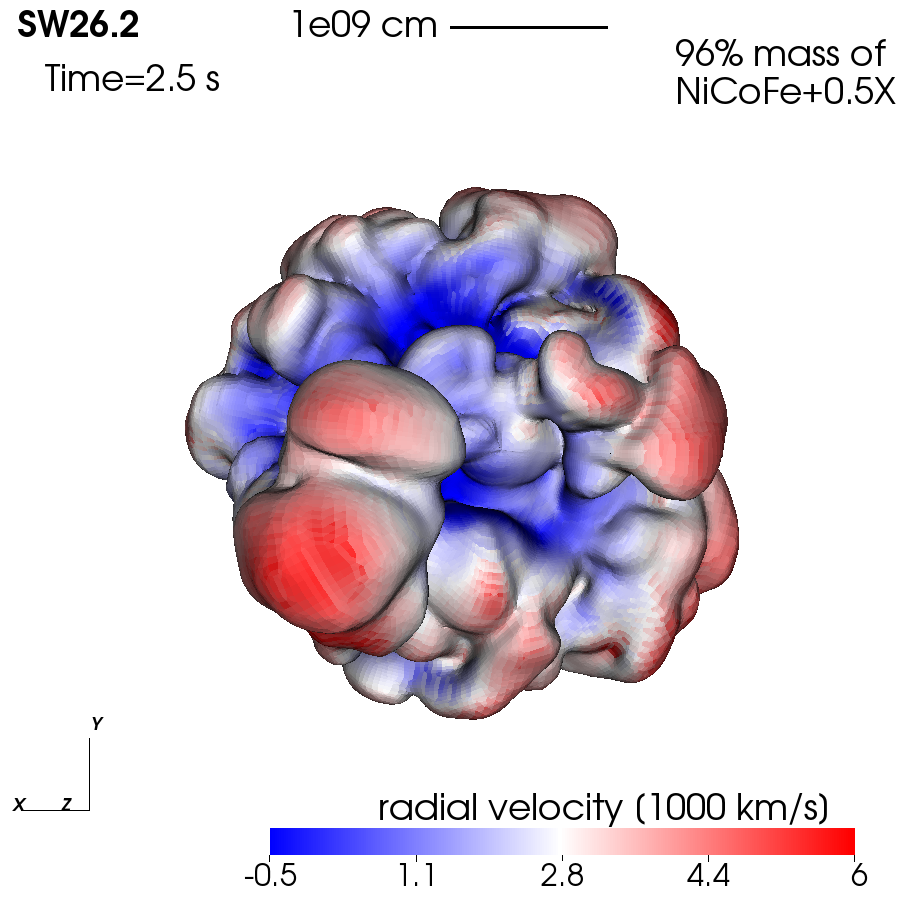}
    \includegraphics[width=0.32\linewidth]{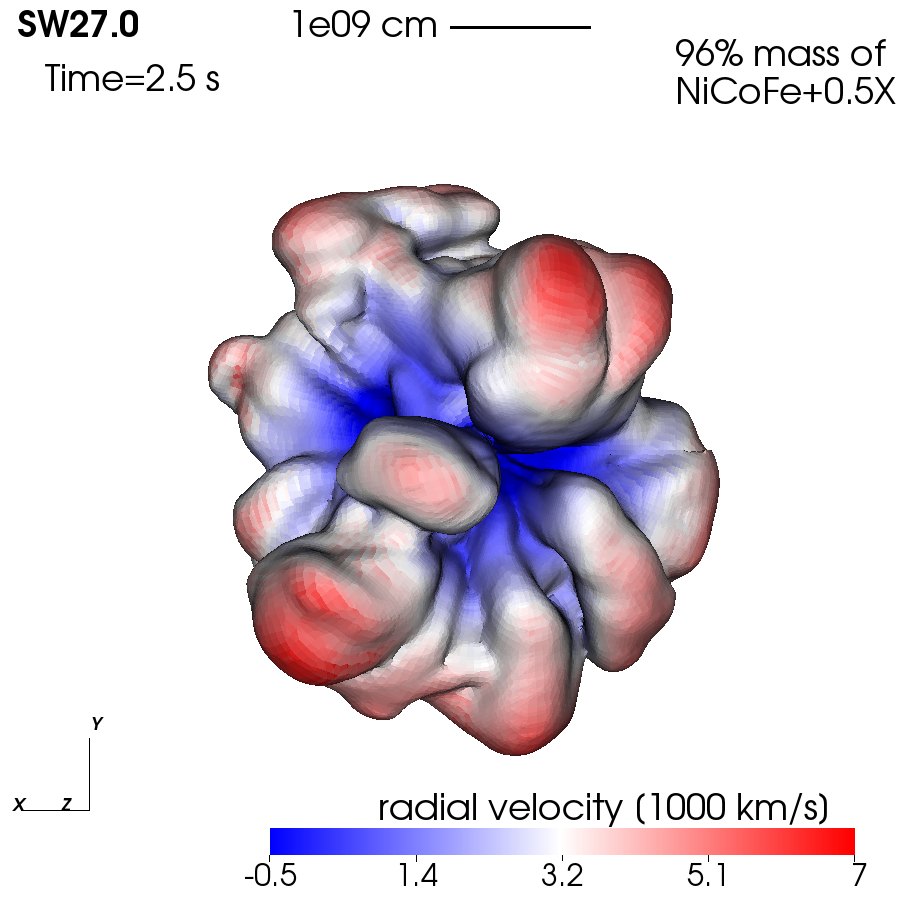}
    \includegraphics[width=0.32\linewidth]{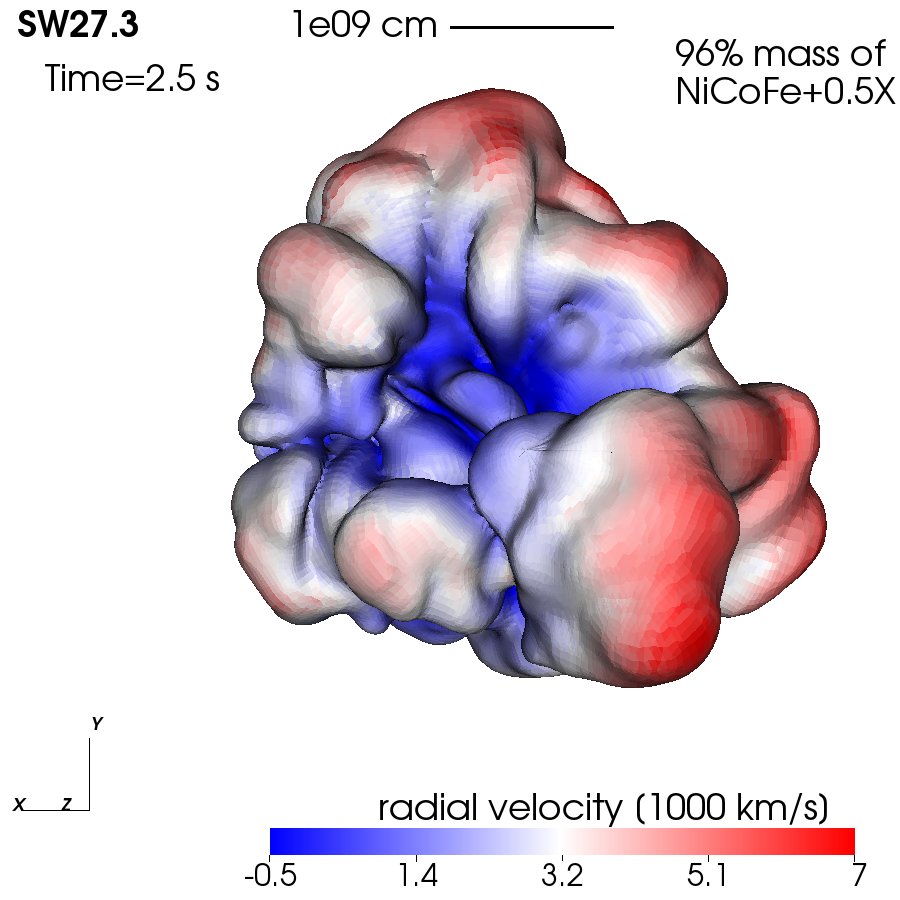}
    \caption{Isosurfaces of the NiCoFe0.5X mass fraction containing $96\%$ of the total mass of NiCoFe0.5X at $t = t_\mathrm{map} = \unit[2.5]{s}$, colour-coded with the radial velocity, for our six models in the HM-HE category with its reference case SW26.2 (lower left panel) and model SW19.8 with a shell-merger progenitor (upper left panel). The mass fraction is typically $X_\mathrm{NiCoFe0.5X} \approx 0.11$ for all the models.
    }
    \label{fig:ni-plumes_cat3}
\end{figure*}

For the computation of the angular momentum, we follow a similar procedure.
Since the progenitor models are non-rotating, any angular momentum acquired by the \ac*{pns} has to be balanced by an equal and opposite angular momentum of the ejected matter.
Deviating from the calculation of the kick velocity, where the integration is performed on the entire grid (equations~\ref{eq:NSvelocity} and~\ref{eq:gasmomentum}), numerical tests showed that it is recommended to constrain the volume of integration of the angular momentum to the immediate vicinity of the \ac*{pns} \citep{wongwathanarat2013}:
\begin{equation}\label{eq:NSangularMomentum}
    \JnsVector(t) = -\left[ \int_{R_\mathrm{NS}}^{r_\mathrm{o}} dV \rho \textbf{j}(t) + \int_0^t dt^{\prime} r_\mathrm{o}^2 \oint_{4\pi} d\Omega\,(\rho \textbf{j} v_r)|_{r_\mathrm{o}} \right] \text{,}
\end{equation}
where $\textbf{j}$ denotes the specific angular momentum, $\Omega$ is the solid angle, and $r_\mathrm{o}$ is chosen to be $\unit[500]{km}$.\footnote{As in \citet{wongwathanarat2013}, we confirmed that the numerical result of the evaluation of equation \eqref{eq:NSangularMomentum} does not depend sensitively on the exact value of $r_\mathrm{o}$, as long as it is in the range between $\unit[500]{km}< r_\mathrm{o}< \unit[1000]{km}$.}
The first term in equation \eqref{eq:NSangularMomentum} accounts for the angular momentum contained in the spherical shell bounded by the \ac*{pns} radius $\Rns$ on one side and by the radius $r_\mathrm{o}$ on the other. 
The second term in the bracket gives the angular momentum that is carried away by the mass leaving the volume of radius $r_\mathrm{o}$ until time $t$.
For all the models of our calculations, $\Jns$ stays roughly constant after about $\unit[2]{s}$ (see, however, \citealt{janka2022} for the possible effects of later fallback and accretion).
In addition to $\Jns$, Table~\ref{tab:3Dexplosion} also provides the relative angle between spin and kick directions, $\alpha_\mathrm{sk}$ (column~13), and the estimated final \ac*{ns} spin period $T_\mathrm{spin}$ (column~14). 
The latter is computed by considering the \ac*{ns} as a rigidly rotating, spherical body. The period is thus given by
\begin{equation}
    T_\mathrm{spin} = \frac{2 \pi I_\mathrm{NS}}{\Jns}\text{,}
\end{equation}
where $I_\mathrm{NS}$ is approximated as the Newtonian moment of inertia of a homogeneous sphere, with final radius of $\Rns = \unit[12]{km}$.
All but one of the models exhibit high angular momenta of $\Jns>\unit[10^{46}]{g \cdot cm^2\,s^{-1}}$. Consequently, the obtained spin periods are on the order of hundreds of milliseconds or below. 
The only deviation from this general behavior is model SW25.6, for which the PNS has a very low momentum ($J_\mathrm{PNS} = 0.32 \cdot \unit[10^{46}]{g \cdot cm^2\,s^{-1}}$) and, hence, a very long spin period ($T_\mathrm{spin} = \unit[3.5]{s}$). These values are also compatible with previous studies (see e.g. \citealt{janka2022}). We attribute the low angular momentum of the \ac*{pns} in model SW25.6 to stochastic effects connected to non-spherical accretion flows that carried angular momentum onto the PNS during and after the phase when violent and chaotic hydrodynamical instabilities enabled the revival of the stalled SN shock. Just by chance the net angular momentum acquired by the PNS in a series of accretion episodes through downflows from different directions was very low when the accretion finally stopped. Note that all of our progenitors had zero angular momentum prior to core collapse. 
The angles $\alpha_\mathrm{sk}$ between the directions of \ac*{pns} spins and kicks are between $37^{\circ}$ and $130^{\circ}$ and do not indicate any trend towards alignment or anti-alignment of the spin and kick directions.

\section{Long-term evolution: instabilities and mixing}
\label{sec:long_times}

In this section, we discuss the evolution of our 13 \ac*{sn} explosion models from $t=\unit[2.5]{s}$ post bounce until 10 days later. Before investigating in detail the radial mixing of the heavy elements created near the explosion centre into the outer layers of the star, we study the (linear) stability of the ejecta with respect to \ac*{rti} in 1D in Section~\ref{sec:RTIs_1d}. We further discuss the morphologies of our 3D simulations and analyse how the mixing can be related to properties of the explosion in Section~\ref{sec:linking}.

\subsection{Starting points of the long-term simulations}
\label{sec:startlongterm}

The morphology of the freshly synthesized, heavy-element ejecta at $t=2.5\,$s is characterized by big buoyant plumes that are created by convective overturn in the neutrino-heated post-shock layer at the onset of the explosion and shortly afterwards. This is shown for our 13 CCSN models in all three model groups in Figs.~\ref{fig:ni-plumes_cat1}--\ref{fig:ni-plumes_cat3}, sorted by category. At the displayed time, the explosion energies have reached their terminal values (see Fig.~\ref{fig:exp}), the $\Ni$ nucleosynthesis is effectively finished (Fig.~\ref{fig:nimass}), and large-scale asymmetries have formed.

Already at this early time, we witness by eye inspection differences in the morphologies. For example, among our LM models, WH12.5 and SW14.2 (Fig.~\ref{fig:ni-plumes_cat1}, left and right panels, respectively) seem to possess a larger dipole asymmetry of their NiCoFe0.5X distributions, and also models SW18.2, SW20.8, and SW21.0 of the HM-LE category (Fig.~\ref{fig:ni-plumes_cat2}, top right and bottom panels, respectively) and models SW25.5 and SW25.6 of the HM-HE group (Fig.~\ref{fig:ni-plumes_cat3}, top middle and right panels, respectively) appear to be more asymmetric with fewer, very prominent and more extended large plumes in preferred directions. In contrast, in particular models SW26.2 (Fig.~\ref{fig:ni-plumes_cat3}, bottom left) and the special case of SW19.8 with its shell-merger progenitor (Fig.~\ref{fig:ni-plumes_cat3}, top left) look more spherical with more isotropically distributed and smaller plumes, at least at this early time and viewed from the perspective displayed in Fig.~\ref{fig:ni-plumes_cat3}.\footnote{We stress that for progenitors with shell mergers, explosion asymmetries in terms of the nickel distribution and radial mixing can be expected to become much more extreme (especially compared to explosions of progenitors with similar masses but without shell mergers) when the 3D core-collapse simulation is started from a 3D pre-collapse model. This can be understood by the fact that oxygen-neon shell mergers in 3D lead to particularly violent convection with large-scale flow asymmetries in the extended O-Ne burning layer already prior to core collapse \citep{Yadav+2020}, thus boosting the growth of extremely strong convection in the neutrino-heated post-shock volume \citep{bollig2021}. All of the SN simulations of the present work, however, were started from 1D progenitors.} These visual asymmetries often, but not always, correlate with the PNS kicks in Table~\ref{tab:3Dexplosion}. Models SW14.2, SW18.2, SW20.8, and SW25.6, have very high kicks, but SW27.0 and SW27.3 with large kicks of nearly 350\,km\,s$^{-1}$ and more than 600\,km\,s$^{-1}$, respectively, do not look particularly asymmetric in their NiCoFe0.5X distributions, whereas WH12.5, SW21.0, and SW25.5 develop average ($\lesssim$\,300\,km\,s$^{-1}$) or even low PNS kicks ($\lesssim$\,100\,km\,s$^{-1}$) despite the asymmetric appearance of their NiCoFe0.5X distributions.

In order to better quantify the explosion asymmetries we performed a spherical harmonics decomposition of the radial momentum density, $f(r,\theta,\phi)=\rho\,X_\mathrm{NiCoFe0.5X}\,v_r$ (with $v_r$ being
the radial velocity), of the entire NiCoFe0.5X distribution at $t_\mathrm{map}$. For doing that we considered the complex spherical harmonics $Y_{\ell m}(\theta,\phi)$ ($\int Y_{\ell m}Y_{\ell'm'}^{*}\,d\Omega=\delta_{\ell\ell'}\delta_{mm'}$). The coefficients in each radial shell are
\begin{equation}
c_{\ell m}(r) \;=\; \int f(r,\theta,\phi)\,Y_{\ell m}^{*}(\theta,\phi)\,d\Omega \,,
\end{equation}
summed coherently over radius (as complex amplitudes, retaining phase, so contributions from different shells
may cancel) to give the total angular decomposition
\begin{equation}
C_{\ell m} \;=\; \sum_r c_{\ell m}(r)\,.
\end{equation}
We computed the rotation-invariant coherent asymmetry amplitudes
\begin{equation}
A_\ell \;=\; \frac{1}{\sqrt{2\ell+1}}\,\frac{\big(\sum_{m=-\ell}^{\ell}|C_{\ell m}|^2\big)^{1/2}}{|C_{00}|}\,,
\end{equation}
with $A_0=1$.  The corresponding non-normalized moments are $a_\ell=A_\ell\cdot a_0$, with
$a_0\equiv|C_{00}|\sqrt{4\pi}$ the monopole (net momentum), so that $a_\ell/a_0=A_\ell$ exactly.

\begin{figure}
\centering
\includegraphics[width=\linewidth]{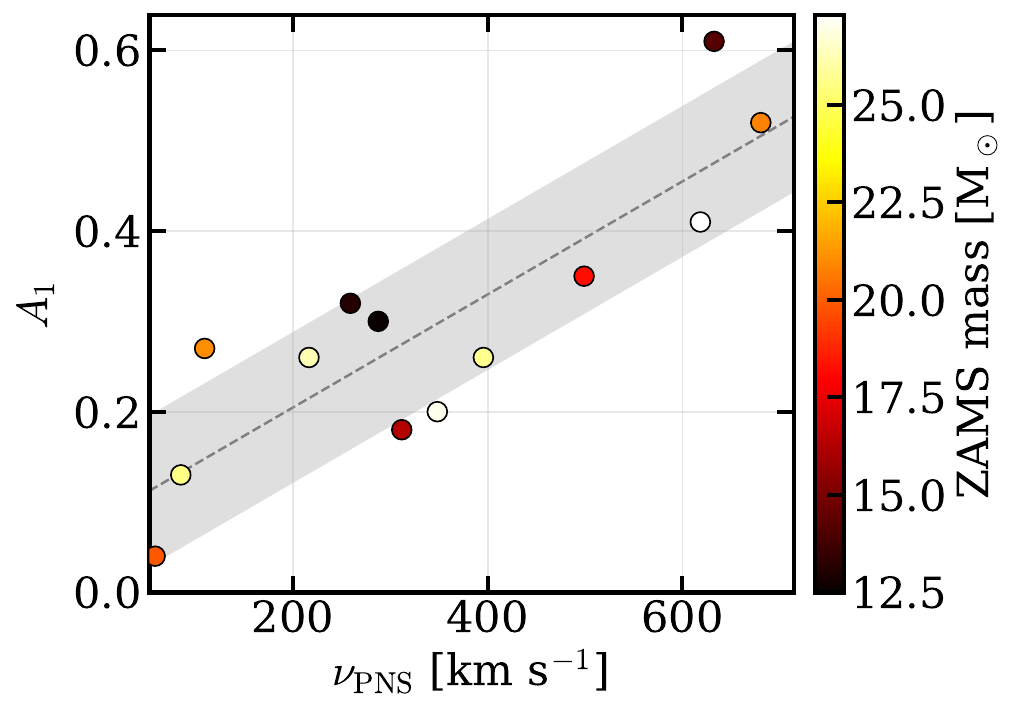}
\caption{Normalized dipole amplitude $A_1$ of the radial momentum of NiCoFe0.5X plotted against NS kick velocity $v_\mathrm{PNS}$ for our 13 models. The colors of the filled circles encode the ZAMS masses of the progenitors using the same color map as in Fig.~\ref{fig:NiEexp}. The dashed line indicates the linear least-squares fit and the grey shaded region is the constant $\pm 1\sigma$ band of the summed (RMS) fit residuals.}
\label{fig:A1_vns}
\end{figure}

The normalized dipole amplitudes $A_1$ of the NiCoFe0.5X distributions are listed in Table~\ref{tab:3Dexplosion} and are roughly correlated with the PNS kicks: Models SW14.2, SW18.2, SW20.8, SW25.6, and SW27.3 display high kick velocities as well as high values of $A_1$. This trend is also highlighted in Fig.~\ref{fig:A1_vns}, where we plot $A_1$ versus $\Vns$. The coefficient of determination is moderately strong $R^2=0.73$. We do not expect a perfect correlation, because the PNS kick depends on the product of a dipole asymmetry parameter \citep[named $\alpha_\mathrm{ej}$ in equation~11 of][]{janka2017} and the explosion energy. Due to their relatively low explosion energies, WH12.5, SW13.1, SW21.0, and SW26.2 possess average or low kicks despite comparatively high dipole asymmetries of their NiCoFe0.5X momentum distributions, whereas SW27.0 with a higher explosion energy shows a fairly large PNS kick of nearly 350\,km\,s$^{-1}$, although its dipole moment $A_1$ is on the lower side. 

Comparing the three model categories, one can witness a slight tendency that all three LM models with their low explosion energies develop a relatively high dipole asymmetry, $A_1 \geq 0.3$, while in the other two categories there are also cases with lower values of the normalized dipole amplitudes. Similarly, in these other two model categories, there seems to exist a trend of lower values of $A_1$ for cases with higher explosion energies. The two smallest values of $A_1$ are found in model SW19.8 (as expected from the visual inspection of Fig.~\ref{fig:ni-plumes_cat3}) and in model SW25.5 (which was classified as a strongly asymmetric case based on Fig.~\ref{fig:ni-plumes_cat3}). In spite of their highest and second-highest explosion energies, SW19.8 and SW25.5 therefore have the two lowest PNS kicks of all of our models. Nevertheless, SW25.5 is among our simulations with the highest normalized quadrupole moments $A_2$, which, however, are not effective in fostering PNS kicks. In contrast, models SW18.2, SW25.6, and SW27.3 with their fairly high explosion energies also develop large dipole asymmetries $A_1$. The scatter in each model group is therefore large, for which reason we feel unable to draw conclusions on how the progenitor properties (e.g., compactness and other structure parameters) and the explosion properties (e.g., shock revival timescale, shock movement after bounce and initial expansion velocity, and explosion energy) might impact the explosion asymmetry caused by the neutrino-driven mechanism. Stochasticity, even for a fixed progenitor, may also play an important role and would require multiple 3D explosion simulations for the same star, started with different initial perturbations, to see how much variation there is. Naturally, also the way how neutrino-driven explosions are initiated (in our simulations parametrically with a defined explosion energy in contrast to models with a fully self-consistent neutrino treatment and possibly stronger and longer accretion by the PNS) may have an influence on a systematic connection between explosion asymmetry and progenitor properties. Finally, in reality, 3D asymmetries (velocity and density perturbations) linked to convective mass motions in the pre-collapse Si- and O-burning shells are likely to have an important influence.

\begin{figure*}
    \centering
    \includegraphics[width=\linewidth]{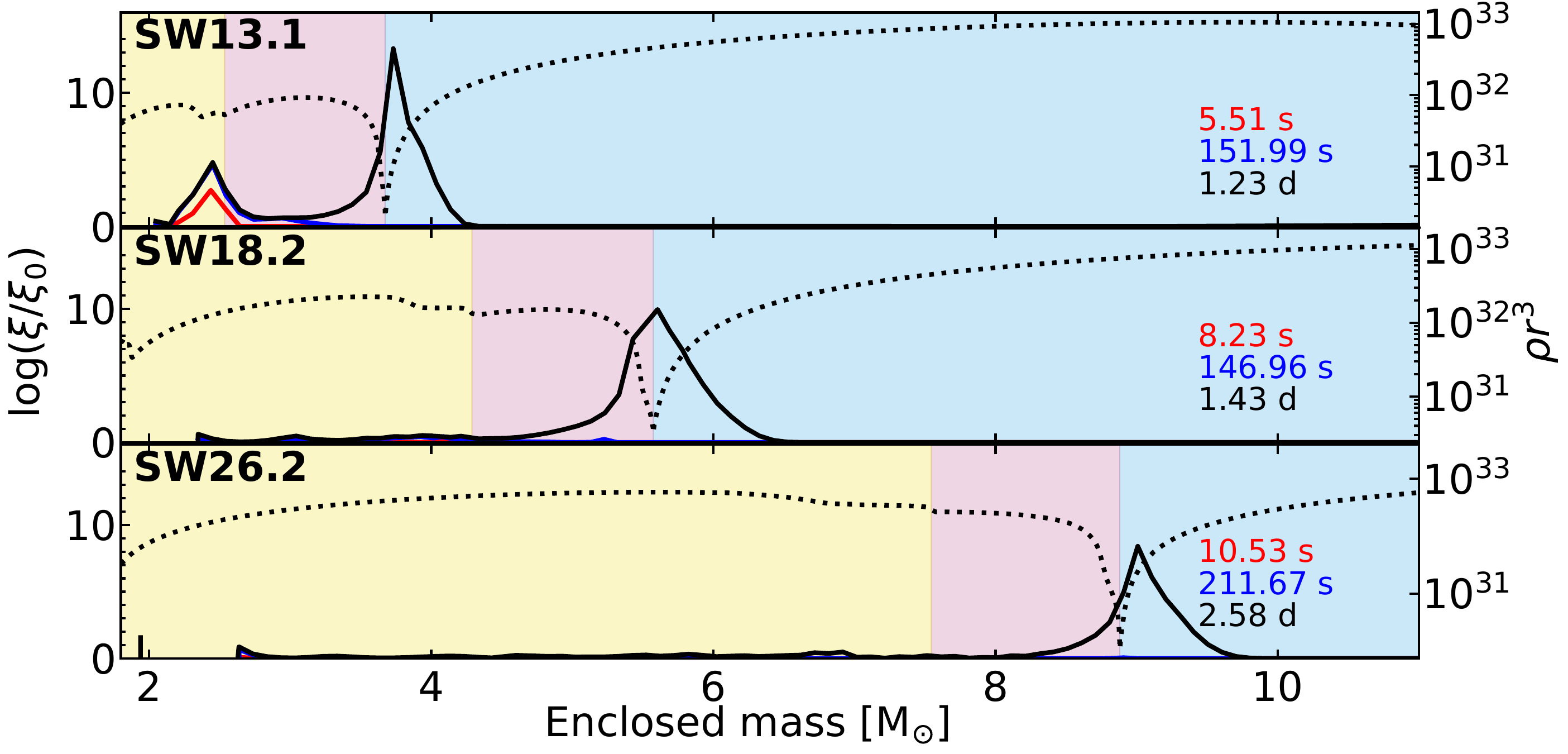} 
    \caption{Growth factors (time-integrated growth rates) of \ac*{rti} as functions of enclosed mass for models SW13.1 (top panel), SW18.2 (middle panel), and SW26.2 (bottom panel).
    They are shown at the time when the SN shock crosses the (C+O)/He interface (red lines), when it crosses the He/H interface (blue lines) and at shock breakout from the stellar surface ($t_\mathrm{sb,1}$, black lines).
    The shaded areas indicate the different shells, with the respective interfaces.
    The yellow area is the C+O shell; the purple area is the He shell; and the blue area depicts the H envelope.
    The dotted lines depict the function $\rho r^3$ for each model with the corresponding scale on the right axis.
    The derivative of $\rho r^3$ of the most massive model SW26.2 does not change sign around the (C+O)/He interface ($\sim7.5\,\Msun$). 
    Therefore, we do not expect significant \acp*{rti} in this region.
    }
    \label{fig:RTgrowth}
\end{figure*}

In any case, the initial asymmetries of the NiCoFe0.5X distribution caused by neutrino-driven post-shock instabilities act as seeds for the long-term growth of extended RT fingers discussed below. Therefore the emerging RT structures retain a memory of the large-scale asymmetries that developed during the earliest stages of the explosion. The main difference between the models in our three categories is connected to the efficiency of the associated radial mixing of heavy elements, which depends on the RT-growth factors at the (C+O)/He and He/H composition interfaces. This will be described in detail in the following.

\subsection{1D models: linear stability analysis}
\label{sec:RTIs_1d}

\begin{figure}
    \centering
    \includegraphics[width=\linewidth]{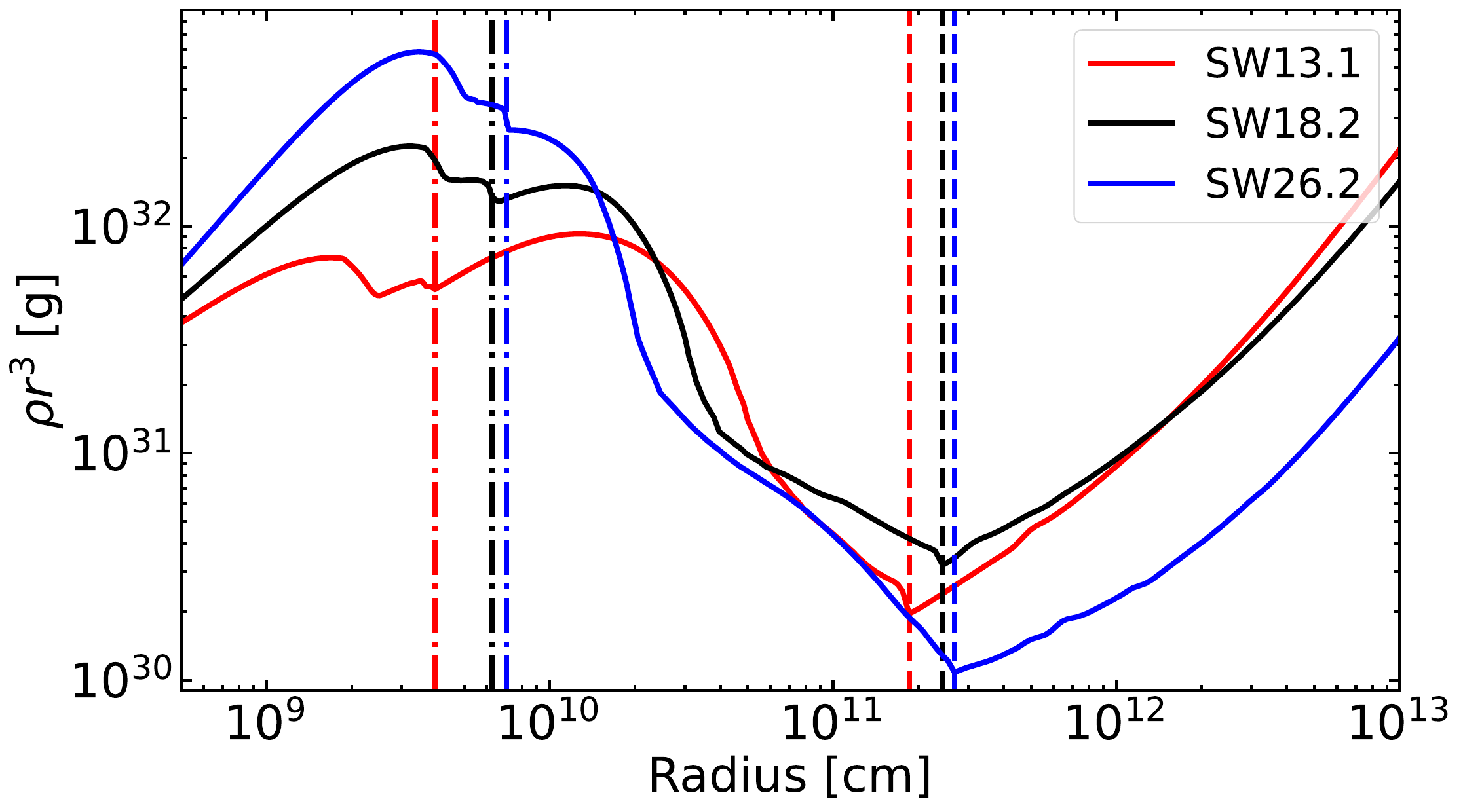}
    \caption{Profiles of $\rho r^3$ as functions of radius for the models SW13.1 (red), SW18.2 (black), and SW26.2 (blue), respectively. The vertical dash-dotted and dashed lines mark the positions of the (C+O)/He and He/H composition interfaces for the three models.} 
    \label{fig:rhor3}
\end{figure}

For simplicity and clarity, we will investigate the linear growth rates of \acp*{rti} first with the aid of 1D simulations. 
During the propagation of a \ac*{sn} shock wave through the star, the velocity of the shock varies significantly. Depending on the density stratification it encounters, a spherically symmetric blast-wave accelerates whenever the density gradient is steeper than $r^{-3}$ (i.e., $d(\rho r^3) / dr < 0$), whereas it decelerates otherwise \citep{sedov1959}.
In stellar progenitors, the density gradients are particularly steep when approaching composition interfaces from the interior and then they become more shallow again in the overlying shell (see Fig.~\ref{fig:densityProfile}). 
Consequently, the value of $\rho r^3$ varies non-monotonically with the radius (Fig.~\ref{fig:RTgrowth}, dotted black lines, and Fig.~\ref{fig:rhor3}) and the shock velocity increases (decreases) when the \ac*{sn} shock approaches (has passed) a composition interface. Deceleration leads to an accumulation of matter, and, correspondingly, the pressure gradient with respect to the radius can change from negative to positive in the post-shock flow when the shock passes a shell interface. In contrast, the density gradient remains negative during the evolution. 

These opposite gradients of pressure and density provide the condition for the growth of \acp*{rti}. The criterion for the development of \acp*{rti} in incompressible fluids as given by \citet{chevalier1976, bandiera1984, benz1990} is
\begin{equation}
    \frac{\de p}{\de r} \frac{\de \rho}{\de r} < 0\text{.}
    \label{eq:RTcriterion}
\end{equation}

To compare the strength of the growth of \acp*{rti} in the explosions of different progenitors, we performed 1D \ac*{sn} simulations for all of the \ac*{rsg} models given in Table~\ref{tab:presn}.\footnote{Doing such an analysis for the RTI growth is, in principle, also possible on grounds of the 3D simulations. However, because of the deformed shock wave, the shock reaches the shell interfaces at different times in different directions. Therefore, the RTI growth parameters would be direction dependent and the analysis would be much more cumbersome without offering new insights. Note that apart from the shock deformation, the propagation of the shock (which is far ahead of the asymmetric inner ejecta) is very similar in 1D and 3D explosions models with effectively the same explosion energy.} 
Each 1D explosion was calibrated such that its energy agrees within 12\% (maximum deviation) with that of the corresponding 3D \ac*{sn} model. The growth strength of \acp*{rti} can be monitored with the time-dependent growth factor (i.e., the growth rates integrated over time) at fixed Lagrangian mass coordinates \citep{hattori1986, wongwathanarat2015, utrobin2019, utrobin2021},
\begin{equation}
    \frac{\xi (t)}{\xi_0} = \exp\left( \int_0^t \sigma_\mathrm{RT}\, d\tau\right)\,.
    \label{eq:RTtime_int}
\end{equation}
This quantity describes how much an initial perturbation with amplitude $\xi_0$ would grow until time $t$. Here, 
$\sigma_\mathrm{RT}$ is the linear RT growth rate for an incompressible fluid, defined as
\begin{equation}
    \sigma_\mathrm{RT} = \frac{1}{\rho} \sqrt{- \frac{\de \rho}{\de r} \frac{\de p}{\de r}}\,\text{.}
    \label{eq:RTgrowth_inc}
\end{equation}
This incompressible RTI growth rate is often used as a lower estimate for stratified flows with buoyancy effects, where the actual growth rate may exceed the incompressible prediction \citep{benz1990}. Therefore, one can conservatively use the incompressible criterion to estimate where RTI may be strongest. However, it may both under- or overestimate the true growth depending on local conditions, which is particularly important in marginally stable regions in some of our models. As we will see later in Section\,\ref{sec:linking}, our analysis is dominated by the major contribution to the integrated, global growth rate, and subtle local changes will not affect our results significantly.

Fig.~\ref{fig:RTgrowth} displays the growth factors for \ac*{rti} as a function of the enclosed mass at different times for our three reference models SW13.1, SW18.2 and SW26.2.
The red, blue and black lines show the growth factors at the time when the SN shock crosses the (C+O)/He interface ($t_\mathrm{CO}\sim4\dots\unit[15]{s}$), 
the He/H interface ($t_\mathrm{He}\sim120\dots\unit[500]{s}$), and when the outermost part of the shock breaks out of the progenitor ($t_\mathrm{sb,1}\sim1\dots\unit[2.6]{d}$), respectively (see Table~\ref{tab:mixing} for these times in all of our 3D simulations). 
The breakout in our \ac*{rsg} models occurs at times roughly one order of magnitude later than those for \ac*{bsg} progenitors. This is due to the fact that \ac*{rsg} stars have, in general, much larger radii than \ac*{bsg} stars with comparable mass \citep{wongwathanarat2015, utrobin2019, gabler2021}.
In each plot three different areas representing different composition shells are colour-shaded: yellow (C+O shell), purple (He shell), and blue (H envelope).

\begin{table*}
\setlength{\tabcolsep}{6pt}
    \centering
    \begin{tabular}{c c c c c c c c c c c c c c c c c c c}
        \hline
        Model & $t_\mathrm{CO}$ & $t_\mathrm{He}$ & $t_\mathrm{sb,1}$ & $\Delta t^\mathrm{CO}_\mathrm{RS}$ & $\Delta t^\mathrm{He}_\mathrm{RS}$ & $R_\mathrm{RS}$ & $\Delta R_\mathrm{CO}$ & $\Delta R_\mathrm{He}$ & \multicolumn{2}{c}{$\log(\xi/\xi_0)$} & 
        $v_\mathrm{XX}^\mathrm{CO}$ & $v_\mathrm{XX}^\mathrm{He}$ & $v_\mathrm{XX}^{10}$ & $\langle v \rangle_\mathrm{XX}^{10}$ & $X_\mathrm{mix}$ & $Y_\mathrm{mix}$ & $M_\mathrm{ej}^\mathrm{3D}$\\
         & [s] & [s] & [d] & [s] & [s] & \multicolumn{3}{c}{[$10^6$ km]} & CO & He & \multicolumn{4}{c}{ [$10^3$ \mbox{km s$^{-1}$}] } & & & [M$_\odot$] \\
        \hline
        WH12.5  &  4.4 & 122 & 1.09 & 522 &  405 &  5.9 & 5.8 & 4.3 & 2.8  & 11.5 & 13.8 & 9.3 & 4.8 & 1.4 & 13.59 & 3.53 & 9.51\\
        SW13.1 &  5.5 & 152 & 1.23 & 657 &  511 &  6.6 & 6.5 & 4.8 & 4.9  & 13.5 & 10.5 & 7.2 & 4.1 & 1.0 & 15.69 & 3.92 & 9.74 \\
        SW14.2 &  6.0 & 182 & 1.20 & 663 &  487 &  6.6 & 6.5 & 4.7 & 3.3  & 13.2 & 11.0 & 8.7 & 4.4 & 1.4 & 15.44 & 3.12 & 10.25 \\
        \hline
        SW16.3 &  6.3 & 166 & 1.33 & 689 &  530 &  6.9 & 6.8 & 4.8 & 1.3  & 10.0 &  8.8 & 6.6 & 2.6 & 0.9 & 8.60 & 2.79 & 12.27 \\
        SW18.2 &  8.2 & 147 & 1.43 & 907 &  768 &  8.7 & 8.6 & 6.2 & 0.8  & 10.0 &  8.5 & 8.0 & 3.6 & 1.3 & 10.64 & 2.70 & 12.92 \\
        SW20.8 & 13.1 & 260 & 2.23 & 1704 & 1457 & 11.8 & 11.7 & 8.8 & 0.3  & 11.9 &  7.6 & 7.8 & 3.8 & 1.5 & 15.79 & 2.54 & 12.79 \\
        SW21.0 & 11.3 & 282 & 1.92 & 1502 & 1232 & 12.0 & 11.9 & 8.8 & 0.4  & 10.7 &  6.6 & 7.2 & 3.4 & 1.2 & 11.13 & 2.94 & 13.82 \\
        \hline
        SW19.8 & 10.2 & 202 & 1.74 & 977 &  786 &  9.9 & 9.8 & 7.2 & 0.7  & 11.4 &  4.3 & 4.4 & 1.7 & 0.7 & 5.84 & 2.47 & 14.42 \\
        SW25.5 & 10.2 & 207 & 2.36 & 1357 & 1161 & 13.0 & 12.9 & 10.5 & 0.1  & 12.8 &  4.9 & 5.5 & 2.5 & 1.2 & 7.89 & 2.12 & 12.27 \\
        SW25.6 & 11.2 & 221 & 2.25 & 1275 & 1066 & 12.5 & 12.4 & 9.7 & 0.3  & 10.5 &  5.0 & 6.0 & 3.0 & 1.2 & 7.10 & 2.48 & 12.10 \\
        SW26.2 & 10.5 & 212 & 2.58 & 1165 &  964 & 12.1 & 12.0 & 9.4 & 0.2  &  8.9 &  4.3 & 4.5 & 2.0 & 1.0 & 4.24 & 2.01 & 11.75 \\
        SW27.0 & 10.9 & 430 & 2.37 & 2502 & 2083 & 24.3 & 24.2 & 18.2 & 0.2  & 11.6 &  4.6 & 5.3 & 2.5 & 1.3 & 7.11 & 2.00 & 11.94 \\
        SW27.3 & 10.7 & 482 & 2.24 & 2348 & 1876 & 25.5 & 25.4 & 17.9 & 0.6  &  8.3 &  5.1 & 6.0 & 2.9 & 1.4 & 5.56 & 2.02 & 11.35 \\
        \hline
    \end{tabular}
    \caption{Parameters of relevance for the NiCoFe0.5X mixing in our set of 3D explosion simulations of \ac*{rsg} progenitors. The subscript XX means NiCoFe0.5X.
    $t_\mathrm{CO}$ and $t_\mathrm{He}$ are the times when the outermost
    \ac*{sn} shock crosses the (C+O)/He and He/H composition interface, respectively. $t_\mathrm{sb,1}$ is the break-out time of the maximum radius of the SN shock.
    $\Delta t^\mathrm{CO}_\mathrm{RS}$ and $\Delta t^\mathrm{He}_\mathrm{RS}$ are the time intervals between reverse-shock formation and the shock passage at the (C+O)/He and He/H interface, respectively (equations~\ref{eq:dtco} and~\ref{eq:dthe}). $R_\mathrm{RS}$ is the radius where the reverse shock forms. $\Delta R_\mathrm{CO}$ and $\Delta R_\mathrm{He}$ are the distances between the formation radius of the reverse shock and the (C+O)/He and He/H interface, respectively. The two columns for $\log(\xi/\xi_0)$ contain the logarithmic maxima of the growth factors of \ac*{rti} near the (C+O)/He interface (time-integrated until the \ac*{sn} shock crosses the He/H interface) and near the He/H composition interface (time-integrated until 10 days after the onset of the explosion), computed from corresponding 1D \ac*{sn} simulations. $v_\mathrm{NiCoFe0.5X}^\mathrm{CO}$, $v_\mathrm{NiCoFe0.5X}^\mathrm{He}$, and $v_\mathrm{NiCoFe0.5X}^{10}$, are the mass-averaged velocities of the fastest 4\% of the NiCoFe0.5X at the time when the \ac*{sn} shock passes the (C+O)/He composition interface in the 3D explosion models, when it passes the He/H interface, and 10 days after the onset of the explosion. $\langle v \rangle_\mathrm{NiCoFe0.5X}^{10}$ is the mass-weighted average velocity of the remaining 96\% (i.e., of the bulk) of the NiCoFe0.5X at day 10. $X_\mathrm{mix}$ is the dimensionless characteristic \ac*{rti} growth parameter defined in equation~\eqref{eq:X_mix} and $Y_\mathrm{mix}$ the dimensionless NiCoFe0.5X mixing parameter defined in equation~\eqref{eq:Y_mix}. $M_\mathrm{ej}^\mathrm{3D}$ are the hydrodynamically evolved ejecta masses that are still on the 3D computational grid at 10 days.}
    \label{tab:mixing}
\end{table*}

As expected, the growth factors for model SW13.1 (top panel of Fig.~\ref{fig:RTgrowth}) become significant only at the shell interfaces, where strong changes of $\rho r^3$ are present.
At $\unit[5.51]{s}$ (red line), the shock has just reached the (C+O)/He interface ($\sim\unit[2.6]{\Msun}$), and, consequently, only there the growth factor has increased significantly. 
Once the shock reaches the He/H interface (blue line), the growth factor at the (C+O)/He interface has already reached its maximum value and the one at the He/H interface ($\sim\unit[3.7]{\Msun}$) starts to grow until reaching its maximum before $t_\mathrm{sb,1}$ (black line). ${\xi}/{\xi_0}$ is larger at the He/H interface, since there the positive and negative gradients of the function $\rho r^3$ are much more pronounced compared to the (C+O)/He interface. This leads to a much stronger deceleration of the \ac*{sn} shock in the H envelope than in the He shell (see also Section~\ref{sec:interactionrev} and figures discussed there).
At around shock breakout, the \ac*{rti} growth factors have reached their saturation values and basically do not change until the end of the simulations. 
The behaviour in model SW13.1 is representative for models with low progenitor masses up to about 16.3\,M$_\odot$.

The growth factors at the (C+O)/He interface for more massive models are much lower or even negligibly small. 
In the central and bottom panels of Fig.~\ref{fig:RTgrowth} we show models SW18.2 and SW26.2 as examples for such cases. 
The reason for this marginally unstable, or even stable behaviour is related to the density profile. 
In the more massive models, $\rho r^3$ either increases only slightly in the He shell or it is a monotonically decreasing function with radius from the (C+O)/He interface through the He shell until it starts increasing at the He/H interface. This is visible in Fig.~\ref{fig:rhor3}, where we show the profiles of $\rho r^3$ as functions of radius.
In contrast to the low-mass models, which exhibit a clear maximum of $\rho r^3$ in the He-shell, only a weak or no maximum is found for models with masses higher than 16.3\,$\Msun$. As a consequence the \ac*{sn} shock hardly decelerates when passing the (C+O)/He interface and travelling into the He shell.

Comparing the HM-LE and HM-HE models, we note that the growth factors of the HM-LE models (see SW18.2 in Fig.\,\ref{fig:RTgrowth}) have a broader peak at the He/H interface than the HM-HE models (like SW26.2). This could indicate favourable conditions for the development of \acp*{rti} in a larger volume of the progenitor.

In some of the models the peak of the time-integrated growth factor does not exactly coincide with the He/H interface but may be slightly shifted into the H envelope (see e.g. model SW26.2 in Fig.~\ref{fig:RTgrowth}). 
This is connected to the location of a pile-up of He-rich matter just behind the \ac*{sn} shock when it decelerates in the H-shell \citep{kifonidis2003}. 
This so-called ``He wall'' pushes against the hydrogen layer without getting unstable to \acp*{rti} immediately. 
The formation of RT unstable conditions is therefore delayed for some time, causing the displacement of the peak of the \ac*{rti} growth factor away from the He/H interface.  

In this section, we have analysed the growth conditions for \ac*{rti} by considering growth factors based on linear perturbation theory applied to the structure of 1D models at different times.
However, once favourable conditions for \ac*{rti} exist, the instability is expected to grow and to quickly enter the non-linear regime in the 3D case, thus significantly changing the structure of the \ac*{sn} ejecta compared to the 1D case. Therefore the consequences of the instability need to be studied by multi-dimensional simulations.

\begin{table*}\centering
\begin{tabular}{lccccccccc}
\hline
Model & H & He & C & O & Si &  NiCoFe1.0X & NiCoFe1.0X ($t=2.5$\,s) & NiCoFe1.0X ($t=10$\,d)\\
&$[M_\odot]$&$[M_\odot]$&$[M_\odot]$&$[M_\odot]$&$[M_\odot]$&$[M_\odot]$ &$[M_\odot]$ &$[M_\odot]$\\
\hline
WH12.5 & 5.30 & 3.48 & 0.12 & 0.37 & 0.019 & 0.064 & 0.071 & 0.065\\
SW13.1 & 5.23 & 3.60 & 0.13 & 0.51 & 0.021 & 0.060 & 0.067 & 0.061\\
SW14.2 & 5.45 & 3.88 & 0.15 & 0.50 & 0.022 & 0.071 & 0.072 & 0.074\\\hline
SW16.3 & 5.82 & 4.34 & 0.21 & 1.15 & 0.096 & 0.119 & 0.138 & 0.122\\
SW18.2 & 5.92 & 4.72 & 0.22 & 1.46 & 0.050 & 0.107 & 0.120 & 0.110\\
SW20.8 & 5.63 & 5.01 & 0.25 & 1.32 & 0.034 & 0.083 & 0.093 & 0.084\\
SW21.0 & 5.62 & 5.00 & 0.25 & 2.01 & 0.051 & 0.085 & 0.110 & 0.089\\\hline
SW19.8 & 5.86 & 5.01 & 0.23 & 2.25 & 0.410 & 0.111 & 0.119 & 0.113\\
SW25.5 & 3.64 & 4.58 & 0.39 & 2.66 & 0.260 & 0.098 & 0.137 & 0.101\\
SW25.6 & 3.70 & 4.57 & 0.34 & 2.58 & 0.230 & 0.099 & 0.124 & 0.100\\
SW26.2 & 3.46 & 4.48 & 0.40 & 2.52 & 0.094 & 0.072 & 0.126 & 0.072\\
SW27.0 & 3.18 & 4.30 & 0.44 & 2.94 & 0.077 & 0.104 & 0.148 & 0.106\\
SW27.3 & 2.88 & 4.17 & 0.44 & 2.83 & 0.080 & 0.115 & 0.145 & 0.118\\\hline
\end{tabular}
\caption{Total masses of selected elements for all models either after shock breakout time $t_\mathrm{sb,2}$ or at $t=2.5$\,s and 10\,d in the last but one and the last column, respectively. The masses of NiCoFe1.0X, calculated as the volume integral over the density of all elements of the iron group $ M_\mathrm{NiCoFe1.0X}=\int \rho\times{\left(X_\mathrm{Ni56}+X_\mathrm{Co56}+X_\mathrm{Fe56}+X_\mathrm{X56}\right)} dV$, decrease for all models between 2.5\,s and $t_\mathrm{sb,2}$ because of the mass loss through the outward shifted inner boundary of the 3D computational grid. The slight increase of the NiCoFe1.0X masses between $t_\mathrm{sb,2}$ and 10\,d is caused by the iron swept up as a part of the CSM (assumed to have the composition of the progenitor's hydrogen envelope) after the SN shock has broken out of the exploding star.}
\label{tab:s5_elements}
\end{table*}

\subsection{Radial mixing}
\label{sec:radmixing}

\begin{figure}
    \includegraphics[width=.99\linewidth]{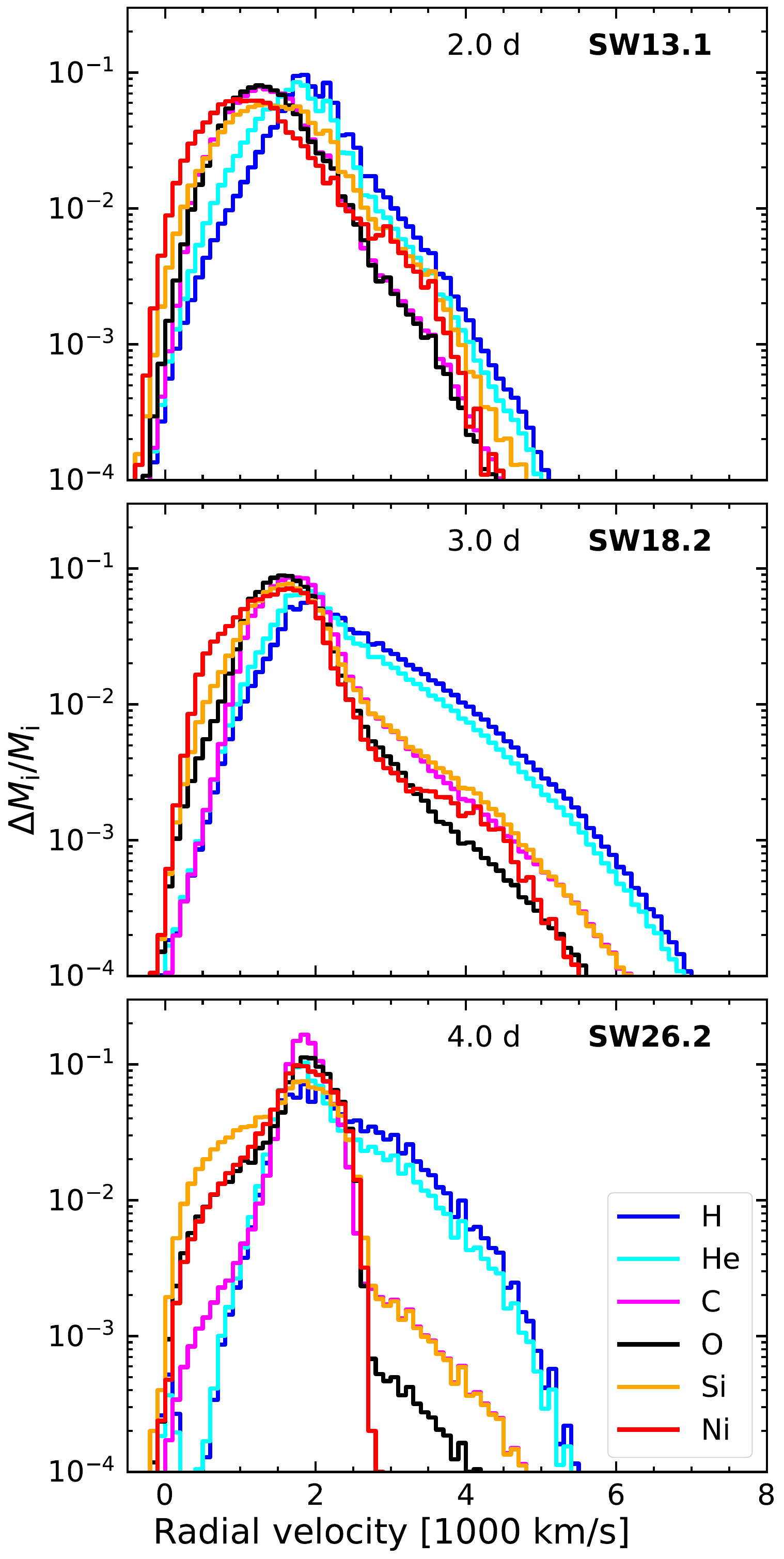}
    \caption{Normalized mass distributions of hydrogen (solid blue line), helium (solid cyan line), carbon (dashed red line), 
    oxygen (dashed black line), silicon (dashed blue line), and NiCoFe0.5X (Ni in the legend, solid red line) versus radial velocity for models 
    SW13.1, SW18.2, and SW26.2 shortly after complete shock breakout.
    The radial velocity bins have a width of $\Delta v_r = \unit[100]{km\,s^{-1}}$. Negative velocities in the innermost ejecta indicate some fallback of gravitationally bound ejecta or ejecta decelerated by the interaction with the reverse shock from the He/H interface.}
    \label{fig:radial_mixing_v}
\end{figure}

\begin{figure}
    \includegraphics[width=.99\linewidth]{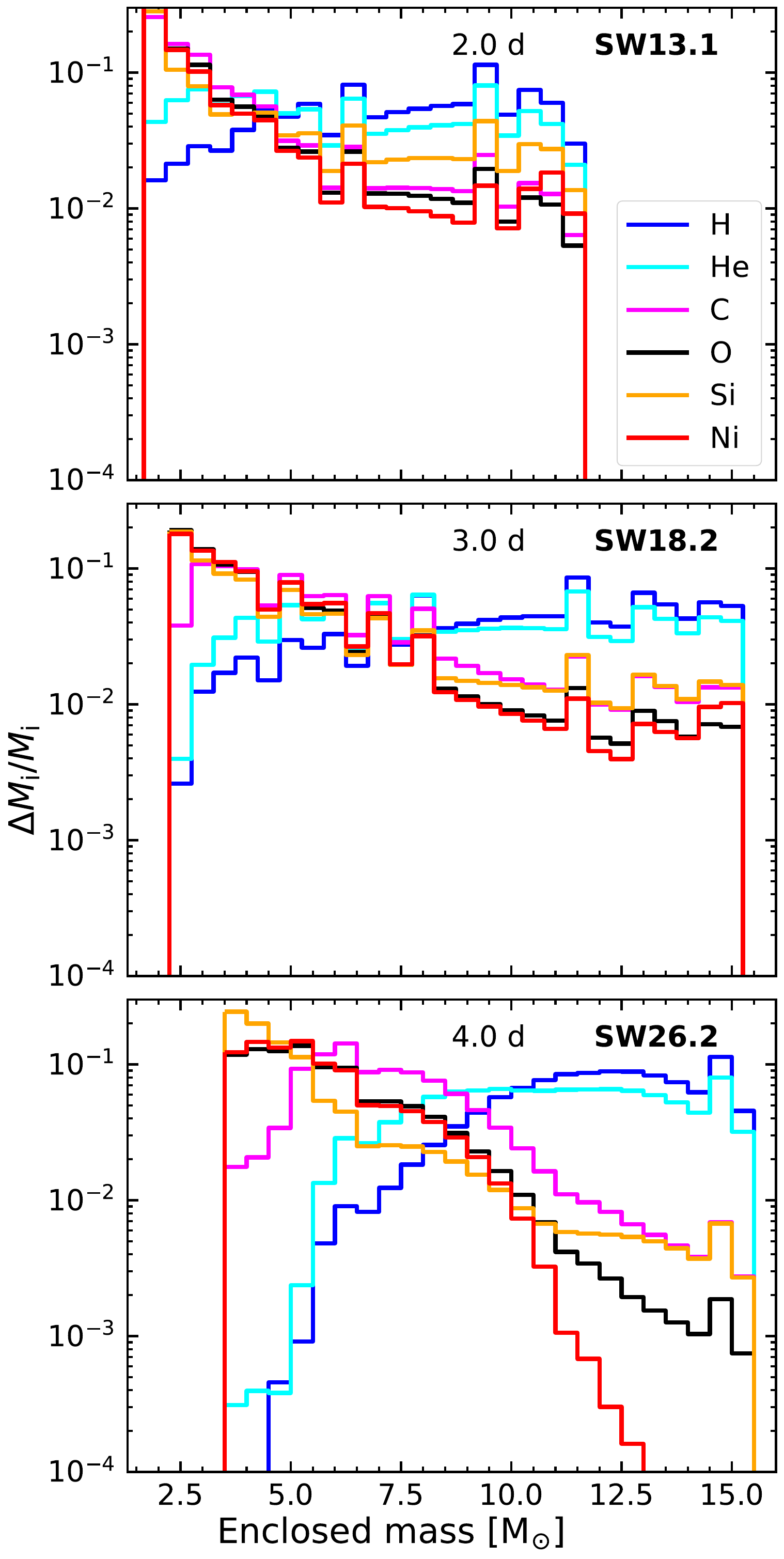}
    \caption{Normalized mass distributions of hydrogen (solid blue line), helium (solid cyan line), carbon (dashed red line), oxygen (dashed black line), silicon (dashed blue line), and NiCoFe0.5X (Ni in the legend, solid red line) versus enclosed mass for models SW13.1, SW18.2, and SW26.2 shortly after complete shock breakout. The mass bins have a width of $\Delta M = \unit[0.4]{\Msun}$. Since some of the low-velocity matter has left the numerical grid through the inner boundary during the simulation, the lines start at enclosed masses that are larger than the \ac*{pns} masses in Table~\ref{tab:3Dexplosion}.}
    \label{fig:radial_mixing_m}
\end{figure}

\begin{figure}
    \centering
    \includegraphics[width=\linewidth]{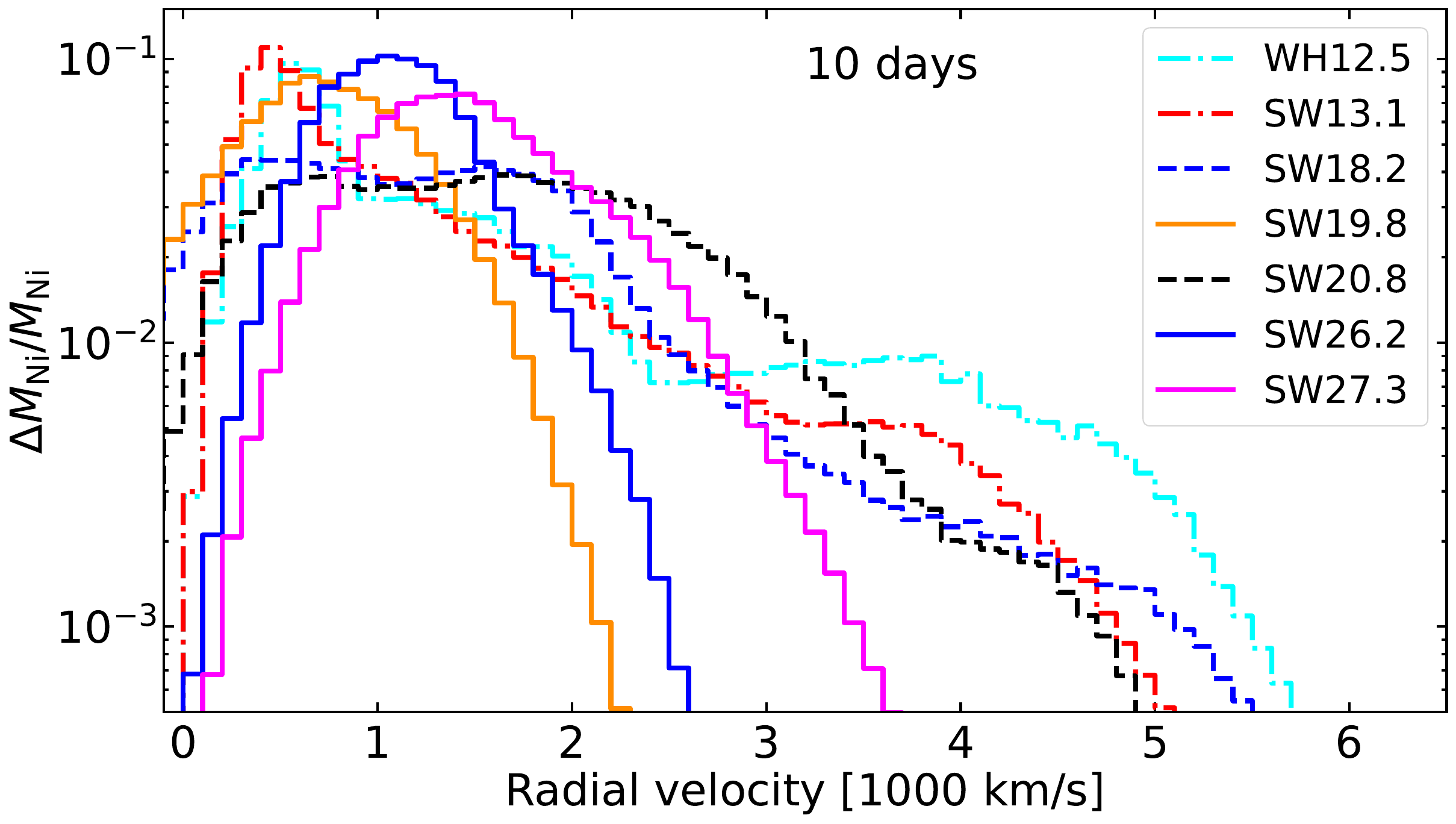}
    \caption{Normalized mass distributions of NiCoFe0.5X versus radial velocity at 10 days after the explosion for different models. Negative velocities indicate some fallback matter in the innermost ejecta.}
    \label{fig:mixing_10days}
\end{figure}

We begin our analysis of the radial mixing in our three-dimensional models by considering the normalized mass distributions for different elements ($\isotope[1]{H}$, $\isotope[4]{He}$, $\isotope[12]{C}$, $\isotope[16]{O}$, $\isotope[28]{Si}$, and NiCoFe0.5X) at a time shortly after the innermost part of the deformed \ac*{sn} shock has left the radius of the star, such that all of the matter of the progenitor has been shocked. 
Figures~\ref{fig:radial_mixing_v} and \ref{fig:radial_mixing_m} display these mass distributions for models SW13.1, SW18.2, and SW26.2 as functions of the radial velocity and enclosed mass, respectively.\footnote{For constructing mass histograms versus radial velocity, we collect matter in the cells of our 3D computational grid (i.e., the masses of different chemical elements) into discrete, predefined bins with a width of $\Delta v_r=100$\,km\,s$^{-1}$ for the values of the radial velocity. Negative velocities correspond to either matter in convective downflows or in fallback flow. The latter is the dominant effect at later evolution times and visible by tails of the mass distributions with slightly negative velocities in Figs.~\ref{fig:radial_mixing_v}, \ref{fig:mixing_10days}, and \ref{fig:mix_evol}. Enclosed masses in our 3D simulations are defined by the total mass enclosed by spherical volumes with certain radii. Therefore, in 3D a thus defined enclosed mass is not an exact Lagrangian quantity.} The normalization factors for the mass distributions can be found in Table~\ref{tab:s5_elements}, where the total ejected masses of the displayed chemical elements are listed at relevant times when the distributions for our models are shown in different plots.
In general, the ejecta can be divided into a group of light elements ($\isotope[1]{H}$ and $\isotope[4]{He}$: blue and cyan solid lines in the figures), intermediate-mass elements that are partially already present in the progenitor ($\isotope[12]{C}$, $\isotope[16]{O}$, and $\isotope[28]{Si}$: red, black, and blue dashed lines) and heavy, iron-group elements represented by NiCoFe0.5X, which is made in the explosion (red solid line). 

Recently, \citet{vartanyan2025b} also provided (normalized) mass distributions of $^{56}$Ni in velocity space, however mass-weighted and projected along arbitrarily chosen $x$-, $y$-, and $z$-directions. These directions are not necessarily representative and do not allow to draw conclusions on the peak velocities and peaks of the mass distributions for all possible viewing directions. Instead, the plots in \citet{vartanyan2025b} yield a rough impression of the clumped structure and asymmetry of the $^{56}$Ni distributions in velocity space as seen from the selected viewing perspectives. Therefore, we can only state that our results show overall and qualitative similarities of the distributions, but a quantitative comparison is hampered by the strong direction-dependent variations of the distributions in velocity space due to ejecta asymmetries and clumping.

At the late stage of the evolution, the former onion-shell structure of the different elements in the progenitor is only partially maintained (Fig.~\ref{fig:radial_mixing_v}). Model SW26.2 (bottom panel) has the least mixing and the high-velocity tails of the distributions reproduce the expected order of the elements best: farthest outside are the lighter elements ($v\lesssim5500\,$km\,s$^{-1}$), followed by intermediate elements ($v\lesssim4500\,$km\,s$^{-1}$) and NiCoFe0.5X is located at the lowest velocities ($v\lesssim3000\,$km\,s$^{-1}$). Note that apparent discrepancies in the fraction of mass per velocity bin can appear due to different total masses. It seems that oxygen is located further inside than carbon or silicon (bottom panel). However, this is an artifact of the rescaling by the total mass of the respective element and the distributions of silicon, oxygen and carbon are actually strongly related. This connection is represented by the same shape of the distribution, not necessarily by the same amplitude. In particular, the massive O-layers of the progenitors imply that smaller fractions of the total O masses are mixed or accelerated to high velocities. This reduces the amplitude of the corresponding high-velocity tails of the displayed normalized distributions. In the other models, SW18.2 (central panel) and SW13.1 (top panel) the mixing of heavy and intermediate mass elements is much more effective and manifests itself in NiCoFe0.5X (red solid lines) reaching or even overtaking the fastest oxygen (black dashed lines).

\begin{table}
    \centering    \setlength{\tabcolsep}{5pt}
    \begin{tabular}{c c c c c c c}
    \hline
    Model & $v_\mathrm{XX, 4\%}^\mathrm{3D}$ & $v_\mathrm{XX, 4\%}^\mathrm{1D}$ & $v_\mathrm{XX, bulk}^\mathrm{3D}$ & $v_\mathrm{XX, bulk}^\mathrm{1D}$ & $v_\mathrm{peak}^\mathrm{3D}$  &  $v_\mathrm{peak}^\mathrm{1D}$ \\
         & \multicolumn{6}{c}{[\mbox{km s$^{-1}$}]}\\\hline 
          WH12.5 & 3538 &  741 & 1499 &  591 & 1450 &  729 \\
          SW13.1 & 2949 & 1184 & 1372 &  984 & 1150 &  953 \\
          SW14.2 & 3541 & 1169 & 1523 &  919 & 1750 & 1036 \\
          \hline
          SW16.3 & 2686 & 2470 & 1796 & 1866 & 1750 & 1727 \\
          SW18.2 & 3139 & 1874 & 1750 & 1661 & 1850 & 1657 \\
          SW20.8 & 2947 & 1950 & 1661 & 1824 & 1650 & 1950 \\
          SW21.0 & 2713 & 2127 & 1741 & 1833 & 1850 & 2127 \\
          \hline
          SW19.8 & 2917 & 3106 & 2144 & 2835 & 2350 & 3004 \\
          SW25.5 & 3004 & 1749 & 2155 & 1484 & 2150 & 1743 \\
          SW25.6 & 3128 & 1872 & 2184 & 1591 & 2350 & 1872 \\
          SW26.2 & 2905 & 2137 & 2002 & 1809 & 2250 & 2010 \\
          SW27.0 & 3209 & 1456 & 2220 & 1174 & 2550 & 1327 \\
          SW27.3 & 3528 & 2130 & 2378 & 1856 & 2550 & 2125 \\
          \hline
    \end{tabular}
    \caption{Different characteristic velocities of NiCoFe0.5X distribution  (XX=NiCoFe0.5X) for 3D and 1D explosions at the time of shock breakout (in the 3D case this is when the minimum radius of the deformed shock breaks out, $t=t_\mathrm{sb,2}$).
    Columns~2~(3) contain the mass-weighted average velocities of the entire fastest 4$\%$ of the NiCoFe0.5X ejecta, $v_\mathrm{NiCoFe0.5X, 4\%}$, and columns~4~(5) the mass-weighted average velocities of the bulk, $v_\mathrm{NiCoFe0.5X, bulk}$, for the 3D (1D) simulations, respectively. In columns~6 and~7, we list the velocities at the peaks of the distributions, $v_\mathrm{peak}$, in 3D and 1D, respectively.
    }
    \label{tab:ni_vels}
\end{table}

\begin{figure*}
    \centering
    \includegraphics[width=0.30\linewidth]{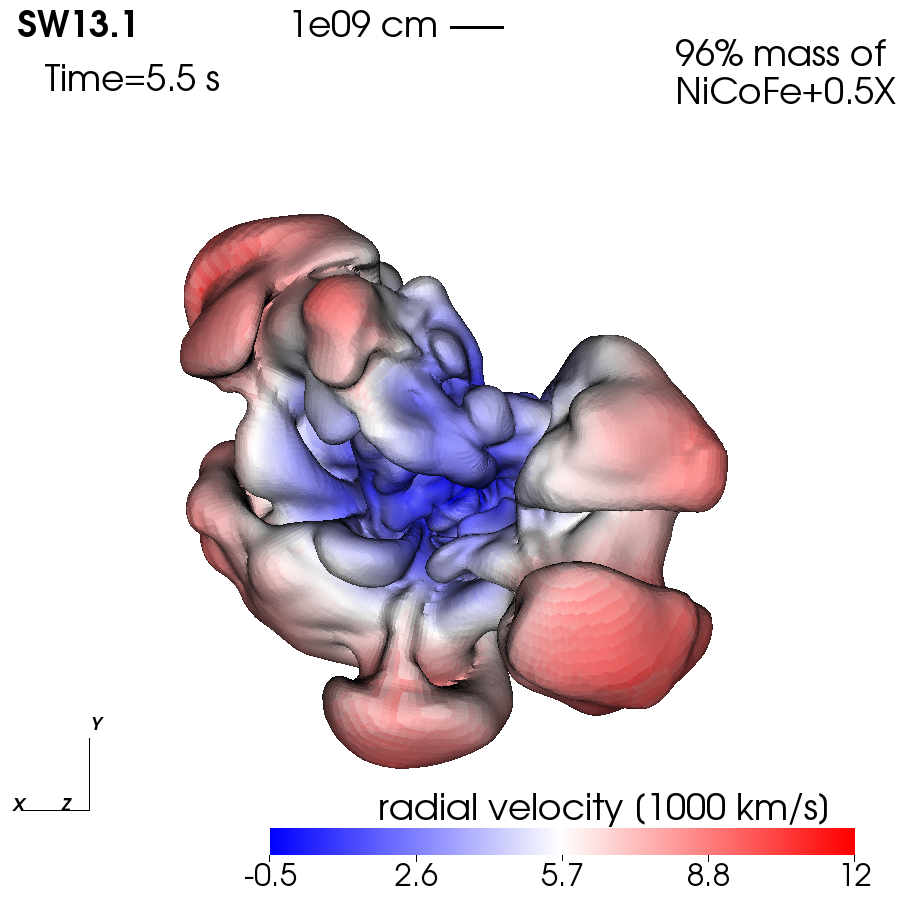} 
    \includegraphics[width=0.30\linewidth]{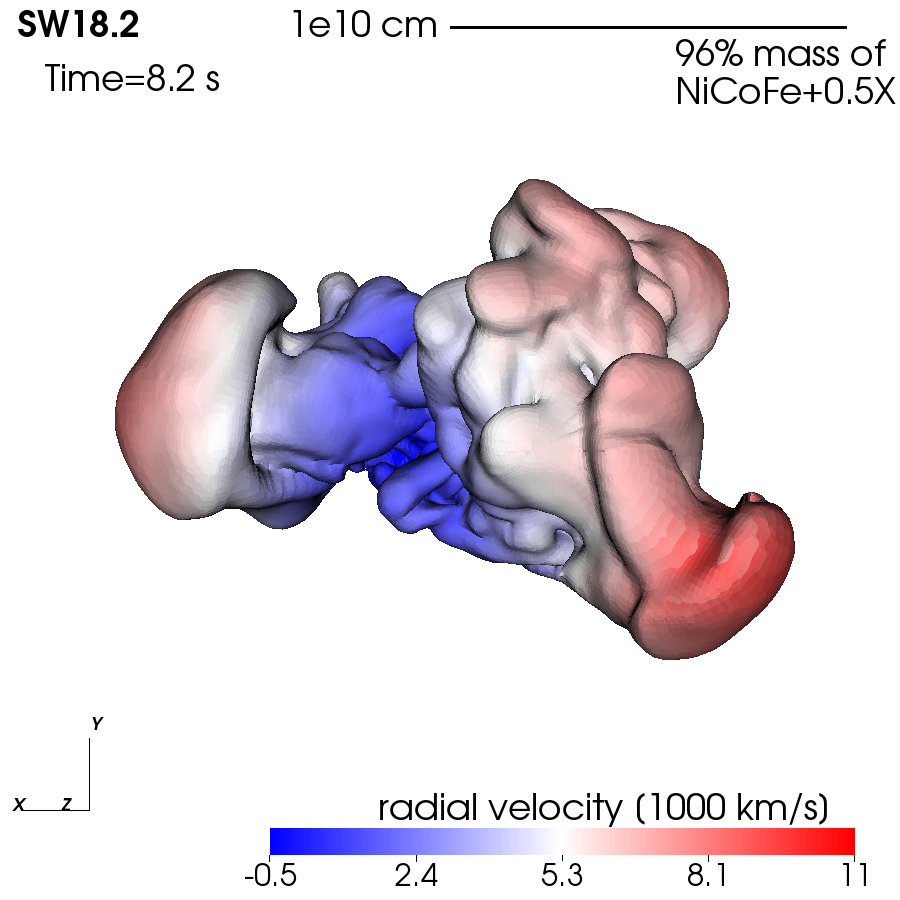}
    \includegraphics[width=0.30\linewidth]{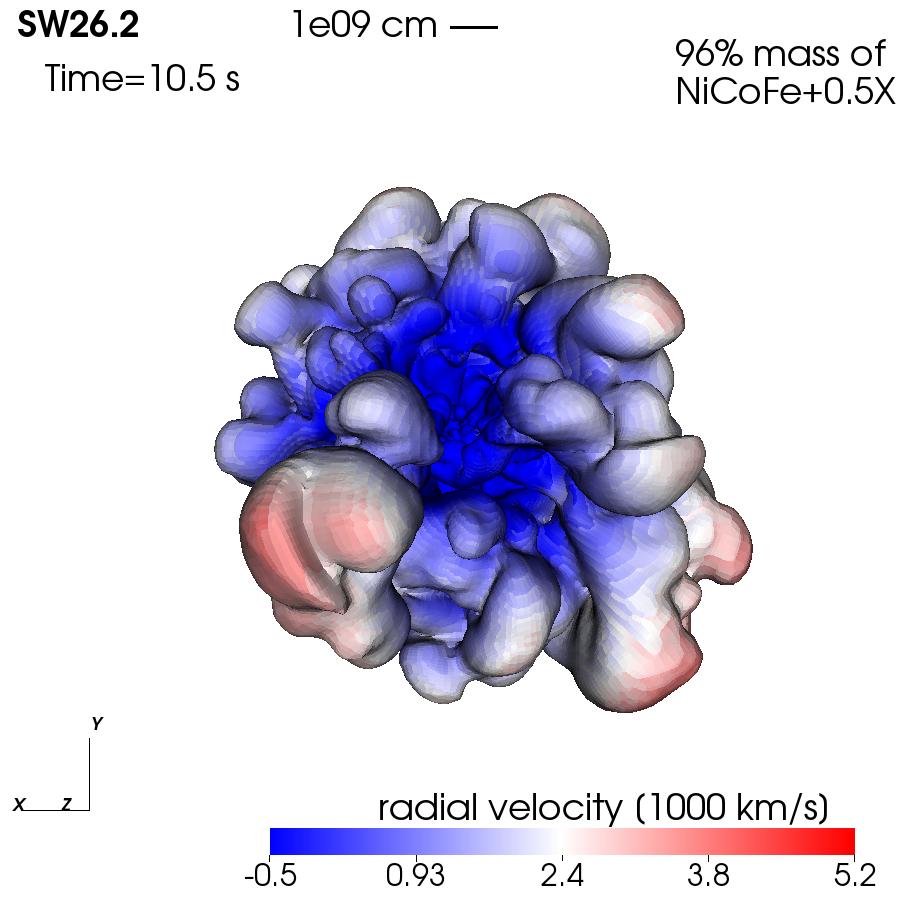}
    \\
    \includegraphics[width=0.30\linewidth]{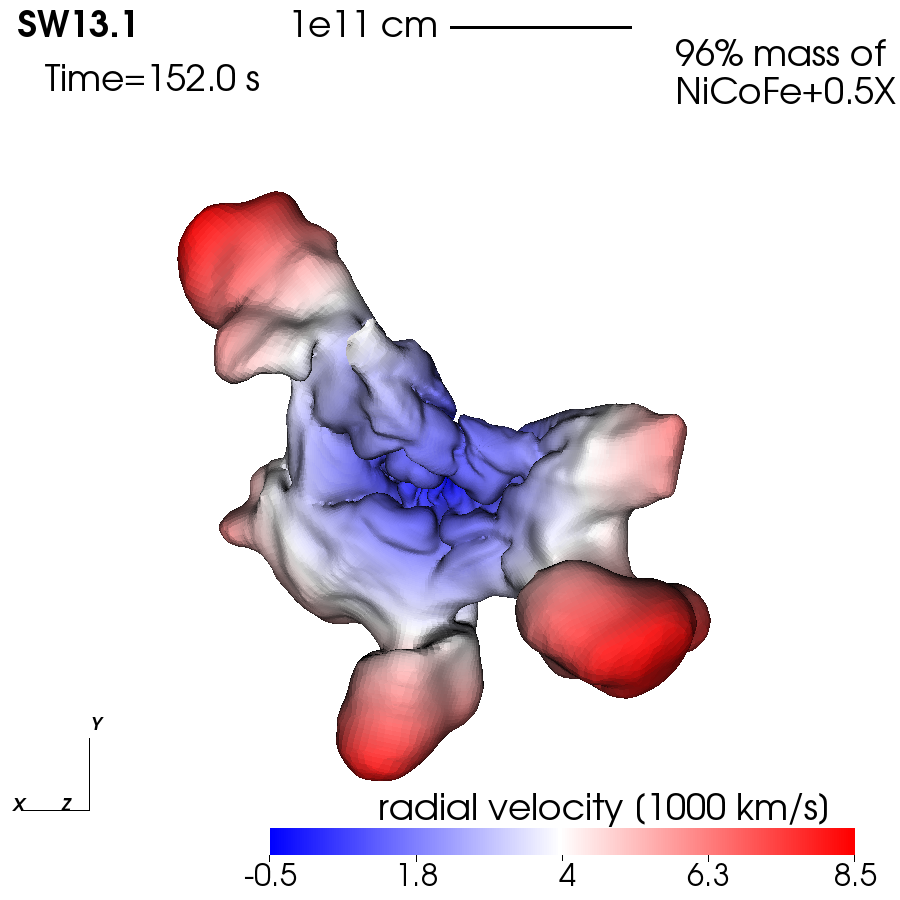}
    \includegraphics[width=0.30\linewidth]{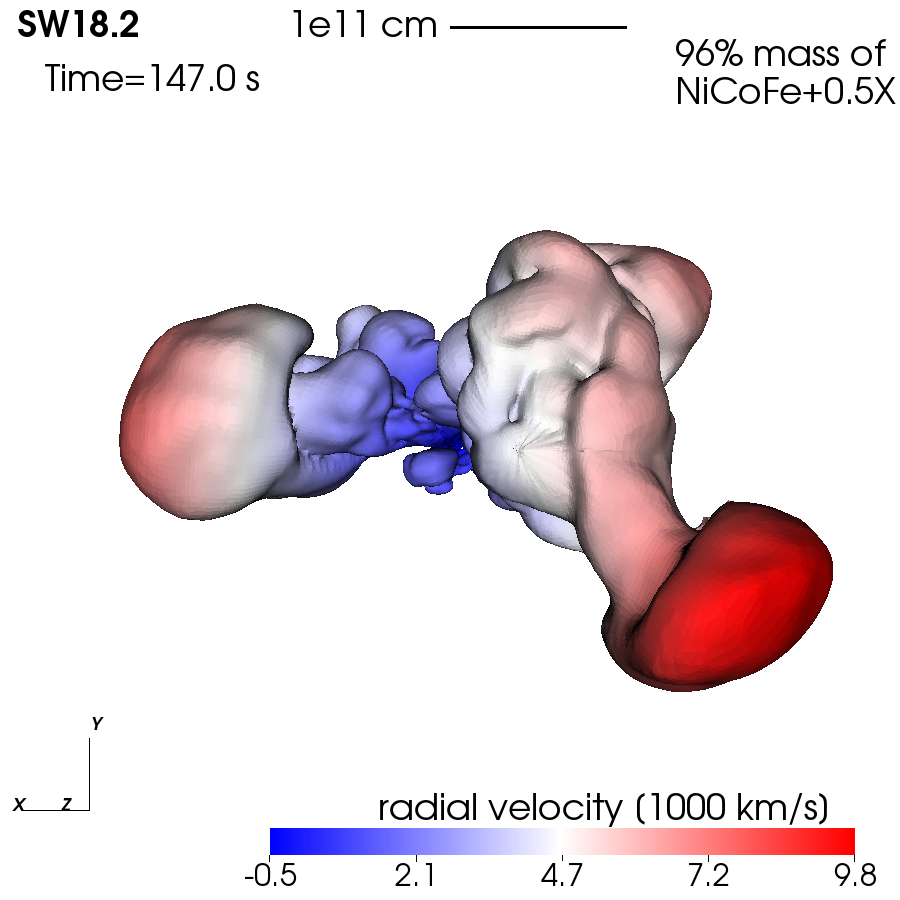}
    \includegraphics[width=0.30\linewidth]{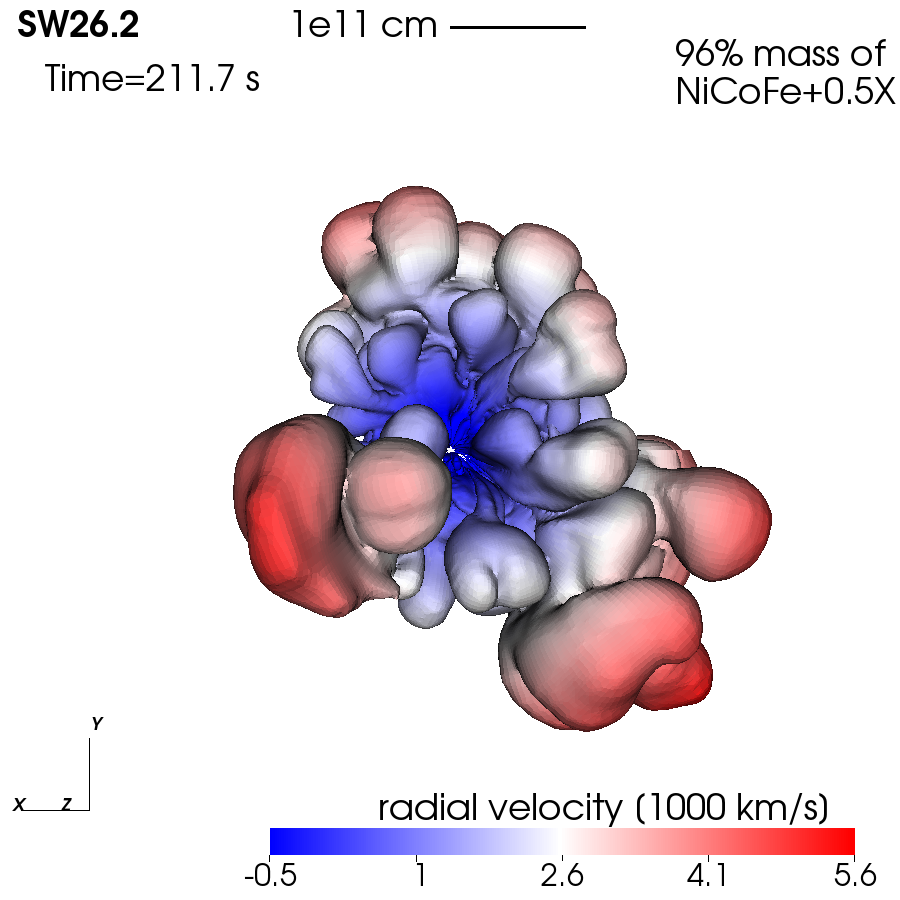}
    \\
    \includegraphics[width=0.30\linewidth]{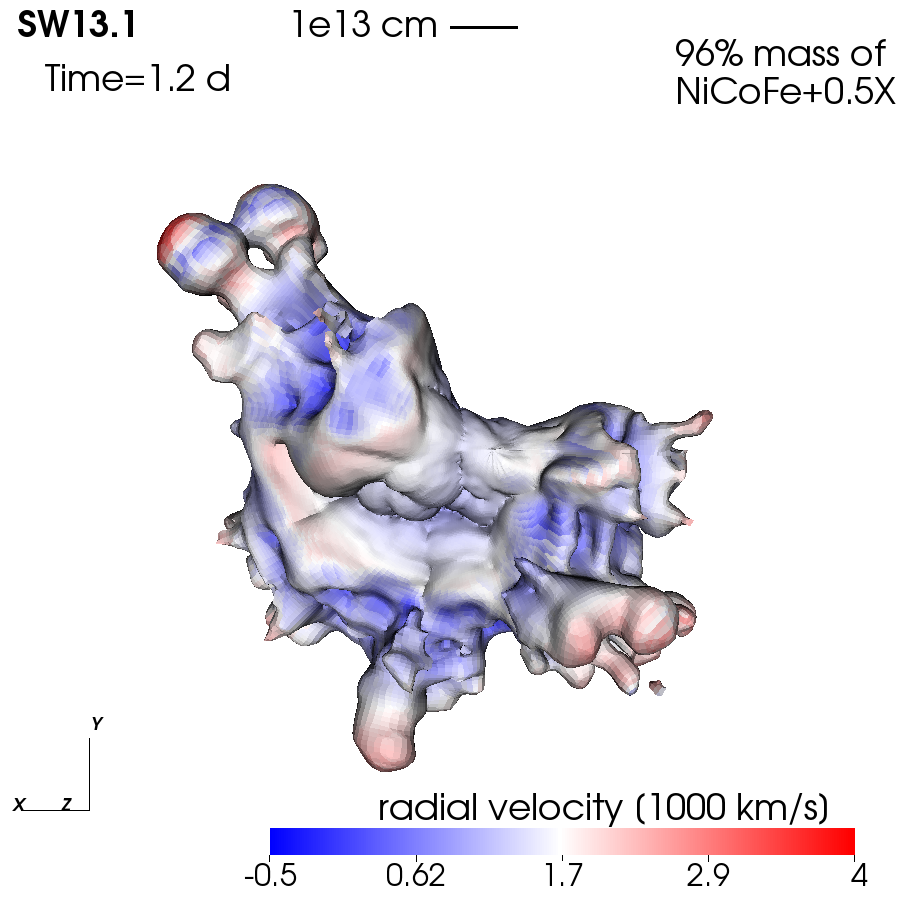} 
    \includegraphics[width=0.30\linewidth]{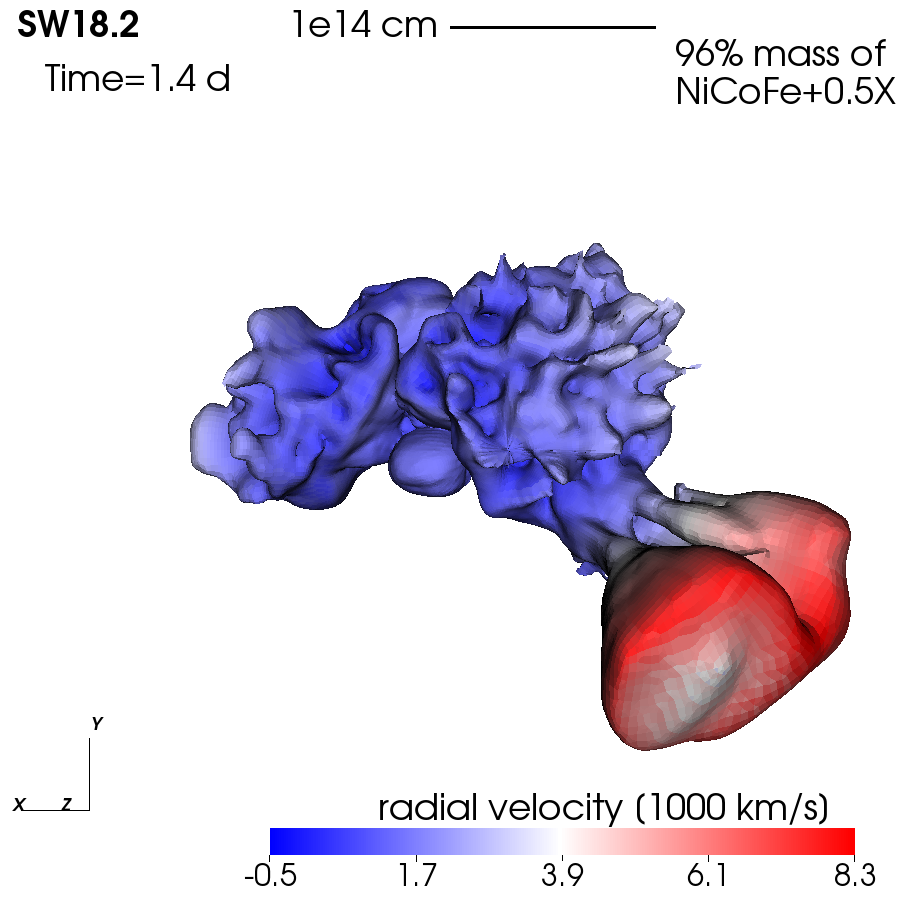}
    \includegraphics[width=0.30\linewidth]{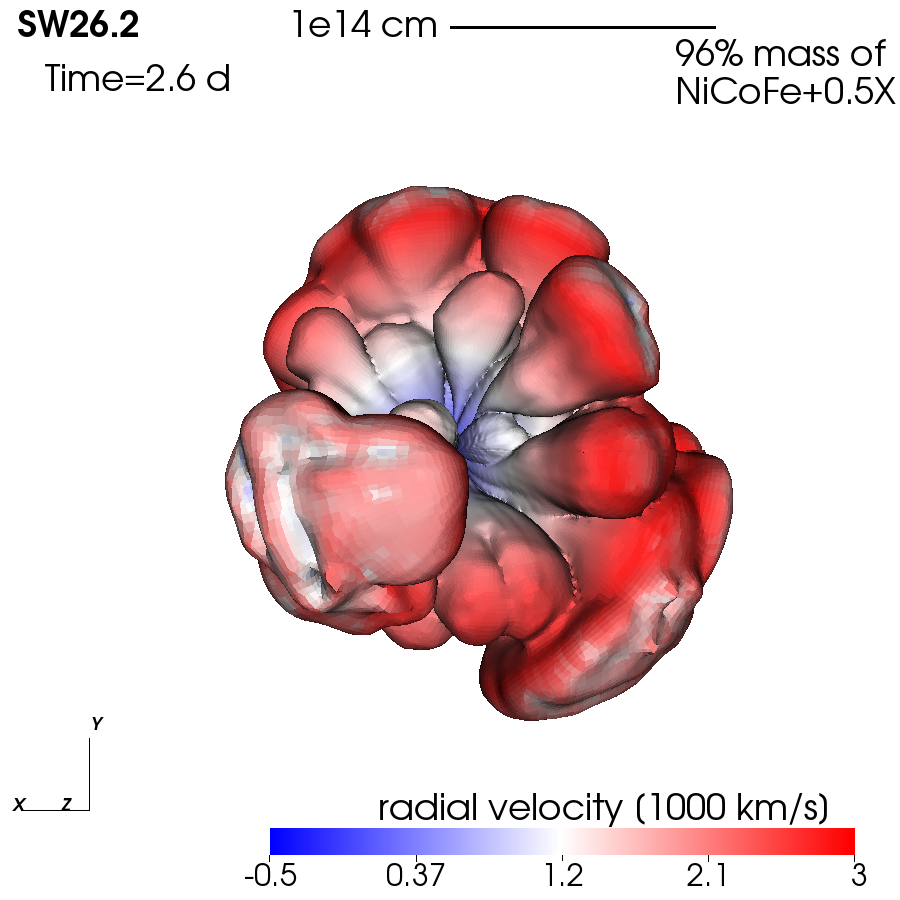}
    \\
    \includegraphics[width=0.30\linewidth]{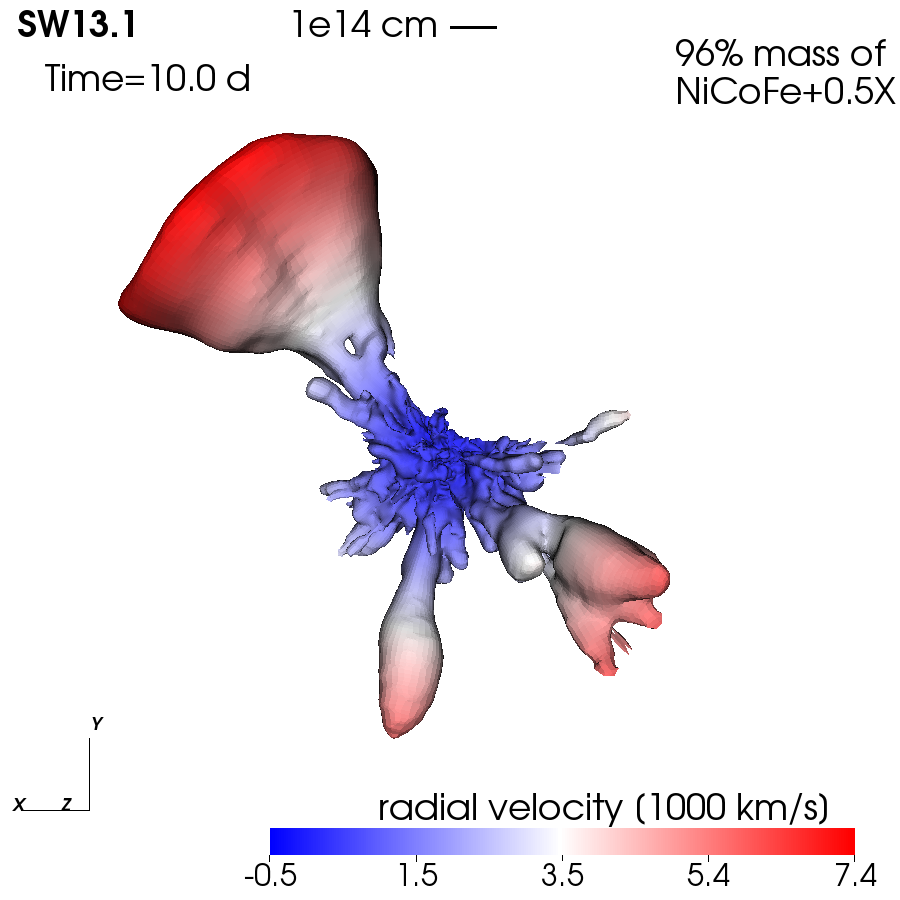} 
    \includegraphics[width=0.30\linewidth]{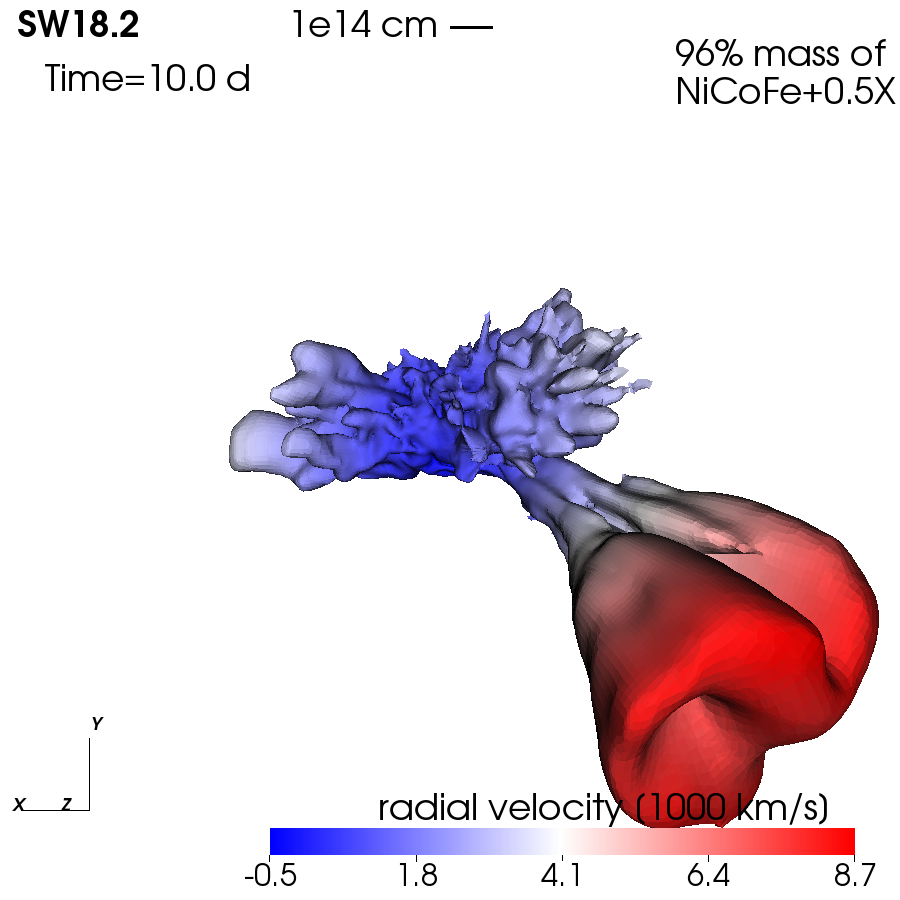}
    \includegraphics[width=0.30\linewidth]{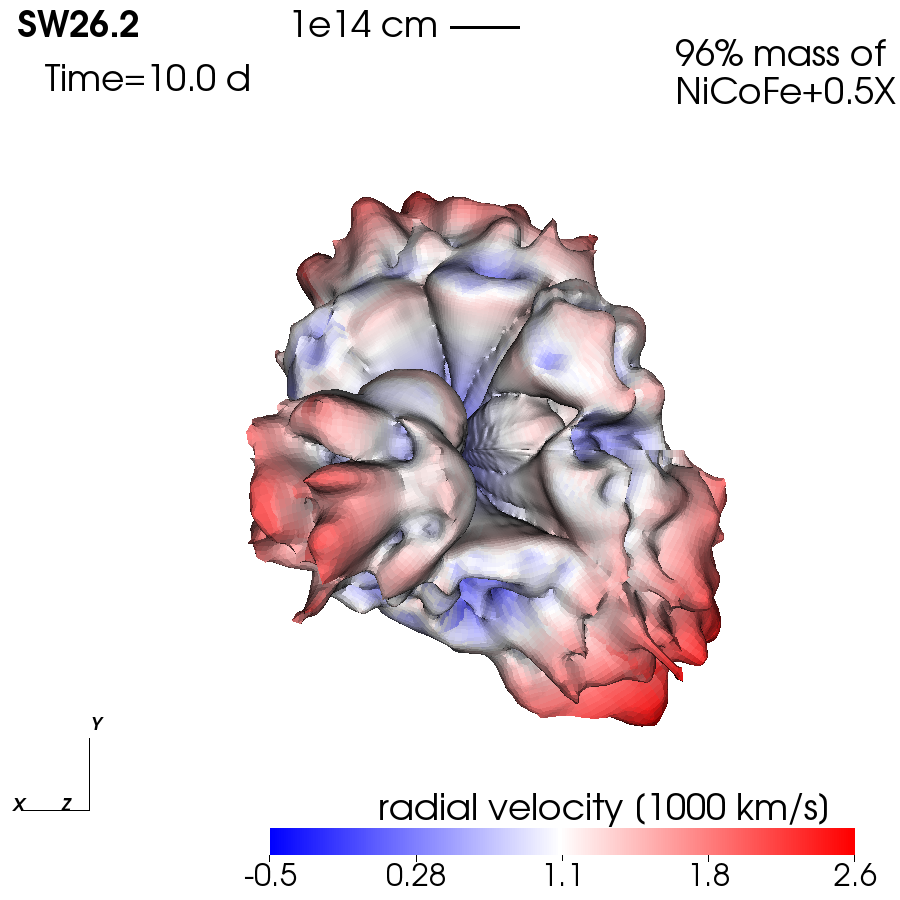}
    \caption{Time evolution of mass-fraction isosurfaces of NiCoFe0.5X that contain 96\% of the total mass of NiCoFe0.5X. The mass fractions typically lie in the interval $0.006 \lesssim X_\mathrm{NiCoFe0.5X} \lesssim 0.10$. The 3D plots are given for the representative models SW13.1, SW18.2, and SW26.2 (left, middle, and right column, respectively). The orientation, i.e., viewing direction, is the same as in Figs.~\ref{fig:ni-plumes_cat1}--\ref{fig:ni-plumes_cat3}.
    Colours show the radial velocities of the material in units of $\unit[1000]{km\,s^{-1}}$. The four selected times correspond to the time when the SN shock crosses the (C+O)/He interface, $t_\mathrm{CO}$ (top row), approximately to the times when it crosses the He/H interface, $t_\mathrm{He}$ (second row), and when the maximum of the shock passes the surface of the star, $t_\mathrm{sb,1}$ (third row), and to the end of the simulations at $\unit[10]{d}$ (bottom row). Model SW13.1 is the most asymmetric one with the most extended \ac*{rti} fingers after 10 days, whereas the ejecta of the most massive model end up with a less extreme deformation. The action of the reverse shocks is visible by bluish patches ahead of deep-red surface regions in the two rightmost panels of the third row, and the bluish region behind the red tips of the RT-fingers in the left plot of the same row.}
    \label{fig:ni_iso}
\end{figure*}

We now discuss the different models in more detail. The velocity distribution of the freshly synthesised NiCoFe0.5X in model SW13.1 (top panel of Fig.~\ref{fig:radial_mixing_v}) is characterized by a maximum around $\approx$\,$\unit[950]{km\,s^{-1}}$. 
The distributions of intermediate-mass and low-mass elements peak at successively higher velocities up to a maximum of $\approx$\,$\unit[2000]{km\,s^{-1}}$ for H. All elements have high-velocity tails reaching up to velocities well above $\unit[4000]{km\,s^{-1}}$. Table~\ref{tab:ni_vels} provides additional quantitative information. There, we give the values of the mass-weighted average velocities of the fastest 4\% of NiCoFe0.5X, $v_\mathrm{NiCoFe0.5X,4\%}$, and of the bulk of NiCoFe0.5X, $v_\mathrm{NiCoFe0.5X, bulk}$. The bulk refers to the 96\% of the mass that remains after excluding the fastest 4\%. In addition, the velocities of the peaks of the distributions are given. These data are taken at the time $t=t_\mathrm{sb,2}$ when the minimum radius of the deformed shock breaks out in 3D, and at the time of shock breakout for the 1D quantities.
The strong mixing, in particular, of the heavy elements from deep inside into the outer layers of the progenitor occurs in the form of extended finger-like structures, which also contain intermediate-mass elements, see Fig.~\ref{fig:ni_iso}, where we plot NiCoFe0.5X isosurfaces at different times for the three representative models. In the mass distribution in Fig.~\ref{fig:radial_mixing_v}, one can notice a slight overabundance of heavier elements like NiCoFe0.5X and $\Si$ compared to the intermediate-mass elements $\C$ and $\isotope[16]{O}$ for velocities in a wider interval around $\sim$\,$\unit[3000]{km\,s^{-1}}$. Since each of the plotted distributions is normalized by the total mass of the corresponding chemical element, this overabundance just means that a larger fraction of the total mass of the element has been mixed into this range of velocities, as mentioned above.
In addition, the lighter elements H and He mix towards low velocities, even negative ones. 
These negative velocities are a consequence of the interaction of inner He and H ejecta with the reverse shock (see Section~\ref{sec:interactionrev}). Later, these ejecta (or a part of them) may be re-accelerated outwards, owing to additional energy input by the $\beta$ decay of radioactive NiCoFe0.5X as well as a possible self-reflection of the reverse shock at the centre \citep{gabler2021}. 
Unfortunately, we cannot track the matter in the innermost regions to study potential fallback onto the compact remnant, because we have to cut out the central part to increase numerical efficiency. The other models with low \ac*{zams} masses, WH12.5, and SW14.2 exhibit qualitatively similar velocity distributions.

In model SW18.2 (Fig.~\ref{fig:radial_mixing_v}, central panel) we find very fast-moving ejecta reaching $\unit[6000]{km\,s^{-1}}$ and more. The much faster ejecta at early times compared to model SW13.1 can usually {\em not} be explained by the higher explosion energies of the HM-LE models, because the energy differences between LM and HM-LE models are relatively small and the ejecta velocities scale only with the square root of this energy. The main reason for the higher velocities is therefore the later times at which the results are evaluated for the HM-LE models, because the breakout of the minimum shock radius ($t_\mathrm{sb,2}$) happens much later in these models. Therefore, the ejecta can re-accelerate after they were slowed down by the reverse shock. In Fig.~\ref{fig:mixing_10days}, we see that the ejecta of the LM models at $\unit[10]{d}$ have comparable or even higher velocities than those of the slightly more energetic HM-LE models, because bigger ejecta masses reduce the expansion velocities ($v_\mathrm{ej} \propto \sqrt{E_\mathrm{exp}/M_\mathrm{ej}}$) and because the LM models have a more efficient RT-caused acceleration. 

As for the mixing, it is important to inspect the differences between different chemical elements in a given model at a given time. Similar to the results for model SW13.1, all elements are mixed up to very high velocities, and we find NiCoFe0.5X up to $\unit[5500]{km\,s^{-1}}$, for example. However, in contrast to the LM model, where all elements are distributed over very similar velocity ranges, we see that H and He, in particular, extend to considerably higher velocities. Moreover, also larger relative fractions of these elements (i.e., higher values of $\Delta M_\mathrm{H}/M_\mathrm{H}$ and $\Delta M_\mathrm{He}/M_\mathrm{He}$) can be found at velocities higher than the peaks of the distributions. These lighter elements also do not mix as efficiently to lower or negative velocities as in the discussed LM case. Model SW18.2, similar to model SW13.1, has well-developed RT fingers that extend far out from the bulk of the element distributions (see panels in the central column of Fig.~\ref{fig:ni_iso}). However, since model SW18.2 has only one region of \ac*{rti} growth at the He/H interface, the inward mixing of He and H into the C+O core is reduced and the outwards rising fingers are slightly less pronounced at 10 days. Similar properties are also observed in models SW16.3, SW20.8, and SW21.0.

Our third mixing case is represented by model SW26.2. For this model, the peaks of the distributions of the heavy elements at $\approx\unit[2000]{km\,s^{-1}}$ are both the most pronounced and found at the highest velocities among the different model classes. These high bulk velocities are caused by the high explosion energies of the HM-HE class. Similar to model SW18.2, the (C+O)/He interface is not RT unstable, and the morphological features in the right panels of the top and second rows of Fig.~\ref{fig:ni_iso} originate from initial asymmetries during the shock revival phase. 
Their structures are present since these very early post-bounce times and hardly change in shape during the subsequent evolution over timescales of hundreds of seconds, indicating that these structures are not related to secondary \acp*{rti} at the (C+O)/He interface.
In contrast to model SW13.1, the velocities of the maxima of the distributions of intermediate-mass and high-mass elements are almost identical. These similar bulk velocities indicate inefficient mixing of NiCoFe0.5X into the outer H and He layers. The ejecta rather stay mostly unmixed, but a fair fraction of the material of the outer progenitor layers has been slowed down by the reverse shock such that it expands with velocities similar to the explosively synthesized inner ejecta. This implies that there is a clear separation of the freshly synthesized higher-mass and intermediate-mass species from the intermediate-mass elements and much of the light elements of the progenitor. The steep drops of the mass distributions of all heavy and intermediate-mass elements at around $v_r\sim\unit[2500]{km\,s^{-1}}$ marks the contact region between the explosively synthesized and reprocessed ejecta and the pre-existing progenitor material. There is essentially no mixing beyond $v_r\sim\unit[2500]{km\,s^{-1}}$. 
This interpretation is also supported by the mass distributions of different chemical elements as functions of enclosed mass for model SW26.2 in the bottom panel of Fig.~\ref{fig:radial_mixing_m}, where one can see that NiCoFe0.5X is mixed outward only up to $\sim\unit[12]{\Msun}$ and the original progenitor composition is present outside of this mass coordinate. At $\unit[12]{\Msun}$ there is a sharp drop in NiCoFe0.5X and a small hump in the other elements, indicating some kind of pile-up of the material. The outermost $\approx\unit[3]{\Msun}$ contain effectively no NiCoFe0.5X. 
In contrast, we find NiCoFe0.5X mixed outwards into the outermost layers of the progenitor in model SW13.1 and, for a slightly lower fraction of the total mass, also in model SW18.2 (top and middle panel of Fig.\ref{fig:radial_mixing_m}). 
This difference can be understood by the fact that the reverse shock in model SW26.2 forms well outside of the NiCoFe0.5X-rich ejecta and does not allow for significant amounts of NiCoFe0.5X to be included in the \ac*{rti} fingers that grow at the He/H interface. 
Therefore, no efficient large-scale radial mixing of NiCoFe0.5X occurs. Consequently, the final geometry of the NiCoFe0.5X ejecta is much less deformed and looks more spherical than that in models SW13.1 and SW18.2 (see bottom right panel of Fig.~\ref{fig:ni_iso}). 
Also, the inward mixing of lighter elements in model SW26.2 is much less significant, and within the innermost $\unit[3]{\Msun}$ there is basically no hydrogen and only very little helium (bottom panel of Fig.\,\ref{fig:radial_mixing_m}). The presence of central helium in regions where there is no hydrogen in this model may be a relic of the nucleosynthesis in the neutrino-heated ejecta. A significant fraction of the alpha particles from the nuclear freeze-out phase may not be able to combine to heavier elements due to the fast expansion of the innermost ejecta. In the lighter models, we see a significant increase of the mass fraction of the $\isotope[1]{H}$ and $\isotope[4]{He}$ at low enclosed mass over time, clearly indicating the inwards mixing of these elements from the outer shells of the progenitors.

In Fig.~\ref{fig:mixing_10days}, we show the mass distributions of the NiCoFe0.5X ejecta as functions of velocity at 10 days for a selection of models to highlight the different behaviours encountered at the same late time. The LM models (WH12.5 and SW13.1) have the sharpest and most narrow peaks of their distributions at the lowest velocities, but their NiCoFe0.5X ejecta also reach the highest velocities in the tails of the distributions. The HM-HE models (SW19.8, SW26.2, and SW27.3) have the overall most narrow distributions with the most prominent peaks at higher velocities and the lowest maximum velocities. And the HM-LE models (SW18.2 and SW20.8) possess the overall flattest distributions with two less pronounced peaks $<2000\,\mathrm{km\,s}^{-1}$ and with tails that tend to lie between those of the other two groups. 

The maximum velocities of the mass distributions at $10\,$d in Fig.~\ref{fig:mixing_10days} anti-correlate roughly,  but not strictly (in particular SW19.8 is a severe outlier due to its shell-merger progenitor), with the progenitor compactness (see Table~\ref{tab:presn}), because of the ordering 
of our three categories of models in terms of their mixing efficiency of NiCoFe0.5X. In contrast, in Table~\ref{tab:ni_vels} the velocities of the fastest $4\%$ of the NiCoFe0.5X at the time of shock breakout $t=t_\mathrm{sb,2}$ do not exhibit a tight correlation or anti-correlation with the compactness. In the same table, $v^\mathrm{3D}_\mathrm{NiCoFe0.5X,bulk}$ for the bulk masses and $v_\mathrm{peak}^\mathrm{3D}$ for the peaks of the mass distributions display a rough correlation with $\xi_{2.5}$, compatible with the increasing explosion energies when going from category LM to categories HM-LE and HM-HE. A correlation of the maximum $^{56}$Ni velocity with core compactness was claimed by \citet{vartanyan2025b}. But our discussion suggests that the degree and sign of the correlation depends on which fraction of the ejecta of NiCoFe0.5X is considered, the time when the velocities are measured, and possibly also on the mass value $M$ where the compactness $\xi_M$ is defined.

\begin{figure}
    \centering
    \includegraphics[width=0.99\linewidth]{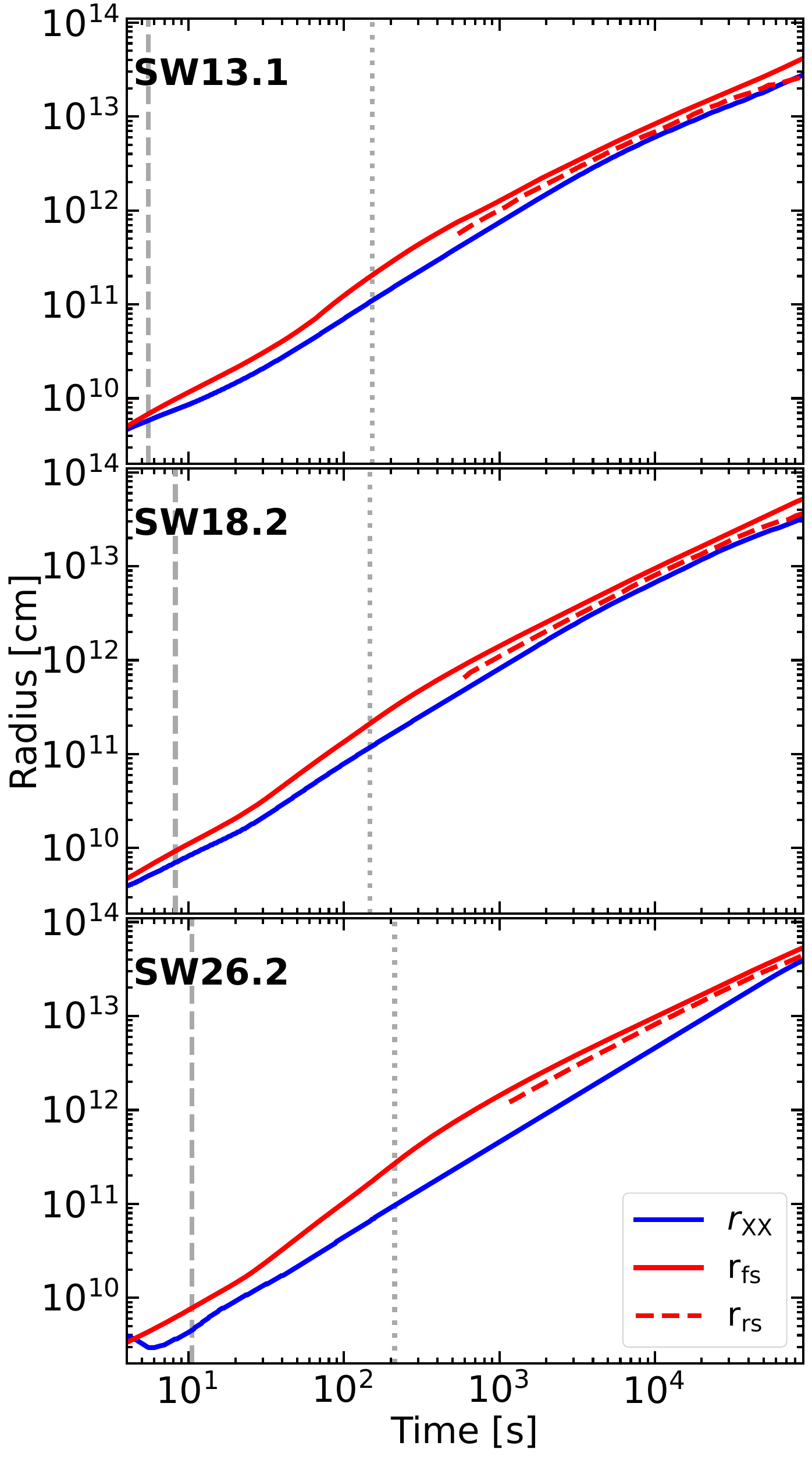}
    \caption{Time evolution until about one day of the angle-averaged radius of the forward SN shock, $r_\mathrm{fs}$ (solid red line), angle-averaged radius of the reverse shock formed by the deceleration of the SN shock in the H envelope $r_\mathrm{rs}$ (dashed red line), and of the inner boundary of the outermost 4\% of NiCoFe0.5X, $r_\mathrm{XX}$ (XX = NiCoFe0.5X, solid blue line), in models SW13.1, SW18.2, and SW26.2. The vertical dashed and dotted grey lines indicate the times when the \ac*{sn} shock passes the (C+O)/He and the He/H interfaces, respectively. Model SW26.2 is the only model among the three in which the shocks and the 4\%-NiCoFe0.5X surface are not very close throughout the entire evolution.}
    \label{fig:rshock_Ni}
\end{figure}
\begin{figure}
    \centering
    \includegraphics[width=0.99\linewidth]{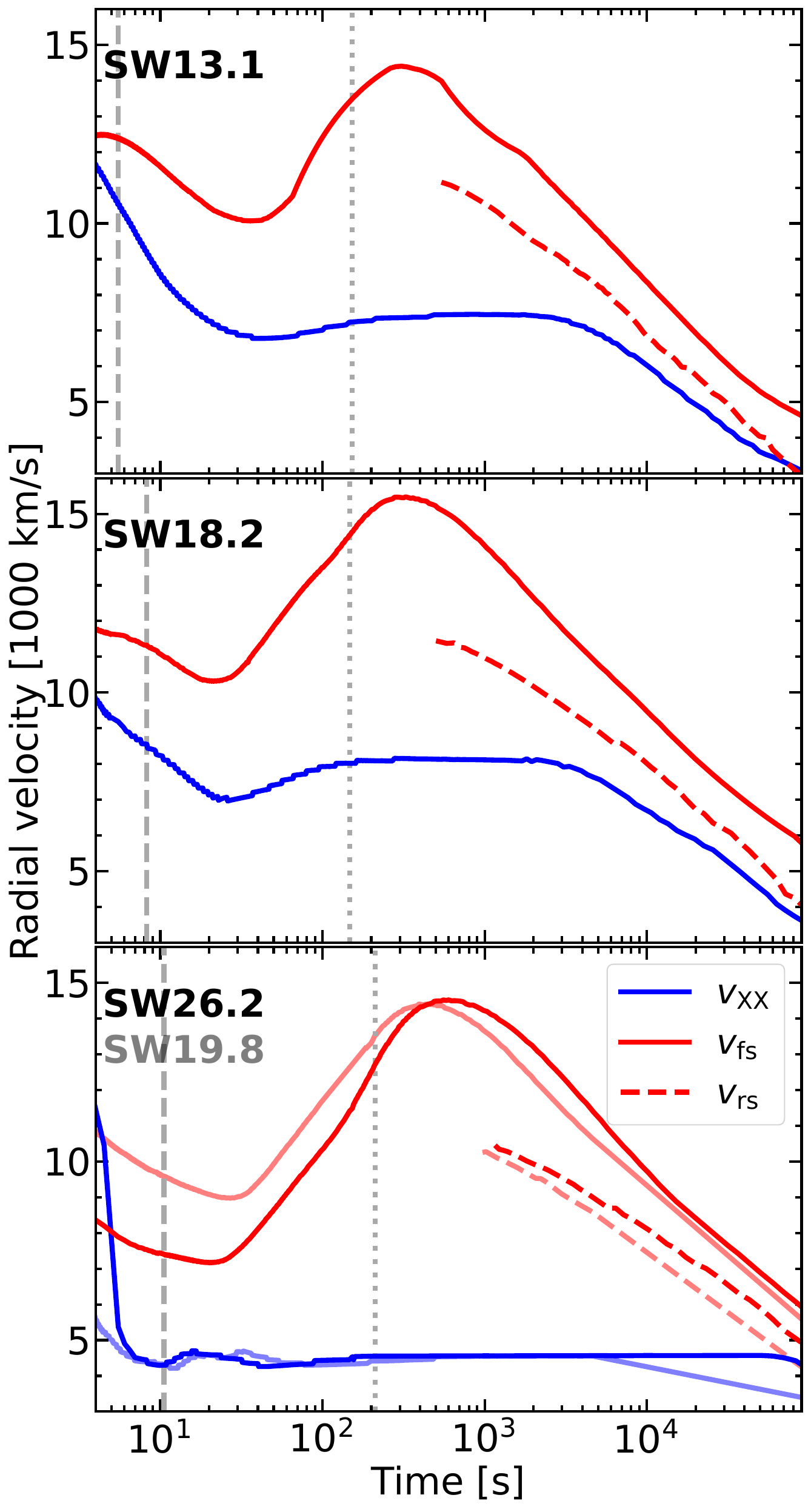}
    \caption{Time evolution until about one day of the velocities of the angle-averaged SN shock, $v_\mathrm{fs}$ (solid red line), of the angle-averaged reverse shock, $v_\mathrm{rs}$ (dashed red line), and of the mass-weighted average velocity $v_\mathrm{XX}$ (XX = NiCoFe0.5X, solid blue line) at the inner boundary $r_\mathrm{XX}$ of the outermost 4\% of NiCoFe0.5X (see Fig.~\ref{fig:rshock_Ni}) for models SW13.1, SW18.2, and SW26.2.
    The vertical dashed and dotted grey lines indicate the times when the \ac*{sn} shock passes the (C+O)/He and the He/H interfaces, respectively.
    In the bottom panel the results of model SW19.8 are also plotted with light-coloured lines. Note that for this additional model the vertical grey lines are not plotted since they are very similar to the respective times for model SW26.2 (see Table~\ref{tab:mixing}). The extreme deceleration of the ejected NiCoFe0.5X in the C+O shells of models SW26.2 and SW19.8 is caused by particularly shallow density profiles in these layers (for more details, see Section~\ref{sec:interactionrev}).
    }
    \label{fig:vshock_Ni}
\end{figure}

\subsubsection{Interaction with the reverse shock from the He/H interface}
\label{sec:interactionrev}

Figs.~\ref{fig:rshock_Ni} and~\ref{fig:vshock_Ni} show the evolution of the angle-averaged positions and velocities of the forward (SN) shock (solid red lines) and the reverse shock (dashed red lines) that is formed when the forward shock passes the He/H interface and gets strongly decelerated in the H envelope of our three sample models. In addition, the radius outside of which 4\% of the total NiCoFe0.5X mass are located as well as the mass-weighted average velocity of these outermost 4\% of the NiCoFe0.5X ejecta are plotted by blue solid lines in the respective figures. It is important to note that these outermost 4\% of the NiCoFe0.5X reside at the largest distances away from the explosion centre at a given time, but they are not necessarily the fastest NiCoFe0.5X ejecta at this instant. The outermost NiCoFe0.5X ejecta may be slowed down by the interaction with the reverse shock, and transiently, some of the NiCoFe0.5X located deeper in the interior may exhibit higher speeds.

The outermost 4\% of the NiCoFe0.5X in SW13.1 (top panel in Fig.~\ref{fig:rshock_Ni}) have a very high initial velocity. At the earliest times, these ejecta even seem to be located outside of the angle-averaged forward shock. This impression, however, is just caused by the different ways to compute the angle-averaged radii, because in each direction NiCoFe0.5X is fully embedded in ejecta behind the \ac*{sn} shock. The corresponding evolution of the velocity is given in the top panel of Fig.~\ref{fig:vshock_Ni}. The initially very fast NiCoFe0.5X ejecta (blue line) slow down to velocities much slower than the averaged forward shock velocity (red solid line) well within the He shell, and then reach similar velocities compared to this shock at later times ($t\gtrsim 3000$\,s) when the \ac*{sn} shock is strongly decelerated in the H envelope. Therefore, when the \ac*{sn} shock has passed the He/H interface (dotted grey line) the outermost NiCoFe0.5X ejecta still follow the average forward shock rather closely. Shortly after this time, the reverse shock forms just slightly in front of the fastest NiCoFe0.5X. Since this NiCoFe0.5X quickly crosses the reverse shock, it is not slowed down significantly, and the reverse shock is considerably faster than the outermost NiCoFe0.5X ejecta only for a short period of time after it has formed. As a result, the fast NiCoFe0.5X-rich matter is involved in the growth of RT fingers, which carry NiCoFe0.5X into the H+He envelope of the progenitor with maintained high velocities, not affected by significant deceleration in the surrounding medium (left panels of the bottom two rows of Fig.~\ref{fig:ni_iso}). The tips of these RT fingers get compressed and correspondingly flattened when they reach the forward shock and get slowed down substantially. 

The evolution of model SW18.2 is very similar. Also here the outermost NiCoFe0.5X remains rather close to the average forward shock all of the time. Once the reverse shock is formed and the outermost NiCoFe0.5X interacts with it, the tips of the fingers containing 96\% of the ejected NiCoFe0.5X mass experience visible flattening (Fig.~\ref{fig:ni_iso}, middle column, third row).

\begin{figure*}
    \centering
    \includegraphics[width=0.4\linewidth]{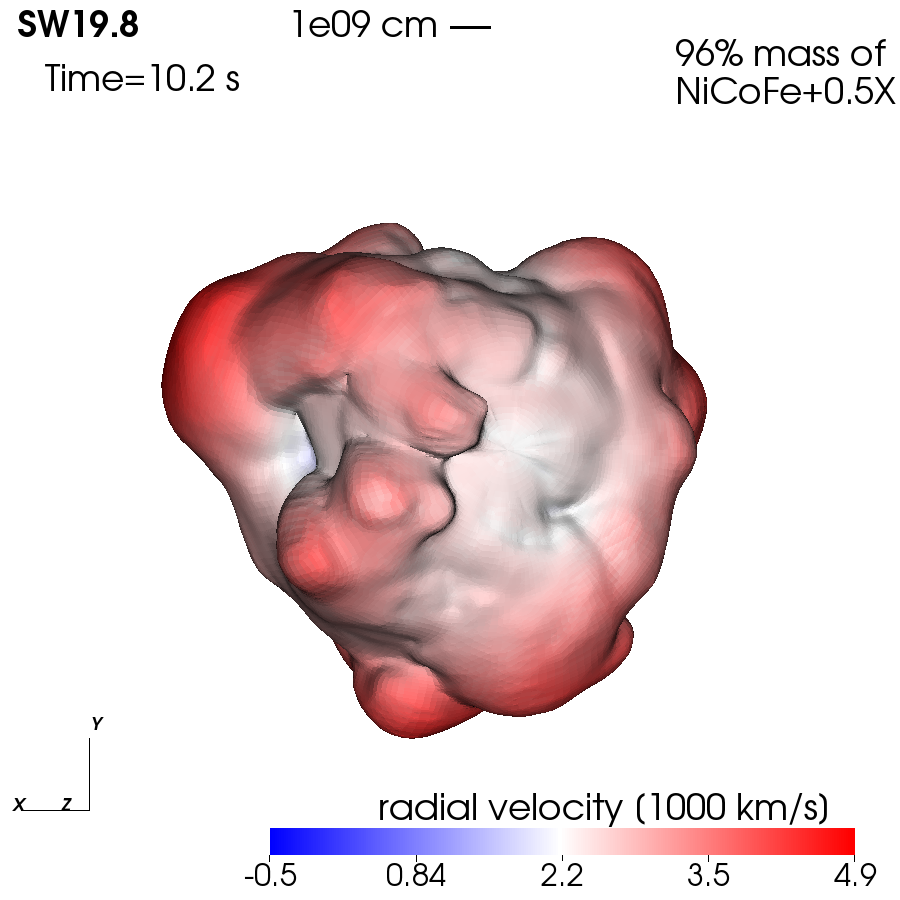}
    \includegraphics[width=0.4\linewidth]{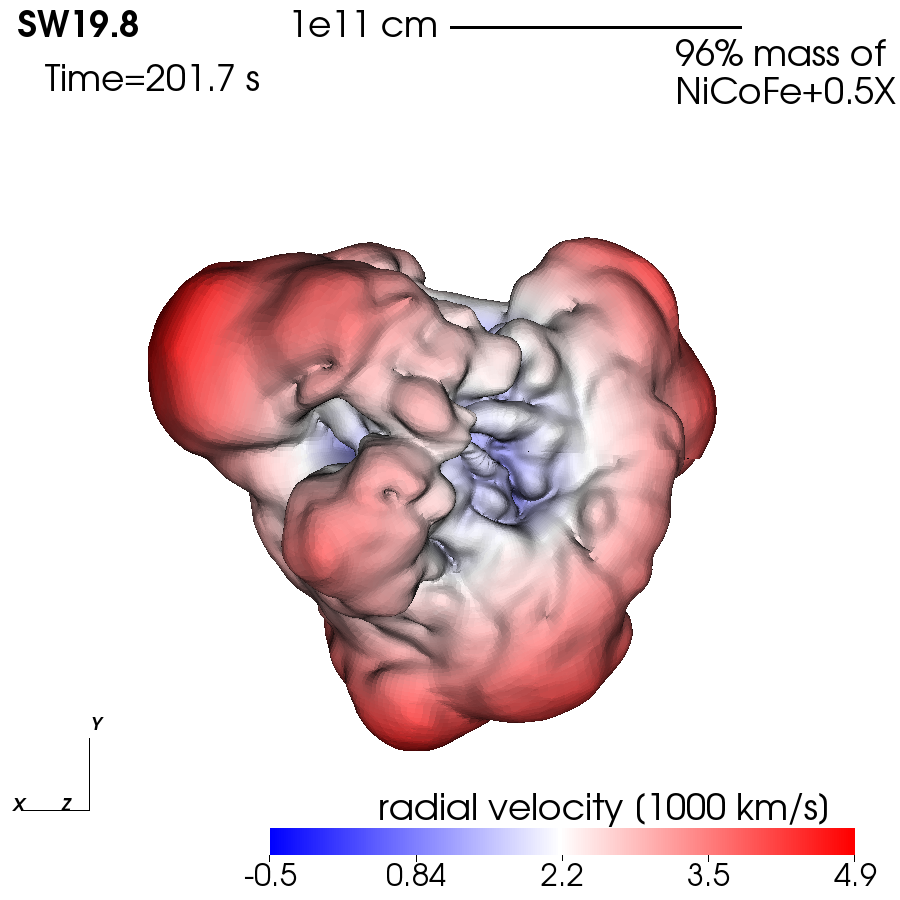}\\
    \includegraphics[width=0.4\linewidth]{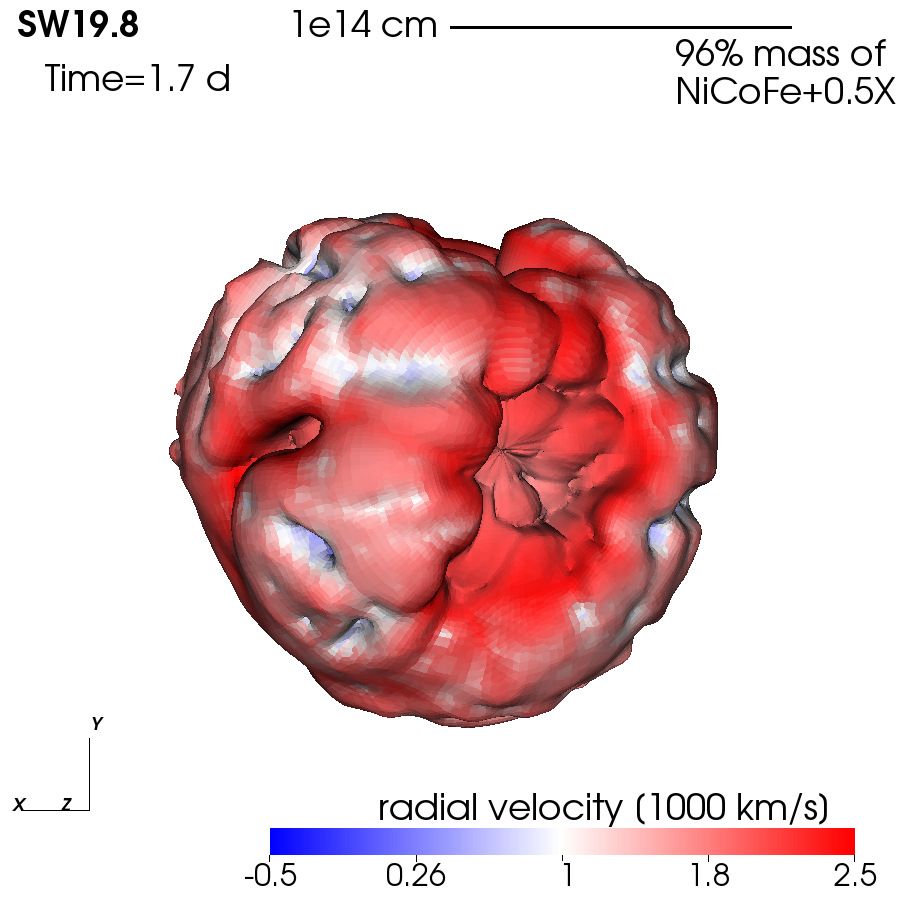}
    \includegraphics[width=0.4\linewidth]{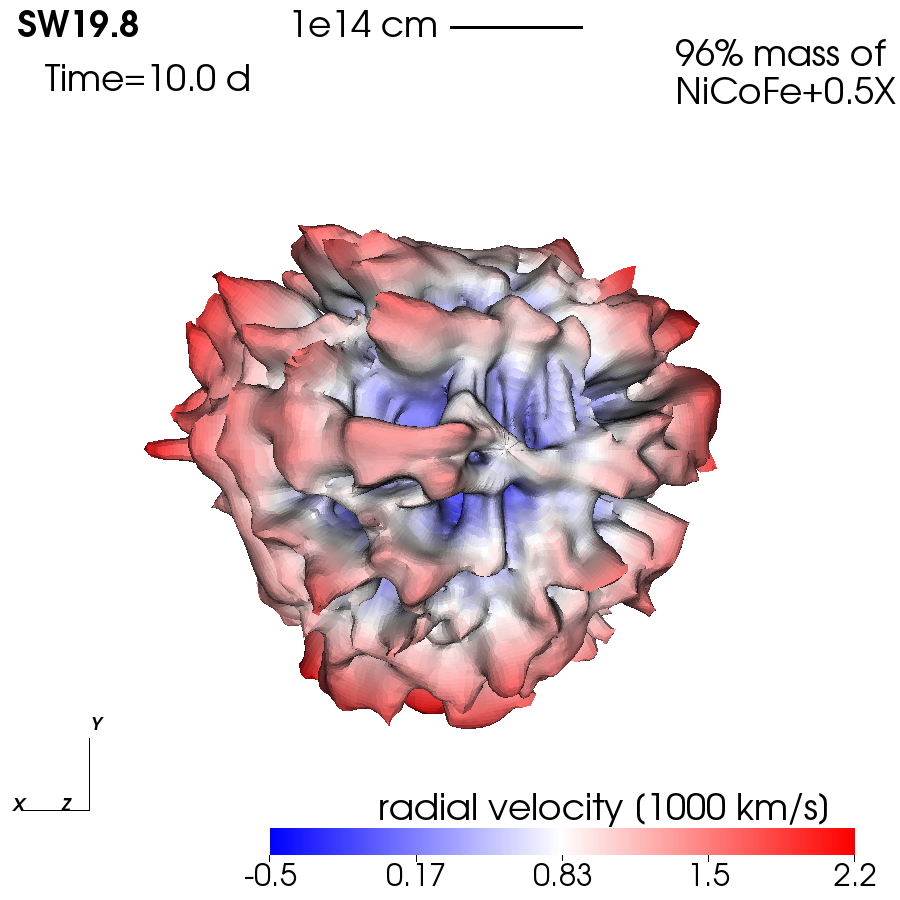}
    \caption{Isosurfaces of mass fraction of NiCoFe0.5X containing $96\%$ of the total mass of NiCoFe0.5X for model SW19.8. The displayed times are selected analogue to Fig.~\ref{fig:ni_iso} and the sequence runs from top left to bottom right. The colour coding represents the radial velocity. The orientation, i.e., viewing direction, is the same as in Fig.~\ref{fig:ni-plumes_cat3}. The mass fractions of the first three images of the sequence lie in the range $0.007 < X_\mathrm{NiCOFe0.5X} < 0.08$, and for the last image at 10\,d the mass fraction is $X_\mathrm{NiCOFe0.5X} \approx 0.007$}
    \label{fig:ni_iso198}
\end{figure*}

In the case of model SW26.2 the situation is different. The outermost 4\% of the NiCoFe0.5X ejecta always stay well behind the forward shock (Fig.~\ref{fig:rshock_Ni}, bottom panel) and expand with a much lower average velocity (Fig.~\ref{fig:vshock_Ni}, bottom panel). Consequently, the reverse shock forms far outside  of these ejecta and the latter approach the shock only after several $10^4$\,s.
Accordingly, the velocities of the two shocks are several 1000\,km\,s$^{-1}$ higher than the average velocity of the outermost 4\% of the NiCoFe0.5X-rich ejecta until about $5\times\unit[10^4]{s}$ (Fig.~\ref{fig:vshock_Ni}). For this reason even the outermost NiCoFe0.5X ejecta are not significantly involved in the growth of the \ac*{rti} at the He/H composition interface, and most of the slow-moving NiCoFe0.5X gets stuck in the dense shell of He (the He wall mentioned in Sect.~\ref{sec:RTIs_1d}) that piles up at the base of the H envelope when the material of the He shell is decelerated by the reverse shock. This pile-up of matter leads to the local peak in the mass distributions at about 15\,M$_\odot$ in the bottom panel of Fig.~\ref{fig:radial_mixing_m}, which also shows that there is hardly any mixing of NiCoFe0.5X beyond this local peak into the H+He envelope of the progenitor. The weak growth of \ac*{rti} fingers including NiCoFe0.5X can also be recognized in Fig.~\ref{fig:ni_iso}, see the third and fourth rows in the right column there.

In the bottom panel of Fig.~\ref{fig:vshock_Ni} we also display the results for model SW19.8 by light-coloured lines. Both the velocity of the shock and the average velocity of the outermost 4\% of the NiCoFe0.5X behave similarly to those in model SW26.2. In particular, the NiCoFe0.5X in both models is strongly decelerated in the very early phase of the explosion ($t \lesssim \unit[5]{s}$). This efficient deceleration is caused by the NiCoFe0.5X interaction with the C+O core, which is much more massive and has a shallower radial density distribution in these models compared to the case of SW18.2 and, in particular, of model SW13.1 (see inset in Fig.~\ref{fig:densityProfile}). Once the fastest NiCoFe0.5X ejecta have been decelerated to velocities around $\unit[5000]{km\,s^{-1}}$, they are not able to re-accelerate later on but keep on moving with an almost constant velocity (bottom panel of Fig.~\ref{fig:vshock_Ni}). Correspondingly, also in model SW19.8 the mixing of NiCoFe0.5X into the H+He envelope is inefficient and the final distribution of NiCoFe0.5X is rather spherical with only small-scale structures associated with weakly developed RT fingers (Fig.~\ref{fig:ni_iso198}). These similarities further justify our inclusion of model SW19.8 in the class of HM-HE models.

\subsubsection{The effects of multi-dimensional instabilities}
\label{sec:multid-effects}

In general, the velocity distributions of the chemical elements in the \ac*{ccsn} ejecta are determined by the explosion energy (setting the overall scale), the density gradients in the progenitor star (allowing for deceleration or acceleration), the $\beta$ decay of radioactive elements \citep[further accelerating NiCoFe0.5X-rich ejecta, see][for a detailed discussion]{gabler2021}, and the development of non-linear and non-radial hydrodynamic instabilities. While the first three effects can be studied to a certain extent in 1D, the genuinely multi-dimensional instabilities affecting the motion and mixing of the ejecta, including in particular NiCoFe0.5X, require full 3D simulations for a reliable assessment of their consequences. To better understand the impact of \acp*{rti}, we thus compare our 3D models to 1D simulations. 

\begin{figure}
    \centering
    \includegraphics[width=.93\linewidth]{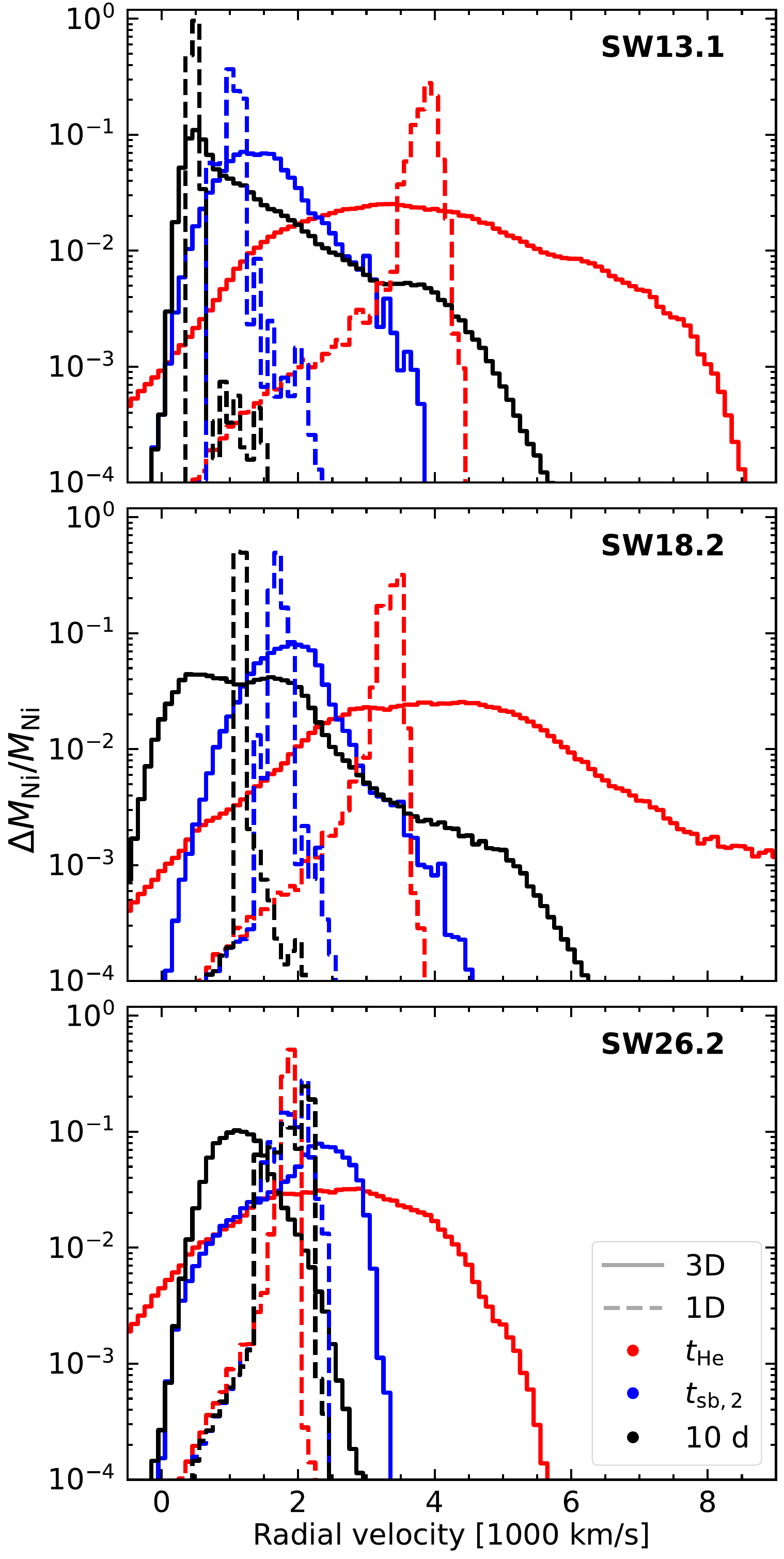}
    \caption{Normalized mass distributions of NiCoFe0.5X versus radial velocity at different times in 1D (dashed lines) and 3D (solid lines) for our reference models.
    The distributions are plotted when the \ac*{sn} shock passes the He/H composition interface ($t_\mathrm{He}$, red lines), at the time when the minimum radius of the shock breaks out of the progenitor star ($t_\mathrm{sb,2}$, blue lines), 
    and at the end of the simulation (black lines). The specific values of these times can be found in Table~\ref{tab:mixing}.
    The 1D data are rescaled by $\sqrt{E^\mathrm{3D}_\mathrm{exp} / E^\mathrm{1D}_\mathrm{exp}}$, 
    where $E^\mathrm{1D}_\mathrm{exp}$ and $E^\mathrm{3D}_\mathrm{exp}$ are given in Table~\ref{tab:3Dexplosion}. Negative velocities indicate some fallback matter in the innermost ejecta.}
    \label{fig:mix_evol}
\end{figure}

\acp*{rti} affect especially the mixing of the fastest of the heavy and intermediate-mass elements. Therefore, we expect the most prominent differences between 1D and 3D to occur in the tails and wings of the mass distributions of these elements, e.g., of NiCoFe0.5X, for models that show particularly strong mixing. The corresponding normalized distributions of the mass fraction of NiCoFe0.5X as functions of radial velocity for our reference models at different times are shown in Fig.~\ref{fig:mix_evol}.

For model SW13.1 (top panel) the peaks of the distributions in 3D and in 1D coincide roughly at all selected times. However, the distributions are much wider and the fastest NiCoFe0.5X is significantly faster in 3D. Initially (when the SN shock passes the He/H interface), the NiCoFe0.5X-rich ejecta have a very wide velocity distribution reaching up to more than $\unit[4000]{km\,s^{-1}}$ ($\gtrsim\unit[8000]{km\,s^{-1}}$) in 1D (3D). During the outward expansion and the interaction with the reverse shock, the ejecta are slowed down significantly until shock breakout ($t_\mathrm{sb,2}$, blue lines), resulting in a much narrower velocity distribution, in particular in 1D. The wider spread of the velocities on both sides of the distribution maximum and the much higher maximum velocities in 3D indicate very strong radial mixing. Note that the high-velocity tail of the distribution of model SW13.1 in 3D (in contrast to 1D) experiences a considerable acceleration from shock breakout up to $\unit[10]{d}$, because a substantial fraction of the NiCoFe0.5X has been mixed into the outermost layers of the progenitor.

As in model SW13.1, the peaks of the normalized mass distributions in model SW18.2 (central panel of Fig.~\ref{fig:mix_evol}) in 3D and 1D are at about the same velocities, but they are much flatter in 3D and the distributions are correspondingly much wider. This is compatible with the strong mixing that occurs also in model SW18.2. Compared to model SW13.1, however, the distribution maxima of the 3D version of SW18.2 are much broader at late times and even develop a double-peak structure. The 1D maximum lies right between these peaks. Similarly to SW13.1, the tail of the distribution in the HM-LE model SW18.2 displays a significant change between shock breakout and $\unit[10]{d}$.   

Different from the other two models, the peak and width of the distributions of model SW26.2 evolve only very little with time in 1D (bottom panel of Fig.~\ref{fig:mix_evol}), mainly because of the lack of significant shock deceleration in the He shell (Fig.~\ref{fig:vshock_Ni}, bottom panel), where the gradient of $\rho r^3$ is negative in the progenitor (Fig.~\ref{fig:rhor3}). In the corresponding 3D simulations, the peak changes its position from about $\sim\unit[3000]{km\,s^{-1}}$ at the time the SN shock crosses the He/H interface to $\sim\unit[1000]{km\,s^{-1}}$ at $t=\unit[10]{d}$. The spread of the distributions is larger in 3D compared to 1D, but the extended tails at high velocities are absent, compatible with the weak mixing of NiCoFe0.5X into the H envelope in model SW26.2.

Therefore, in contrast to the 1D simulations, 3D models with efficient RT mixing exhibit some acceleration of the fastest NiCoFe0.5X from $t=t_\mathrm{sb,2}$ to $10\,$d. Such a late-time acceleration can be caused by four effects: acceleration in a steep density gradient of the CSM around the progenitor star, heating due to $\Ni$ decay, re-acceleration by the self-reflected reverse shock from the He/H interface, and ongoing RTI growth. While the first three effects are also present in 1D and, in particular, $\beta$ decay and self-reflected shock acceleration should affect more the bulk of the ejecta, late-time acceleration due to RTI is the most likely cause of the velocity increase of the fastest ejecta. This reasoning is supported by the last two rows of panels for models SW13.1 and SW18.2 in Fig.~\ref{fig:ni_iso}. There, we see an increase in the maximum velocity of the colour scale with increasing time, and the RT fingers extend much further after shock breakout. Since the NiCoFe0.5X-rich matter in model SW26.2 does not show strong RTI, the absence of late-time acceleration in this model is consistent with this line of argument.

In Table~\ref{tab:ni_vels} we provide the characteristic velocities computed from the mass distributions of NiCoFe0.5X for the 1D and 3D versions of all of our models at the time when the shock (minimum radius of the shock, $t_\mathrm{sb,2}$) breaks out from the star in 1D (3D): the mass-weighted average velocities of the fastest 4\% of the NiCoFe0.5X, $v_\mathrm{NiCoFe0.5X,4\%}$, the mass-averaged bulk velocities of the remaining 96\%, $v_\mathrm{NiCoFe0.5X,bulk}$, and the velocities of the peaks of the mass distributions, $v_\mathrm{peak}$,. In particular, for the 1D simulations, the LM models display a trend to lower bulk velocities than the HM-LE and HM-HE models. 
In 3D, the bulk velocities of LM models are considerably higher than in 1D due to the strong mixing, and the separation to the HE models does not appear as extreme. 
Among the 3D HM models the HM-HE tend to have higher velocities of the bulk and of the peak of the distributions than the HM-LE models. While there are some exceptions, like the low $v_\mathrm{peak}^\mathrm{1D}$ of model SW27.0, these trends are slightly clearer for the peak velocity compared to the bulk velocities. 
Almost all LM and HM-HE models show higher peak and bulk velocities in 3D compared to 1D, while the HM-LE models show comparable 1D and 3D velocities without a clear trend towards faster (bulk) ejecta in 3D.

In 1D, there is a clear correlation between the bulk velocity and $v_\mathrm{NiCoFe0.5X, 4\%}^\mathrm{1D}$. The LM models have the lowest velocities for both, and the velocities of HM-LE models are significantly faster. The absolute velocities of the HE-HM models show a larger dispersion, however, the correlation between $v_\mathrm{NiCoFe0.5X, bulk}^\mathrm{1D}$ and $v_\mathrm{NiCoFe0.5X, 4\%}^\mathrm{1D}$ remains. In 3D, this correlation does not hold. The LM models have still the lowest bulk velocities, while, in contrast, their $v_\mathrm{NiCoFe0.5X, 4\%}^\mathrm{3D}$ are amongst the highest. 
Due to their stronger mixing, the LM models exhibit very high velocities of the fastest NiCoFe0.5X. 
Though some of the models (e.g. SW16.3) show similar velocities of the distribution peaks in 1D and in 3D, other models do not agree that well, and the differences between the velocities of the peaks of the distributions in 3D and 1D are generally comparable to the differences between the corresponding bulk velocities. Therefore, we will continue our analysis with the bulk velocities as reference values. In addition, $v_\mathrm{NiCoFe0.5X, bulk}^\mathrm{3D}$ was also considered by \citet{utrobin2019} and \citet{utrobin2021}, whose discussion we partly extend to \acp*{rsg} here. 

We point out somewhat larger differences between the 3D and 1D results of model SW27.0 compared to our other simulations for progenitors with similar masses. 
This peculiarity of model SW27.0 is connected to a strong initial expansion of the stalled shock, which breaks down because of insufficient push by neutrino heating and is followed by a shock-contraction phase before the explosion of the 1D model ultimately sets in. Such a dynamical behaviour of the shock with one or more transient expansion and contraction episodes preceding the final explosion is a feature that was witnessed in 1D models before \citep[see, e.g.,][]{buras2006,ertl2016}, but occurs very rarely in 3D. As a consequence, the explosion of model SW27.0 in 3D sets in earlier than its 1D counterpart and its explosion energy becomes about 10\% higher than that of model SW27.0 in 1D (Table~\ref{tab:3Dexplosion}). Although the long-term evolution of the shock velocity and of the outer ejecta is similar in the 1D and 3D simulations, the region where most of the NiCoFe0.5X forms is located at slightly smaller radii in the 1D simulation of SW27.0. During the later evolution this leads to considerably lower NiCoFe0.5X velocities in the 1D case despite the fact that the total kinetic energy of all ejecta of the two models is comparable. Since the explosion dynamics of SW27.0 in 3D resembles that of the other progenitors with comparable masses, the results for the NiCoFe0.5X velocities of these 3D models are more similar to the other models of comparable mass.

\subsection{Evolution of 3D morphologies}
\label{sec:morph}

Fig.~\ref{fig:ni_iso} shows snapshots of the time evolution of the isosurfaces that contain 96\% of the total NiCoFe0.5X mass with the highest mass fraction in our representative models SW13.1, SW18.2, and SW26.2. 
The mass fractions enclosing $96\%$ of the mass are typically lower for later times and reach from $X_\mathrm{NiCoFe0.5X}\sim0.1$ at $2.5\,$s (Figs.~\ref{fig:ni-plumes_cat1}--\ref{fig:ni-plumes_cat3}) to about $X_\mathrm{NiCoFe0.5X}\gtrsim0.005$ at $10\,$d. This decrease is related to the expansion and mixing of the NiCoFe0.5X-rich ejecta and stalls once the expansion is homologous. The colour coding represents the radial velocities of the matter on this surface. 

The NiCoFe0.5X morphology at the times when the SN shock crosses the (C+O)/He shell interface (top row of panels in Fig.~\ref{fig:ni_iso}) still reflects the 3D NiCoFe0.5X distribution shortly after the onset of the explosion when $\Ni$ nucleosynthesis is finished (see Figs.~\ref{fig:ni-plumes_cat1}--\ref{fig:ni-plumes_cat3}). The subsequent changes of the geometry of the chosen surface are affected by the shock deceleration and acceleration phases in the stellar composition shells, the \acp*{rti} at the composition-shell interfaces, and the reverse shock that is formed when the SN shock is decelerated upon travelling though the extended \ac*{rsg} envelope. When the SN shock slows down in the He shell, the most elongated initial NiCoFe0.5X plumes can transiently get compressed before they grow again by \acp*{rti} at the (C+O)/He interface. This can be seen when comparing the morphology at subsequent times in the different rows of Fig.~\ref{fig:ni_iso}. The time increases from top to bottom.

In model SW13.1, the most extended NiCoFe0.5X fingers are clearly connected to the largest initial NiCoFe0.5X plumes, but their shapes change over time. At the time when the shock crosses the He/H interface (second row of panels in Fig.~\ref{fig:ni_iso}), these large plumes have already undergone one phase of RTI growth and extend farther from the central ejecta. The initial mushroom-head-like structures are less clearly defined. In contrast, in models SW18.2 and SW26.3 the NiCoFe0.5X plumes at about the same time (roughly 150\,s--220\,s post bounce) are still more similar to those at the earlier times in the top row. The mushroom-like NiCoFe0.5X fingers have actually become more stretched and possess more prominent mushroom heads, because these characteristic features could grow (nearly) unaffected by the effects of SN shock deceleration in the He shell and by the growth of \ac*{rti} at the (C+O)/He interface, both of which are weak in these two models (Section~\ref{sec:RTIs_1d}).  

At $t\sim t_\mathrm{sb,2}$ (third row in Fig.~\ref{fig:ni_iso}) the global asymmetries imprinted on the NiCoFe0.5X distribution by the generically anisotropic explosion mechanism can still be witnessed in all cases, but the effects of the \ac*{rti} at the He/H interface and that of the reverse shock building up between $\sim$500\,s and $\sim$1000\,s (see Fig.~\ref{fig:rshock_Ni}) are clearly visible now. The most prominent NiCoFe0.5X mushrooms exhibit clear signs of fragmentation to smaller structures and of corrugation due to a second phase of RT growth in all three models. Mushroom ``splitting'' occurs in models SW13.1 and SW18.2 and mushroom merging and surface roughening is visible in model SW26.2. Moreover, the biggest NiCoFe0.5X plumes in SW26.2 are particularly strongly affected by compression due to the fact that the reverse shock forms far ahead of the outermost NiCoFe0.5X (see Section~\ref{sec:interactionrev} and Fig.~\ref{fig:rshock_Ni}). In model SW18.2, the reverse shock has massively decelerated the weaker and initially slower of the two most prominent initial NiCoFe0.5X fingers (at the 9 o'clock position in the middle panel of the third row in Fig.~\ref{fig:ni_iso}), whereas in model SW13.1, the reverse shock has only just started to interact with the matter behind the most extended NiCoFe0.5X plumes. Some of these plumes have already penetrated into the H envelope through the growth of \ac*{rti} at the He/H interface, while others still exhibit deceleration at the tips of their heads with faster moving matter pushing from below (see also top panel in Fig.~\ref{fig:rshock_Ni}). We note that the fastest NiCoFe0.5X finger in model SW18.2 at the 5 o'clock position contains extremely little mass (compatible with Figs.~\ref{fig:radial_mixing_v}, \ref{fig:radial_mixing_m}, and~\ref{fig:mix_evol}), which explains why its velocity is still nearly 8000\,km\,s$^{-1}$ at shock breakout, although the fastest 4\% of the NiCoFe0.5X in this model should have much lower velocities at this time when extrapolating the velocity evolution seen in Fig.~\ref{fig:vshock_Ni}.

At 10 days (bottom row in Fig.~\ref{fig:ni_iso}), prominent, very elongated fingers have reappeared in models SW13.1 and SW18.2 in the directions of the largest NiCoFe0.5X plumes existing at the earliest times (top row). These structures have grown again because of the action of \ac*{rti} at the He/H interface and possibly partial acceleration of the outermost NiCoFe0.5X by radioactive heating though the decay of $\Ni$ to $\Co$. At the same time the reverse shock has compressed the central ejecta such that the RT fingers stick out even more pronounced. The outermost NiCoFe0.5X in models SW13.1 and SW18.2 reaches the SN shock, which deforms the caps of these fingers into anvil-like shapes. This effect is not visible in model SW26.2, where NiCoFe0.5X is only weakly mixed into the H envelope (Section~\ref{sec:radmixing} and Fig.~\ref{fig:radial_mixing_m}). The strong compression and deceleration by the reverse shock have led to a more spherical shape of the NiCoFe0.5X ejecta at 10\,d in this model, although their global geometry is still preserved from the early times. Considerable fragmentation has taken place due to the \ac*{rti} at the He/H boundary and has created fine structure and smaller clumps in the final NiCoFe0.5X distribution. 

These effects observed in model SW26.2 are even more extreme in model SW19.8 (Fig.~\ref{fig:ni_iso198}). In this latter model, NiCoFe0.5X is similarly strongly decelerated to maximum velocities of only around 5000\,km\,s$^{-1}$ when the SN shock slows down while propagating through the (C+O) layer (Fig.~\ref{fig:vshock_Ni}). In this very early and energetic explosion, the initial NiCoFe0.5X plumes are more isotropically distributed than in the other discussed reference models, and the overall shape of the NiCoFe0.5X isosurface therefore looks more spherical (Figs.~\ref{fig:ni-plumes_cat1}--\ref{fig:ni-plumes_cat3}). In the absence of very prominent seed asymmetries and due to the lack of substantial RTI growth at the (C+O)/He interface, no extended NiCoFe0.5X fingers form until $\sim$$\unit[200]{s}$ (upper right panel of Fig.~\ref{fig:ni_iso198}) and the ejecta display a less extreme asphericity than in the other models at about the same time (Fig.~\ref{fig:ni_iso}). 
Subsequently, after the SN shock has entered the H envelope and once again experiences a phase of deceleration, the reverse shock forms far ahead of the slowly moving NiCoFe0.5X (Figs.~\ref{fig:rshock_Ni} and~\ref{fig:vshock_Ni}). The reverse shock is particularly strong because of the energetic explosion of model SW19.8. After about 1 day it begins to directly interact with the outermost NiCoFe0.5X and compresses the heads of the mildly developed $\Ni$-rich RT mushrooms. This is visible as the slower whitish-bluish NiCoFe0.5X ahead of faster, red-coloured material deeper inside in the lower left panel of Fig.~\ref{fig:ni_iso198}. 
The massive He wall that builds up at the base of the H envelope and the late growth of \ac*{rti} at the He/H interface, which takes place quite deep inside the H shell in this model, does not permit efficient NiCoFe0.5X mixing into the outer layers of the progenitor. The final NiCoFe0.5X surface in Fig.~\ref{fig:ni_iso198} (lower right panel) is therefore closer to a sphere that is nearly isotropically perturbed only by small NiCoFe0.5X fingers of roughly the same size in all directions.              

\begin{table*}
    \centering
    \begin{tabular}{c|ccccccc}
    \hline
        Model & $t_\mathrm{sb,1}$ & $t_\mathrm{sb,2}$ & $\Delta t_\mathrm{sb}$ & $\Delta t_\mathrm{sb}\,t_\mathrm{sb,1}^{-1}$&$R_\mathrm{shock}^\mathrm{max}/R_\mathrm{shock}^\mathrm{min}$ &$A_\mathrm{shock,1}$ &$A_\mathrm{shock,2}$\\
        & \multicolumn{3}{c}{[d]} &&&&\\
        \hline
        WH12.5 & 1.09 & 1.56 & 0.47 &0.43 &1.32 &0.026 &0.0046 \\
        SW13.1 & 1.23 & 1.51 & 0.28 &0.23 &1.21 &0.019 &0.0015 \\
        SW14.2 & 1.20 & 1.75 & 0.55 &0.46 &1.40 &0.048 &0.0105 \\
        \hline
        SW16.3 & 1.33 & 1.53 & 0.20 &0.15 &1.15 &0.018 &0.0025 \\
        SW18.2 & 1.43 & 1.96 & 0.53 &0.37 &1.29 &0.027 &0.0052 \\
        SW20.8 & 2.23 & 3.04 & 0.81 &0.36 &1.41 &0.049 &0.0144 \\
        SW21.0 & 1.92 & 2.58 & 0.66 &0.34 &1.24 &0.029 &0.0054 \\
        \hline
        SW19.8 & 1.74 & 1.84 & 0.10 &0.06 &1.06 &0.005 &0.0017 \\
        SW25.5 & 2.36 & 2.61 & 0.25 &0.11 &1.08 &0.008 &0.0039 \\
        SW25.6 & 2.25 & 2.83 & 0.58 &0.26 &1.20 &0.019 &0.0046 \\
        SW26.2 & 2.58 & 2.89 & 0.31 &0.12 &1.10 &0.010 &0.0020 \\
        SW27.0 & 2.37 & 2.73 & 0.36 &0.15 &1.12 &0.016 &0.0020 \\
        SW27.3 & 2.24 & 2.79 & 0.55 &0.25 &1.17 &0.022 &0.0019 \\
        \hline
    \end{tabular}
    \caption{ 
    Shock breakout times (in days) in all models listed in column~1, for the maximum radius of the forward shock ($t_\mathrm{sb,1}$, column~2), the minimum radius of the
    forward shock ($t_\mathrm{sb,2}$, column~3), and the difference between the two ($\Delta 
    t_\mathrm{sb}$, column~4). The fifth column gives the ratio of the columns 4 and 2, and sixth, seventh, and eighth columns list the ratio of maximum to minimum shock radius and the normalized dipole and quadrupole moments of the SN shock at time $t_\mathrm{sb,1}$ and thus before the shock gets distorted by running into the CSM.}
    \label{tab:sb_times}
\end{table*}

\subsection{Asymmetric shock breakout}\label{sec:sb}

Early-time observations of a growing number of different kinds of \acp*{ccsn} have been able to capture the shock-breakout phase \citep[e.g.,][]{garnavich2016,barbarino2017,foerster2018,huang2018,singh2019,sounagnac2019,xiang2019} and thus provide important insights into the near-surface structure of the progenitors and of their immediate surroundings, which were shaped by the mass loss that had taken place shortly before the stars collapsed \citep[e.g.,][]{dessart2017,rubin2017,bersten2018,kozyreva2020}. Effects associated with the intrinsically 3D structure of fully convective RSG envelopes \citep[e.g.,][and references therein]{chiavassa2011,chiavassa2024} and the lower-density ``halo'' of material outside of the traditional photosphere in corresponding 3D models \citep{goldberg2022} as well as inhomogeneities in the close circumstellar environment \citep{fryer2020} can have an important impact on the rise time, duration, luminosity, and direction dependence of the shock-breakout signal and on conclusions that can be drawn on the stellar radius. A similarly important role can be expected for the asphericity of the SN shock due to the generically asymmetric explosions \citep{stockinger2020,kozyreva2022,vartanyan+2025}.

For this reason we provide the times (in days) for the breakout of the deformed shocks in all of our 3D SN simulations in Table~\ref{tab:sb_times}, specifically the breakout times of the maximum radius of the forward 
shock, $t_\mathrm{sb,1}$, those of the minimum radius of the forward shock, $t_\mathrm{sb,2}$, and the differences between these two instants, $\Delta t_\mathrm{sb}$. In all models the outermost parts of the shock arrive at the stellar surface within $\unit[1]{d} < t_\mathrm{sb,1} < \unit[3]{d}$, which is in 
line with previous findings for energetic explosions of \acp*{rsg} \citep{wongwathanarat2015,gabler2021,vartanyan+2025}, and the slowest parts of the shock reach the surface between 0.1\,d and $\sim$0.8\,d later.

Generally, the breakout times $t_\mathrm{sb}$ of our models tend to correlate with the progenitor radii, and for our set of RSGs, also with the progenitor masses (see the overall trend of increasing values of $t_\mathrm{sb,1}$ and $t_\mathrm{sb,2}$ from top to bottom in Table~\ref{tab:sb_times}). Because of 
\begin{equation}
    t_\mathrm{sb} \sim \frac{R_\mathrm{prog}}{\langle v_\mathrm{sh}\rangle} \propto R_\mathrm{prog}\,\sqrt{\frac{M_\mathrm{ej}}{E_\mathrm{exp}}} \,,
    \label{eq:tsbexpression}
\end{equation} 
where we have used that the time-averaged shock velocity depends on the explosion energy per ejecta mass, $\langle v_\mathrm{sh}\rangle \propto \sqrt{E_\mathrm{exp} M_\mathrm{ej}^{-1}}$, the exact correlation is proportional to the stellar radius and the square root of the ejecta mass, and inversely proportional to the square root of the explosion energy. Since the ejecta masses differ by at most 50\% between the different models (Table~\ref{tab:mixing}), mainly the other two competing terms determine the behaviour. On the one hand, the explosion energy of our 3D models varies by up to a factor $\sim$2.5 (between 0.69\,B and 1.73\,B; Table~\ref{tab:3Dexplosion}) between LM and HM-HE cases, and the average shock velocity scales roughly with the square root of this energy. On the other hand, the progenitor stars of the LM cases possess smaller radii (roughly half the values of the HM-HE cases; Table~\ref{tab:presn}) and the shocks reach the stellar surfaces more quickly. Therefore our LM models typically possess smaller breakout times than the models with higher masses, but because of the two opposing effects the correlation is not perfect. For example, model SW20.8 has a significantly larger breakout time ($t_{\mathrm{sb},1}$ as well as $t_{\mathrm{sb},2}$ in Table~\ref{tab:sb_times}) than models with similar masses and radii (e.g., SW19.8 and SW21.0), because the explosion energy of SW20.8 is particularly low (Table~\ref{tab:3Dexplosion}). Similarly, SW13.1 has a larger $t_{\mathrm{sb},1}$ than SW14.2, although its radius is smaller. And also SW26.2 has larger $t_{\mathrm{sb},1}$ and $t_{\mathrm{sb},2}$ than SW27.0 and SW27.3 despite its smaller progenitor radius, which again is a consequence of the lower explosion energy of model SW26.2.

\begin{figure}
    \begin{center}
    \begin{minipage}{0.30\textwidth}
      \includegraphics[width=\linewidth]{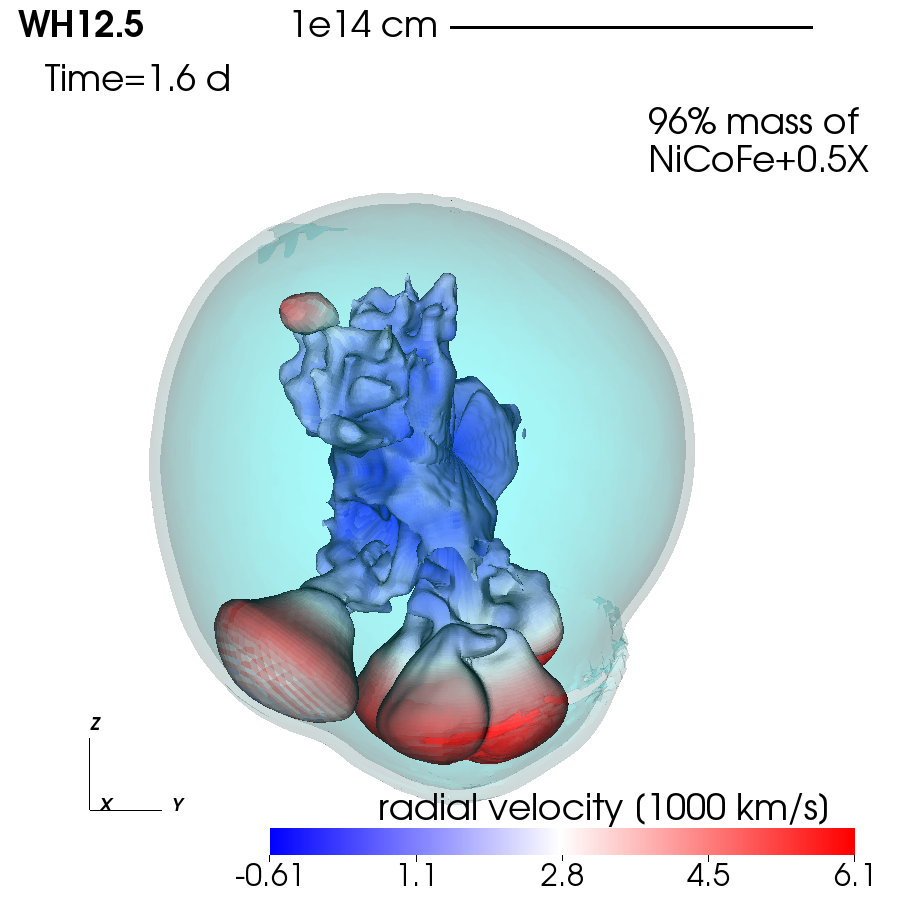}
    \end{minipage}
  \end{center}
  \caption{Isosurface of the NiCoFe0.5X mass fraction that contains 96\% of the total ejecta mass of NiCoFe0.5X, colour-coded with the radial velocity at time $t_\mathrm{sb,2}$ for model WH12.5. The cyan surface depicts the location of the forward shock. Note that the viewing angle is chosen such that the largest asymmetries are well captured.}
  \label{fig:3dshock_WH12.5}
\end{figure}

\begin{figure*}
    \begin{center}
    \begin{minipage}{0.30\textwidth}
      \includegraphics[width=\linewidth]{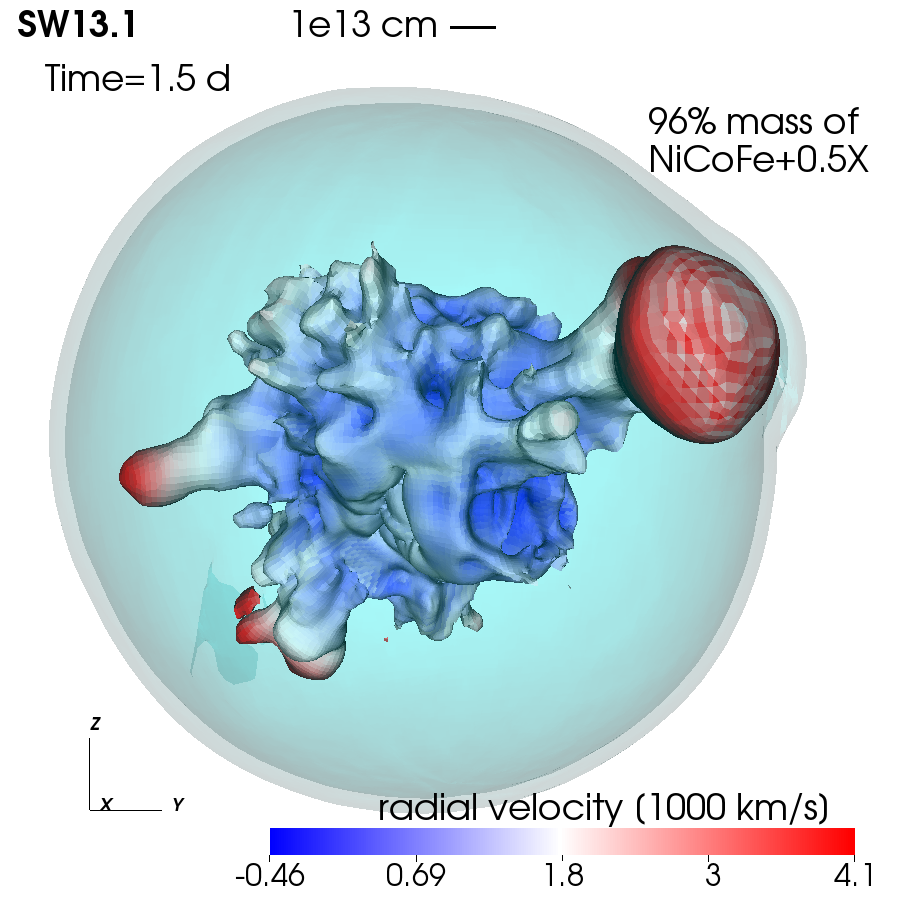}
    \end{minipage}
    \begin{minipage}{0.3\textwidth}
      \includegraphics[width=\linewidth]{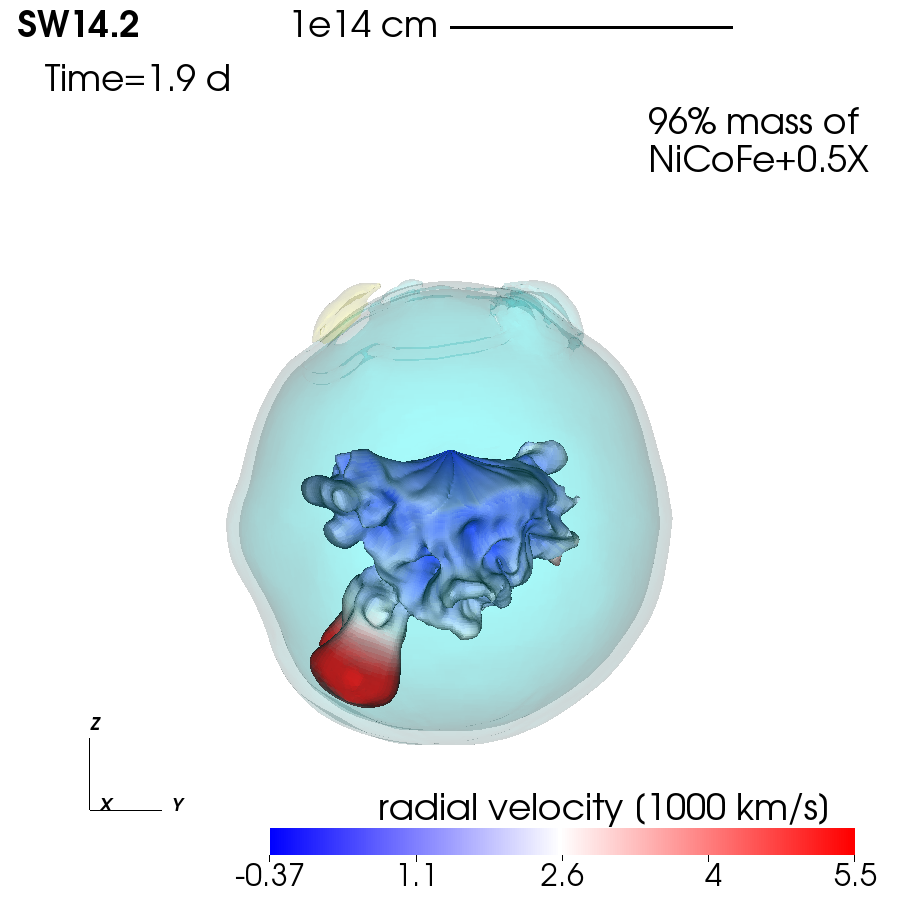}
    \end{minipage}
    \begin{minipage}{0.3\textwidth}
      \includegraphics[width=\linewidth]{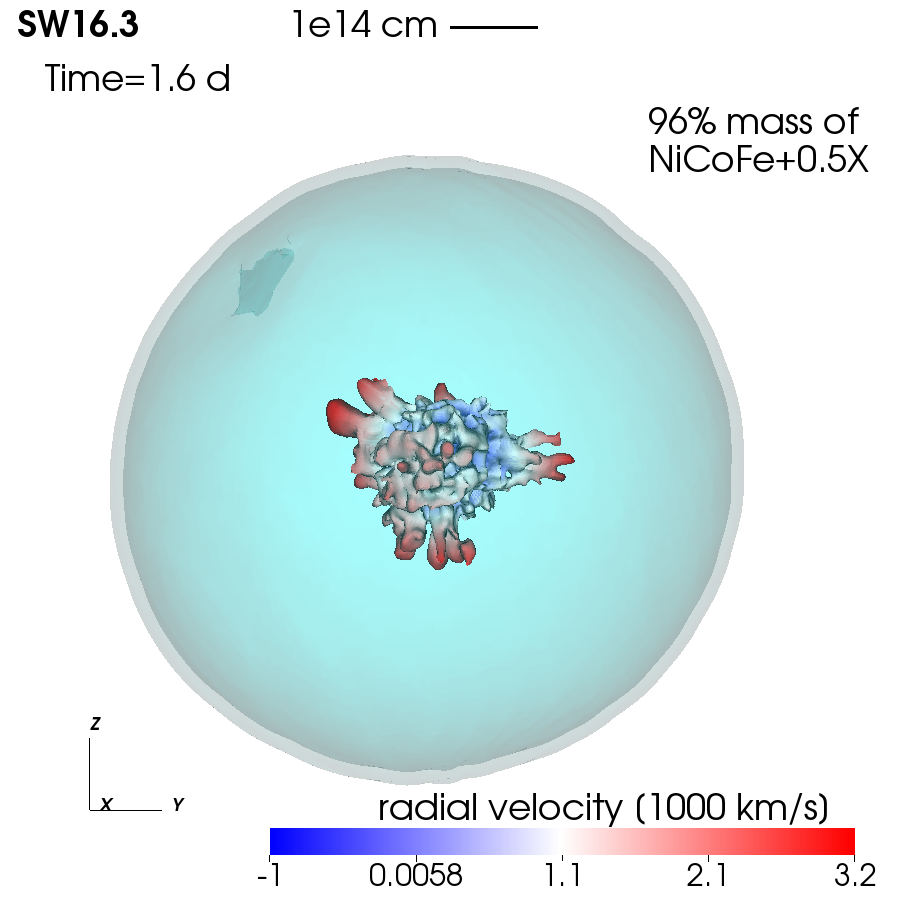}
    \end{minipage}
    \begin{minipage}{0.3\textwidth}
      \includegraphics[width=\linewidth]{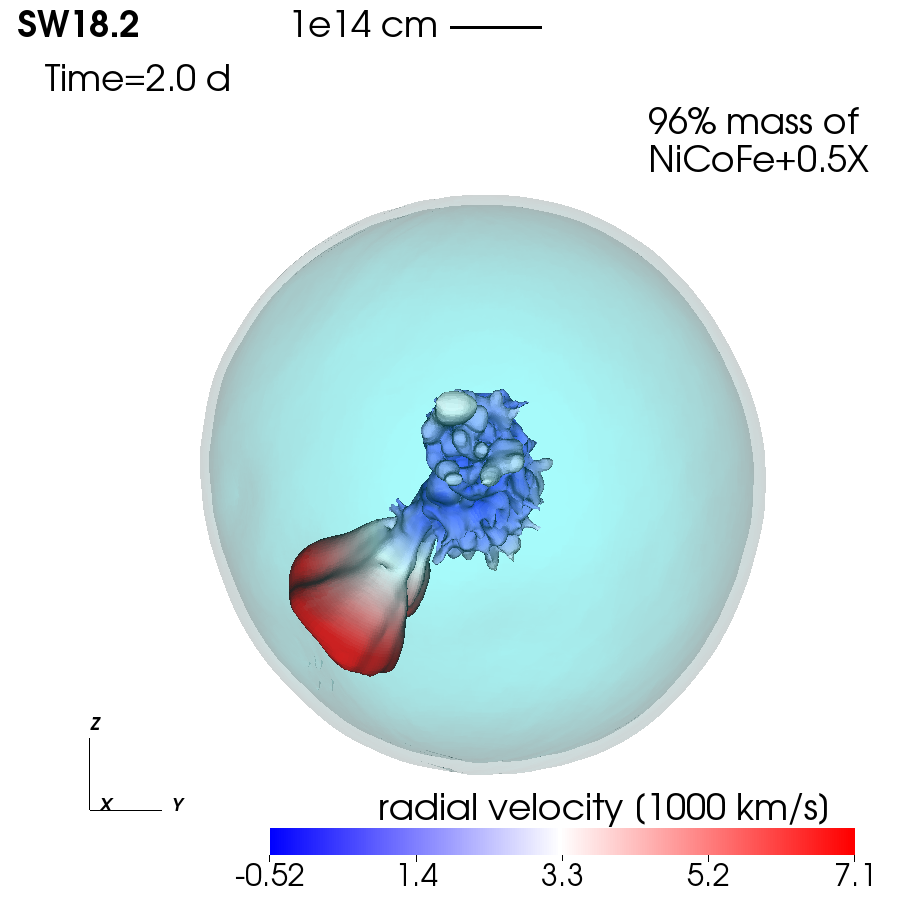}
    \end{minipage}
    \begin{minipage}{0.33\textwidth}
      \includegraphics[width=\linewidth]{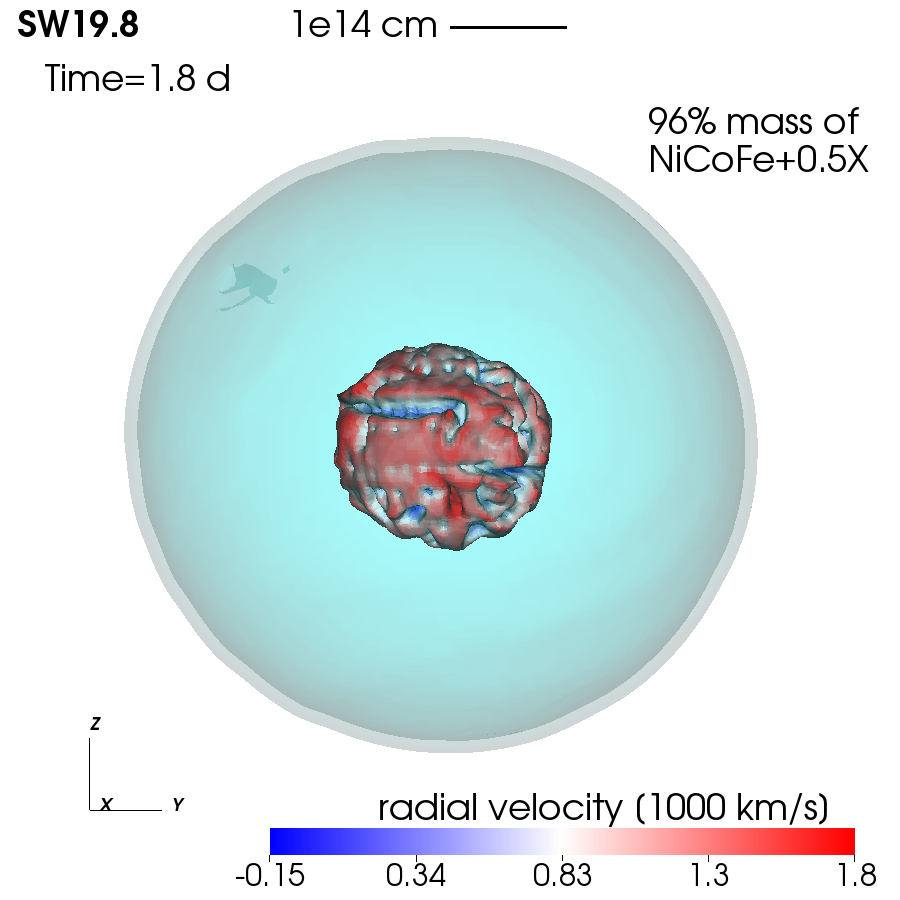}
    \end{minipage}
    \begin{minipage}{0.3\textwidth}
      \includegraphics[width=\linewidth]{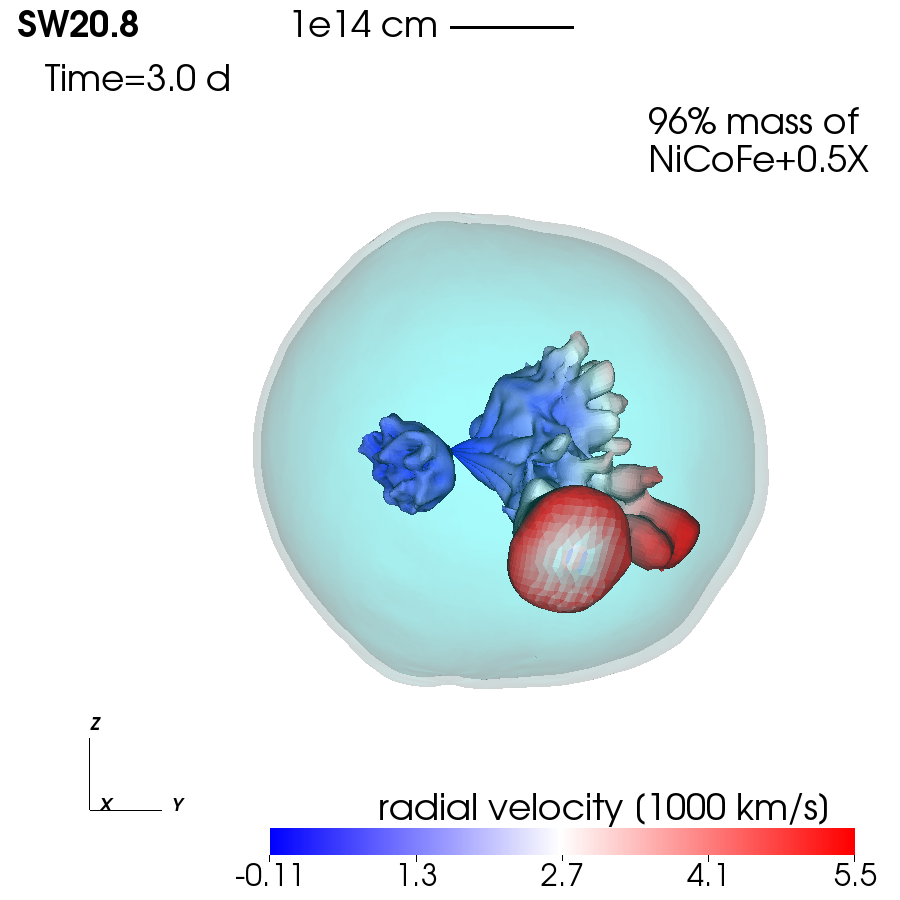}
    \end{minipage}
    \begin{minipage}{0.30\textwidth}
      \includegraphics[width=\linewidth]{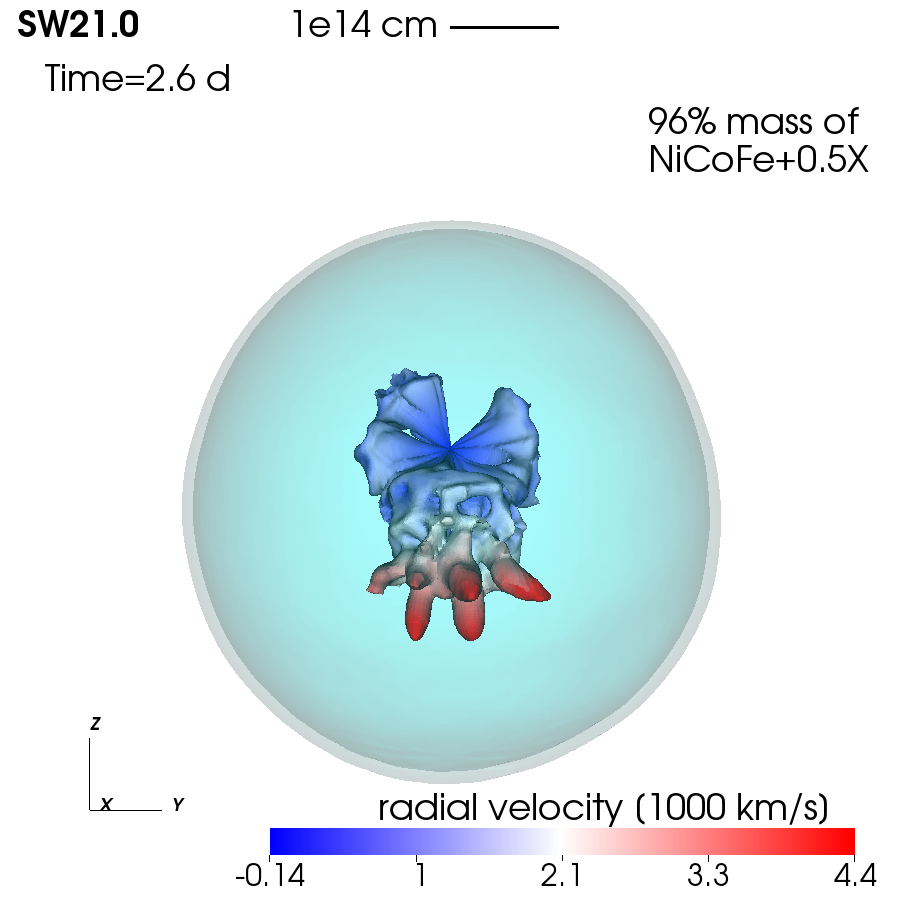}
    \end{minipage}
    \begin{minipage}{0.3\textwidth}
      \includegraphics[width=\linewidth]{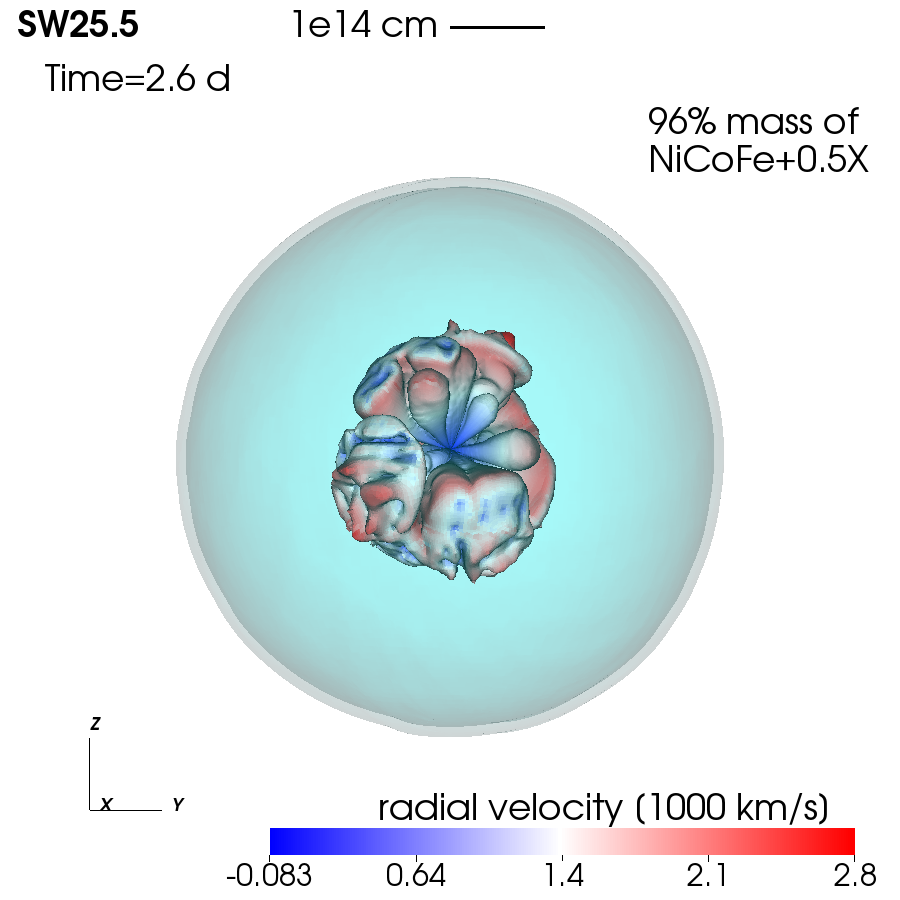}
    \end{minipage}
    \begin{minipage}{0.3\textwidth}
      \includegraphics[width=\linewidth]{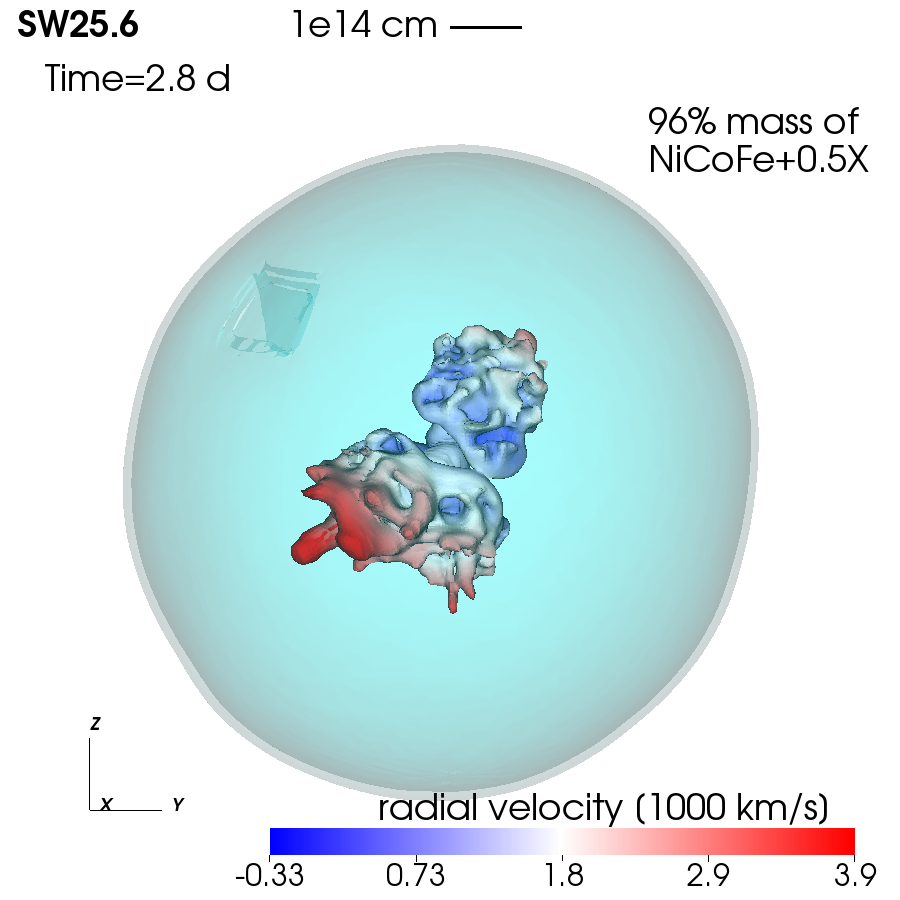}
    \end{minipage}
    \begin{minipage}{0.3\textwidth}
      \includegraphics[width=\linewidth]{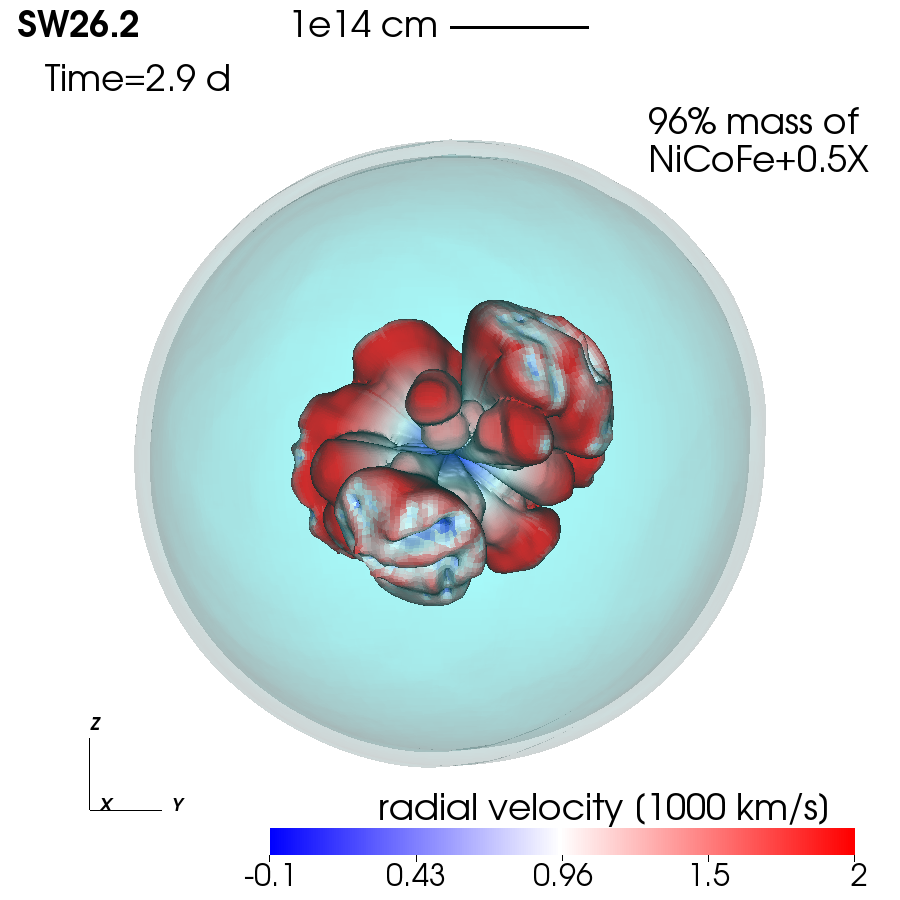}
    \end{minipage}
    \begin{minipage}{0.3\textwidth}
      \includegraphics[width=\linewidth]{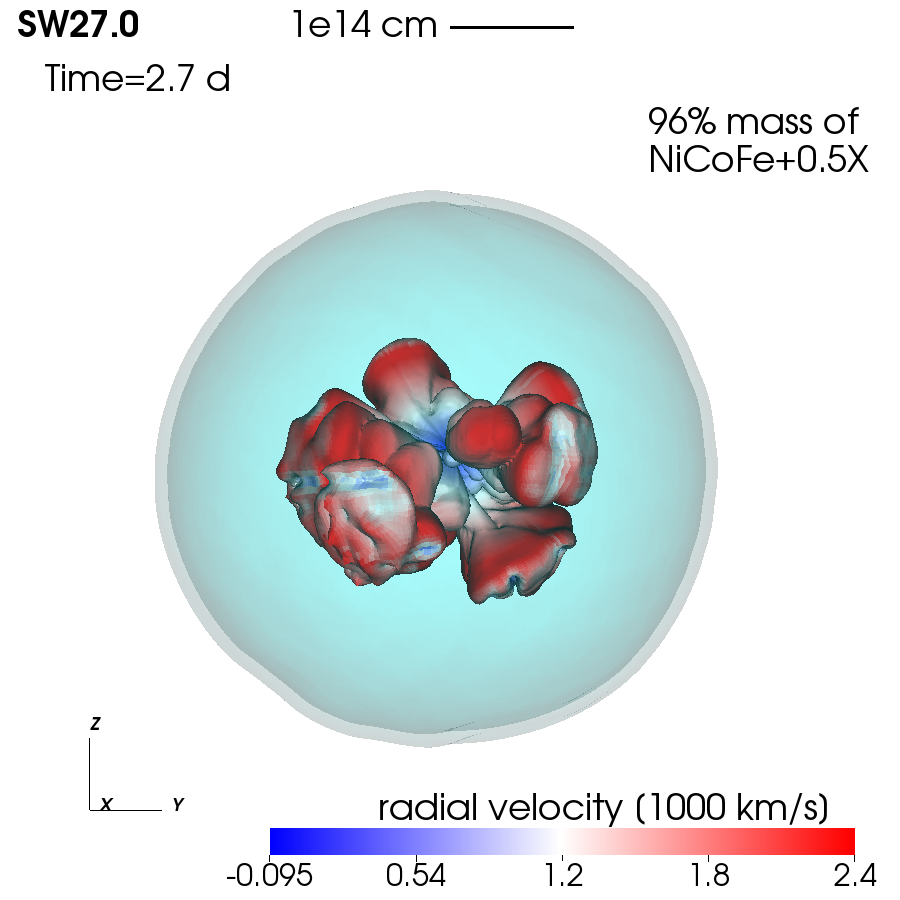}
    \end{minipage}
    \begin{minipage}{0.3\textwidth}
      \includegraphics[width=\linewidth]{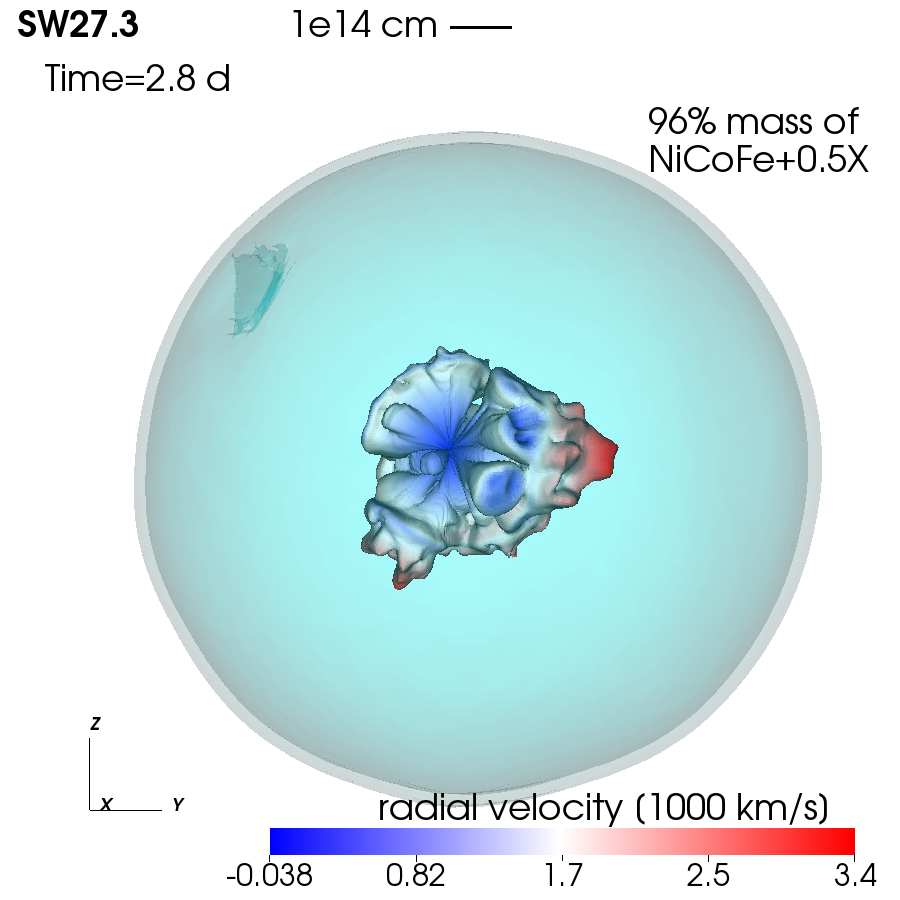}
    \end{minipage}
  \end{center}
  \caption{Isosurfaces of the NiCoFe0.5X mass fraction that contain 96\% of the total mass of ejected NiCoFe0.5X, colour-coded with the radial velocity at time $t_\mathrm{sb,2}$ for all the models except WH12.5, which is shown in Fig.~\ref{fig:3dshock_WH12.5}. The cyan surfaces depict the locations of the forward shocks. Note that the viewing angles are the same as in Fig.~\ref{fig:3dshock_WH12.5}.}
  \label{fig:3dshock}
\end{figure*}

\begin{figure*}
    \centering
    \includegraphics[width=0.75\linewidth]{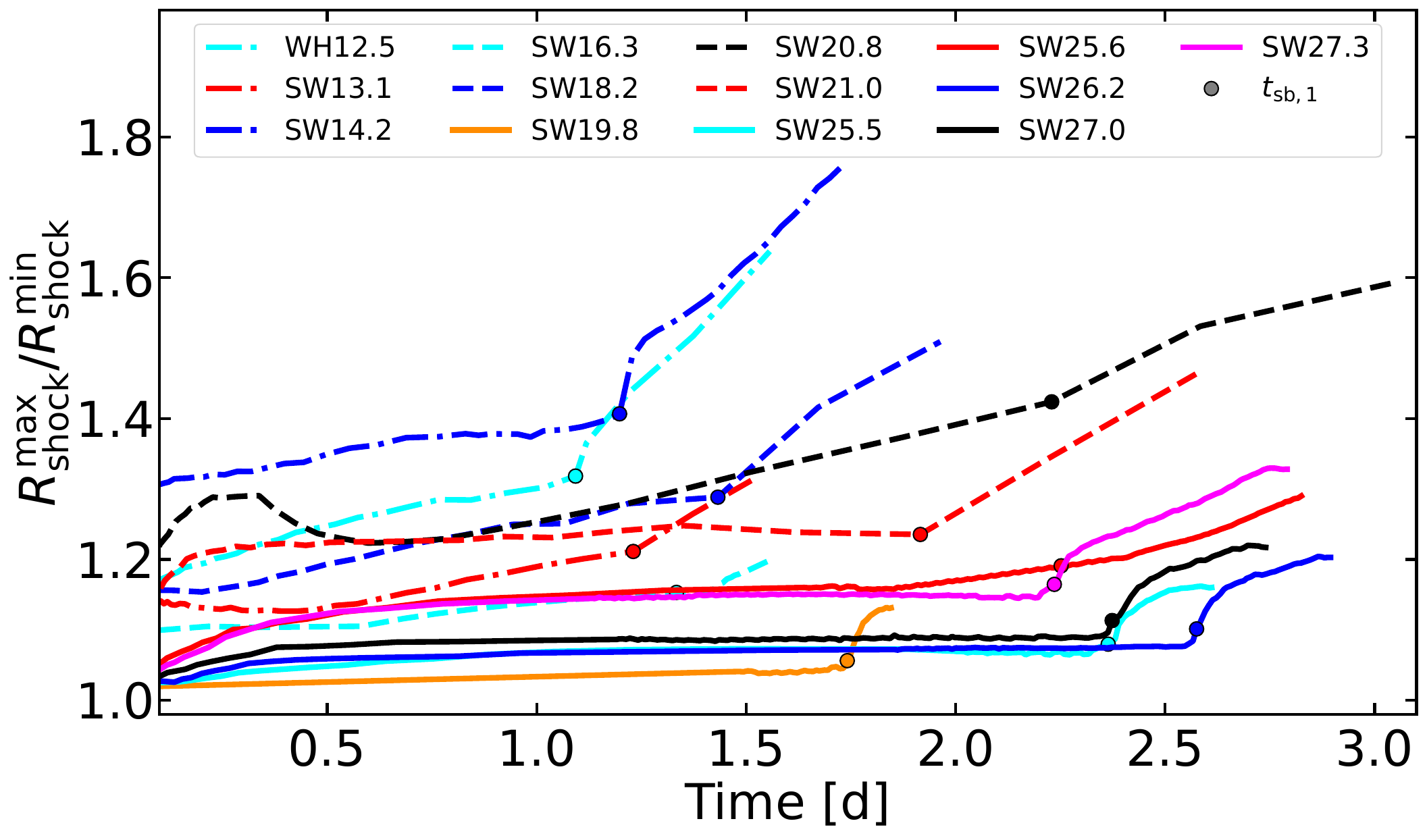}
    \caption{Asymmetry of the forward shock in all 3D SN models as function of time, expressed by the dimensionless ratio of the maximum shock radius and the minimum shock radius. The coloured bullets mark the times when
    the maximum shock radius reaches the surface of the stars, corresponding to the times $t_\mathrm{sb,1}$ given in Table~\ref{tab:sb_times}. All lines stop at the times when the minimum shock radius arrives at the surface of the progenitors ($t_\mathrm{sb,2}$ in Table~\ref{tab:sb_times}).
    }
    \label{fig:shock_ratio}
\end{figure*}

Moreover, models WH12.5, SW14.2, SW18.2, and SW20.8, whose explosions start extremely asymmetrically (see the NiCoFe0.5X momentum-dipole amplitudes in Table~\ref{tab:3Dexplosion}) and which retain the large degree of asymmetry of their NiCoFe0.5X distribution in the H envelope due to efficient NiCoFe0.5X mixing into the outer stellar layers, also show the highest spread of the shock-breakout times in terms of $\Delta t_\mathrm{sb}\,t_\mathrm{sb,1}^{-1}$. In these models fast, very extended NiCoFe0.5X fingers in mostly one hemisphere reach out to the shock and push it from behind (Figs.~\ref{fig:3dshock_WH12.5} and~\ref{fig:3dshock}). They foster the development of most extreme shock deformation at the time of shock breakout. This is evident from Fig.~\ref{fig:shock_ratio}, which displays the time evolution of the ratio of the maximum to the minimum shock radius as a measure of the shock asphericity, and from the corresponding values of $R_\mathrm{shock}^\mathrm{max}/R_\mathrm{shock}^\mathrm{min}$ as well as the large shock-dipole amplitudes $A_\mathrm{shock,1}$ in Table~\ref{tab:sb_times}. A visual impression is provided in Figs.~\ref{fig:3dshock_WH12.5} and~\ref{fig:3dshock}, where the shock deformation and the asymmetric NiCoFe0.5X distribution are shown at shock-breakout time $t_\mathrm{sb,2}$ for all of our models. All models explode asymmetrically, although the degree of initial asymmetry can be more or less extreme (Figs.~\ref{fig:ni-plumes_cat1}--\ref{fig:ni-plumes_cat3}, \ref{fig:ni_iso}, and \ref{fig:ni_iso198}). However, even models that have very large, asymmetrically distributed NiCoFe0.5X plumes initially are not guaranteed to preserve a strong asymmetry of the shock until the latter reaches the progenitor surface. Only in those cases where efficient mixing of NiCoFe0.5X at the He/H interface takes place, the ratio of $R_\mathrm{shock}^\mathrm{max}/R_\mathrm{shock}^\mathrm{min}$ is still high after the shock has entered the H envelope (i.e., at early times in Fig.~\ref{fig:shock_ratio}, where the time axis starts at $t \sim 10^4$\,s). Then it generally remains high or even grows noticeably until the maximum radius of the shock breaks out from the stellar surface. In contrast, in the 3D explosions of the more massive progenitors, in particular those of the HM-HE category, whose NiCoFe0.5X mixing at the He/H interface is much weaker (Sections~\ref{sec:RTIs_1d} and~\ref{sec:radmixing}; Figs.~\ref{fig:RTgrowth}, \ref{fig:radial_mixing_m}, \ref{fig:rshock_Ni}, and~\ref{fig:vshock_Ni}), the ratio $R_\mathrm{shock}^\mathrm{max}/R_\mathrm{shock}^\mathrm{min}$ is closer to unity at $t \sim 10^4$\,s (Fig.~\ref{fig:shock_ratio}) and stays nearly constant or increases only little when the shock continues to propagate through the H envelope until $t_\mathrm{sb,1}$. Model SW19.8 has a particularly low SN shock asymmetry (radius ratio near 1) because of its quite roundish ejecta morphology already before the shock crosses the (C+O)/He interface (see Fig.~\ref{fig:ni_iso198}). After first shock breakout, the ratio $R^\mathrm{max}_\mathrm{shock} / R^\mathrm{min}_\mathrm{shock}$ begins to rapidly rise in all the models, because the slower parts of the shock still travel in the H envelope, whereas the outermost sections of the shock are already accelerating in the steep density decline outside of the star which connects the surface of the pre-collapse progenitor model to the CSM assumed around the progenitor.

The degree of asymmetry of a SN explosion can be described and quantified by a variety of parameters, for example by the magnitude of the PNS kick (or of the linear momentum of the SN ejecta mass, which is anti-aligned to the PNS kick; equation~\ref{eq:gasmomentum}); by the dipole moment of the corresponding spatial distribution of NiCoFe0.5X or of its radial momentum (which also points opposite to the PNS kick for large kicks; Section~\ref{sec:startlongterm}); by the ratio $R_\mathrm{shock}^\mathrm{max}/R_\mathrm{shock}^\mathrm{min}$, which leads to the time difference $\Delta t_\mathrm{sb}$ of the shock breakout; or by the dipole moment of the deformed shock surface. Which measure is the best or most suitable depends on the context. Some of them are roughly correlated.

Because of its correspondence to the previous spherical harmonics analysis of the NiCoFe0.5X momentum distribution in Section~\ref{sec:startlongterm}, we also perform a multipole decomposition of the forward-shock surface $R_\mathrm{shock}(\theta,\phi)$ at shock-breakout time $t_\mathrm{sb,1}$. For doing so we use the same $Y_{\ell m}(\theta,\phi)$ as before. However, here no radial sum is required and the coefficients are given by
\begin{equation}
C_{\ell m} \;=\; \int R_\mathrm{shock}(\theta,\phi)\,Y_{\ell m}^{*}(\theta,\phi)\,d\Omega \,.
\end{equation}
The mean shock radius is $a_0=C_{00}/\sqrt{4\pi}$, and the deformation amplitudes are
\begin{equation}
A_\mathrm{shock,\ell} \;=\; \frac{1}{\sqrt{2\ell+1}}\,\frac{\big(\sum_{m=-\ell}^{\ell}|C_{\ell m}|^2\big)^{1/2}}{C_{00}} \,,
\end{equation}
with $A_\mathrm{shock,0}=1$ and $a_\ell=A_\mathrm{shock,\ell}\cdot a_0$ [cm] being the non-normalized
dipole ($\ell=1$) and quadrupole ($\ell=2$) shock displacements.

The corresponding values of the normalized dipole and quadrupole amplitudes $A_\mathrm{shock,1}$ and $A_\mathrm{shock,2}$, respectively, are listed in Table~\ref{tab:sb_times}. We notice a tight correlation between the values of all discussed deformation measures for the SN shock, $\Delta t_\mathrm{sb}$, $\Delta t_\mathrm{sb}t_\mathrm{sb,1}^{-1}$, $R_\mathrm{shock}^\mathrm{max}/R_\mathrm{shock}^\mathrm{min}$, and $A_\mathrm{shock,1}$. But for the last three quantities this correlation holds mainly within our three model categories separately and much less between models that belong to different categories. Furthermore, the correlation between shock deformation, represented by any of the above quantities, and the size of the NS kick velocity is weaker than the correlations between the four shock deformation measures. It seems that the connection between the kick magnitude and the shock deformation at the time of shock breakout is strongly affected by the different efficiencies of the radial element mixing in the three model groups. For instance, models WH12.5 and SW21.0 possess quite large shock-deformation measures at shock breakout, although the NS kicks in these two cases are not very high.

Let us deepen this discussion a bit further. A very rough tendency of correlation, even beyond the boundaries of the three model groups, holds for the magnitude of the PNS kick velocity and $\Delta t_\mathrm{sb}$, but it is not particularly strong. While the direction of the more powerful explosion and thus faster shock expansion is opposite to the direction of the PNS kick, we would actually not expect any tight correlation of the absolute value of the kick velocity and $\Delta t_\mathrm{sb}$. The reason for that is two-fold: A large NS kick correlates with an initially high dipole asymmetry of the explosion and a correspondingly pronounced shock deformation at early stages, and also with a large asymmetry of the (radial-momentum) distribution of $^{56}$Ni (see Section~\ref{sec:startlongterm}). But when the SN shock travels outward, its shape and geometry are influenced by subsequent effects. On the one hand, the shock sweeps up the spherical mass shells of the progenitor, thus becoming more and more spherical (around the center of the explosion). This effect is stronger when more progenitor matter is swept up, i.e., in the more massive stars of the HM-LE and HM-HE categories. On the other hand, in the lower-energy explosions of the LM progenitors, which possess lower values of $E_\mathrm{exp}/M_\mathrm{ej}$, the shock is more extremely decelerated in the extended hydrogen envelope, and NiCoFe0.5X gets less decelerated in the C+O shell (see Fig.~\ref{fig:vshock_Ni} and Table~\ref{tab:mixing}) so that the most extended $^{56}$Ni plumes can catch up with the SN shock and deform it at the time it breaks out from the stellar surface (as in model WH12.5 of Fig.~\ref{fig:3dshock_WH12.5} and models SW13.1 and SW14.2 of Fig.~\ref{fig:3dshock}). This effect is much less strong in the models of our two HM categories (see Fig.~\ref{fig:3dshock}). Only the plumes in two of the HM-LE models (SW18.2 and SW20.8) reach the forward shock and cause a significant shock deformation, while the other two models of the same group (SW16.3 and SW21.0) show substantial NiCoFe0.5X asymmetries, which, however, by far do not extend enough to get close to the SN shock. None of the models of the HM-HE group has NiCoFe0.5X fingers that reach the forward shock, and some of these models even display a fairly spherical NiCoFe0.5X morphology compatible with their relatively low NS kicks (SW19.8, SW25.5, and SW26.2).

\begin{figure*}
    \centering
    \includegraphics[width=0.75\linewidth]{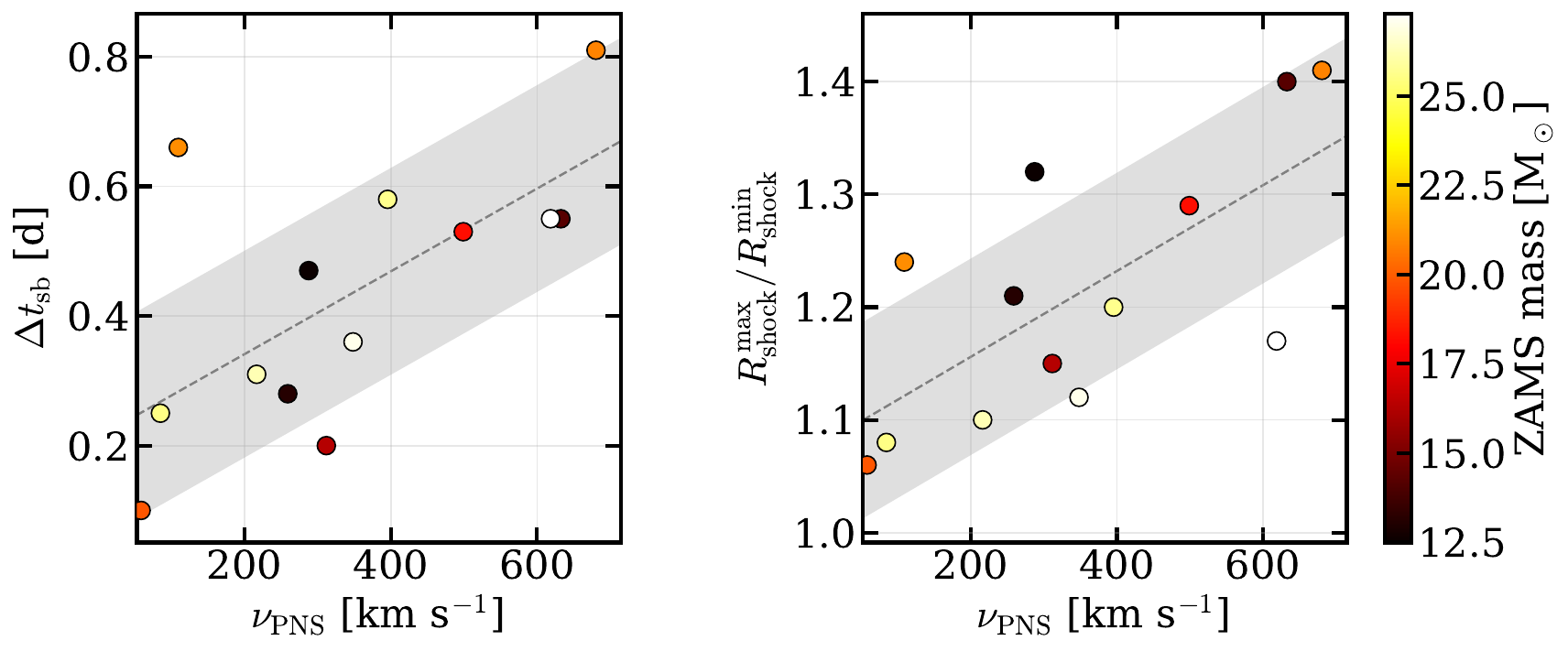}
    \caption{Correlations of the NS kick velocity with the breakout time difference $\Delta t_\mathrm{sb}$ (left panel) and the ratio of maximum to minimum shock radius (right panel) for all of our 13 RSG models. The correlations are much weaker than the one between $v_\mathrm{PNS}$ and the radial momentum of the NiCoFe0.5X shown in Fig.~\ref{fig:A1_vns}.}
    \label{fig:asymm_correlation}
\end{figure*}

Consequently, the NS kick correlates only loosely with shock-asymmetry measures such as $\Delta t_{\rm sb}$ and $R_{\rm shock}^{\rm max}/R_{\rm shock}^{\rm min}$ (see Fig.~\ref{fig:asymm_correlation}), and the trend is far from being uniform across the sample. Rough correlations between $v_\mathrm{PNS}$ (Table~\ref{tab:3Dexplosion}) and the shock-deformation measures of Table~\ref{tab:sb_times} can be witnessed only for stars within each of our three categories, but hardly between models that belong to different categories. For example, model SW13.1 has a larger $\Delta t_{\rm sb}$ than SW16.3 but a lower NS kick, and a $\Delta t_{\rm sb}$ similar to that of SW25.5 despite a much higher kick; SW14.2 and SW18.2 have almost the same $\Delta t_{\rm sb}$ although the kick of the former is significantly larger. Models SW14.2, SW20.8 and SW27.3 all have similarly large kicks, yet SW20.8 stands out by far with the largest $\Delta t_{\rm sb}$ of these models. Even among models within the same category the scatter is large: WH12.5 and SW13.1 have similar $\nu_{\rm PNS}$, but the former has a considerably larger $\Delta t_{\rm sb}$, and the NS kick of SW25.5 is only about $45\%$ higher than that of SW19.8, but its $\Delta t_{\rm sb}$ is 2.5 times larger.

Model SW21.0 is a more drastic outlier and special in several respects, also showing the influence of stochasticity in the development of the initial explosion asymmetries by hydrodynamic instabilities in the post-shock region during and shortly after the phase of shock revival. On the one hand, SW21.0 has several large $^{56}$Ni plumes that are similarly extended and fairly symmetrically distributed in different spatial directions (Figs.~\ref{fig:ni-plumes_cat2} and~\ref{fig:3dshock}), which means that higher-order multipoles dominate its ejecta asymmetry (see the large quadrupole amplitude compared to the dipole of the radial momentum of NiCoFe0.5X in this model in Table~\ref{tab:3Dexplosion}). Therefore its NS kick is small. On the other hand, the explosion energy of SW21.0 is relatively low and its ejecta mass is among the largest of all models, which implies a small value of $E_\mathrm{kin}/M_\mathrm{ej}$ and a small average value of the shock velocity. Combined with a comparatively large progenitor radius, both $t_\mathrm{sb,1}$ and $t_\mathrm{sb,2}$ are quite big (see equation~\ref{eq:tsbexpression}) and accordingly $\Delta t_\mathrm{sb}$ is big, too.

All these examples therefore confirm our expectation that a correlation between the PNS velocity and $\Delta t_\mathrm{sb}$ is weak, because the outward propagating shock decouples from the NiCoFe0.5X distribution in most cases and its geometry is strongly influenced by the spherical shape of the progenitors.

\subsection{The link between 3D mixing and RTIs}
\label{sec:linking}

As discussed in Section~\ref{sec:multid-effects}, the velocity of the fastest 4\% of NiCoFe0.5X and its relation to the bulk velocity of the NiCoFe0.5X can be used to characterize the degree of chemical-element mixing that takes place in hydrodynamic 3D explosion models.
This measure was considered by \citet{utrobin2019} and \citet{utrobin2021} to link the NiCoFe0.5X mixing to the growth of \acp*{rti} in a larger set of 3D SN simulations of \ac*{bsg} progenitors. Here, we extend the discussion of those papers to \acp*{sn} of \ac*{rsg} progenitors and generalize the relations found for the explosions of \acp*{bsg}. The values of model-specific quantities needed in our analysis of the present section are listed in Table~\ref{tab:mixing}.

Let $v$ be the maximum velocity of NiCoFe0.5X. 
The contribution of the growth of \acp*{rti} to the evolution of the velocity can be roughly described by the following differential equation:
\begin{equation}
    \frac{dv}{dt} = \beta \sigma_\mathrm{RT} v_\mathrm{0}\,,
    \label{eq:ni_vel_eq}
\end{equation}
where $\beta$ is an empirical buoyancy coefficient, $v_\mathrm{0}$ is the initial radial velocity of NiCoFe0.5X, and $\sigma_\mathrm{RT}$ is the linear RT growth rate defined in equation~\eqref{eq:RTgrowth_inc}.
The solution of equation~\eqref{eq:ni_vel_eq} is given by
\begin{equation}
    v(t) = \beta v_0 \int_0^t \sigma_\mathrm{RT}(\tau) d\tau = \beta v_0 \ln \left(\frac{\xi(t)}{\xi_0}\right) \,,
    \label{eq:v_ni_sol}
\end{equation}
where $\xi(t)/\xi_0$ is the time-integrated \ac*{rti} growth factor defined in equation~\eqref{eq:RTtime_int}.
Depending on the model, we have up to two main phases of the growth of \acp*{rti}. 
In order to estimate the total effect on the velocity evolution, we split the integral in equation~\eqref{eq:v_ni_sol}:
\begin{align}
    v(t) &= \beta v_0 \left(\int_0^{t_\mathrm{He/H}} \sigma_\mathrm{RT}^\mathrm{CO}(\tau) d\tau + \int_{t_\mathrm{He/H}}^t \sigma_\mathrm{RT}^\mathrm{He}(\tau) d\tau \right)\nonumber\\
    &= \beta v_0 \left[\ln \left(\frac{\xi(t_\mathrm{He/H})}{\xi_0}\right)^\mathrm{CO} + \ln \left(\frac{\xi(t)}{\xi_0}\right)^\mathrm{He} \right] \,,
    \label{eq:v_ni_sol_sum}
\end{align}
where $t_\mathrm{He/H}$ is the time when the shock passes the He/H interface, $\sigma_\mathrm{RT}^\mathrm{CO}$ and $\sigma_\mathrm{RT}^\mathrm{He}$ are the \ac*{rti} growth rates at the respective interfaces, and $(\xi / \xi_0)^\mathrm{CO}$ and $(\xi / \xi_0)^\mathrm{He}$ the respective RT growth factors. Since the contribution to $\xi(t) / \xi_0$ from the (C+O)/He interface saturates roughly when the shock passes the He/H interface, we can limit the first integral to that time. Given that there is no growth of \acp*{rti} at the He/H interface before the shock passes this interface, we further approximate the second term by
\begin{equation}
    \int_{t_\mathrm{He/H}}^t \sigma_\mathrm{RT}^\mathrm{He}(\tau) d\tau \approx \int_0^t \sigma_\mathrm{RT}^\mathrm{He}(\tau) d\tau\,.
    \label{eq:sigmaint}
\end{equation}
Thus we can directly employ the values of the analysis in Section~\ref{sec:RTIs_1d} at the respective interfaces in the current analysis. For the calculations in this section, only the \ac*{rti} growth factors are computed from the 1D counterparts of the 3D \ac*{sn} simulations as described in Section~\ref{sec:RTIs_1d}. All other quantities relevant for the NiCoFe0.5X mixing such as the velocities of NiCoFe0.5X and the times at which the (outermost) part of the \ac*{sn} shock passes the different composition interfaces, are computed from the 3D data.

The interaction with a reverse shock can hinder the mixing of NiCoFe0.5X that is facilitated by the growth of \acp*{rti}, potentially even completely suppress such mixing \citep{wongwathanarat2015}. 
As we have seen in Figs.\,\ref{fig:rshock_Ni} and \ref{fig:vshock_Ni}, the reverse shock usually forms below the He/H interface in the expanding \ac*{sn} ejecta after the forward shock has entered the hydrogen envelope. Therefore it can strongly affect the transport of NiCoFe0.5X into the RT unstable region at the He/H composition interface. 
The strength of this influence depends on the ratio of two different timescales \citep{utrobin2019, utrobin2021}, namely the time it takes for the reverse shock to form and the time needed by the fastest NiCoFe0.5X to arrive at the location where the reverse shock builds up. The larger this ratio is, the more can the \acp*{rti} grow before the fastest NiCoFe0.5X is hit by the reverse shock. For large ratios the obstructive impact of the reverse shock is diminished and NiCoFe0.5X is permitted to be encompassed by the \ac*{rti} at the He/H interface and thus to be carried deep into the H envelope with high velocities. Otherwise, if the timescale ratio is small, the reverse shock forms much more rapidly than the outermost NiCoFe0.5X is able to catch up, and therefore the reverse shock and the associated accumulation of dense He in its downstream region become a serious obstacle to the propagation of NiCoFe0.5X into the RT unstable region at the He/H interface. As a consequence, NiCoFe0.5X is severely slowed down and its mixing into the H envelope is weak. 

The timescale in the numerator can be estimated as the time between reverse shock formation, $t_\mathrm{RS}$, and the time when the \ac*{sn} shock passes a given interface
\begin{align}
\Delta t^\mathrm{CO}_\mathrm{RS}&=t_\mathrm{RS}-t_\mathrm{CO}\,.\label{eq:dtco}\\
\Delta t^\mathrm{He}_\mathrm{RS}&=t_\mathrm{RS}-t_\mathrm{He}\,,\label{eq:dthe}
\end{align}
The travel time that is required by the fastest NiCoFe0.5X to reach the location of the reverse-shock formation is given by the distance to the reverse shock divided by the mass-weighted average of the fastest 4\% of NiCoFe0.5X, $v_\mathrm{NiCoFe0.5X}^i$, when the \ac*{sn} shock passes the respective interface ($i = \mathrm{CO, He}$ for the (C+O)/He and He/H interface, respectively). For example, for the C+O/He interface this distance is $\Delta R_\mathrm{CO}=R_\mathrm{RS} - R_\mathrm{CO/He}$, and the time for the NiCoFe0.5X to hit the radius $R_\mathrm{RS}$ is expressed by the ratio ${\Delta R_\mathrm{CO} }/{v_\mathrm{NiCoFe0.5X}^\mathrm{CO}}$.

Combining the timescale ratio, $\Delta t^\mathrm{CO}_\mathrm{RS} \times \left(\Delta R_\mathrm{CO} / v_\mathrm{NiCoFe0.5X}^\mathrm{CO}\right)^{-1}$, with the corresponding RT growth factor $\log(\xi/\xi_0)^\mathrm{CO}$ and adding up the contributions from both the C+O/He and the He/H interfaces yields a dimensionless property that we denote by $X_\mathrm{mix}$:
\begin{equation}\label{eq:X_mix}
X_\mathrm{mix} \equiv \sum_{i=\mathrm{CO,He}} \Delta t^i_\mathrm{RS} \times \left(\frac{\Delta R_i}{v_\mathrm{NiCoFe0.5X}^i}\right)^{-1} \times \log \left(\frac{\xi}{\xi_0}\right)^i \,.
\end{equation}
This quantity characterizes the impact of the growth of \acp*{rti} on the radial NiCoFe0.5X mixing in a considered 3D explosion model.\footnote{We note that \citet{utrobin2019, utrobin2021} used the product of the growth factors $\log(\xi / \xi_0)^\mathrm{He} \times \log(\xi / \xi_0)^\mathrm{CO}$ instead of their sum as we do here. However, the sum appears to be a natural choice based on the discussion around equation~\eqref{eq:v_ni_sol_sum}. Moreover, some of the \ac*{rsg} progenitors have very small growth factors at the (C+O)/He interface. For these models, the product would thus be unreasonably small, too. This problem did not occur in the explosion models considered in \citet{utrobin2019, utrobin2021}, so the use of the product of the logarithms of the growth factors seems plausible for their case. Additionally, we refine their criterion by considering the radial distance between the shock formation position and the composition-shell interface in each term of the sum specifically instead of just using the radius $R_\mathrm{RS}$.}

Further following \citet{utrobin2019, utrobin2021}, we estimate the computationally determined extent of NiCoFe0.5X mixing by the dimensionless ratio of the mass-weighted average velocity of the fastest 4\% of NiCoFe0.5X to the mass-weighted velocity of the bulk of the remaining 96\% of the mass of NiCoFe0.5X as obtained in our 3D \ac*{sn} explosion simulations. Different from the previous papers, we consider these velocities at $t=\unit[10]{d}$ after core collapse, because at later times the energy input due to $\beta$ decay can lead to significant late-time acceleration of the NiCoFe0.5X-rich plumes \citep{gabler2021}. The relevant velocities are therefore denoted by $v_\mathrm{NiCoFe0.5X}^{10}$ and $\langle v \rangle_\mathrm{NiCoFe0.5X}^{10}$, respectively, and their ratio defines our NiCoFe0.5X mixing parameter: 
\begin{equation}\label{eq:Y_mix}
Y_\mathrm{mix}\equiv \frac{v_\mathrm{NiCoFe0.5X}^{10}}{\langle v \rangle_\mathrm{NiCoFe0.5X}^{10}} \,.
\end{equation}
As a velocity ratio, this quantity is not strongly dependent on the explosion energy.

\begin{figure}
    \centering
    \includegraphics[width=\linewidth]{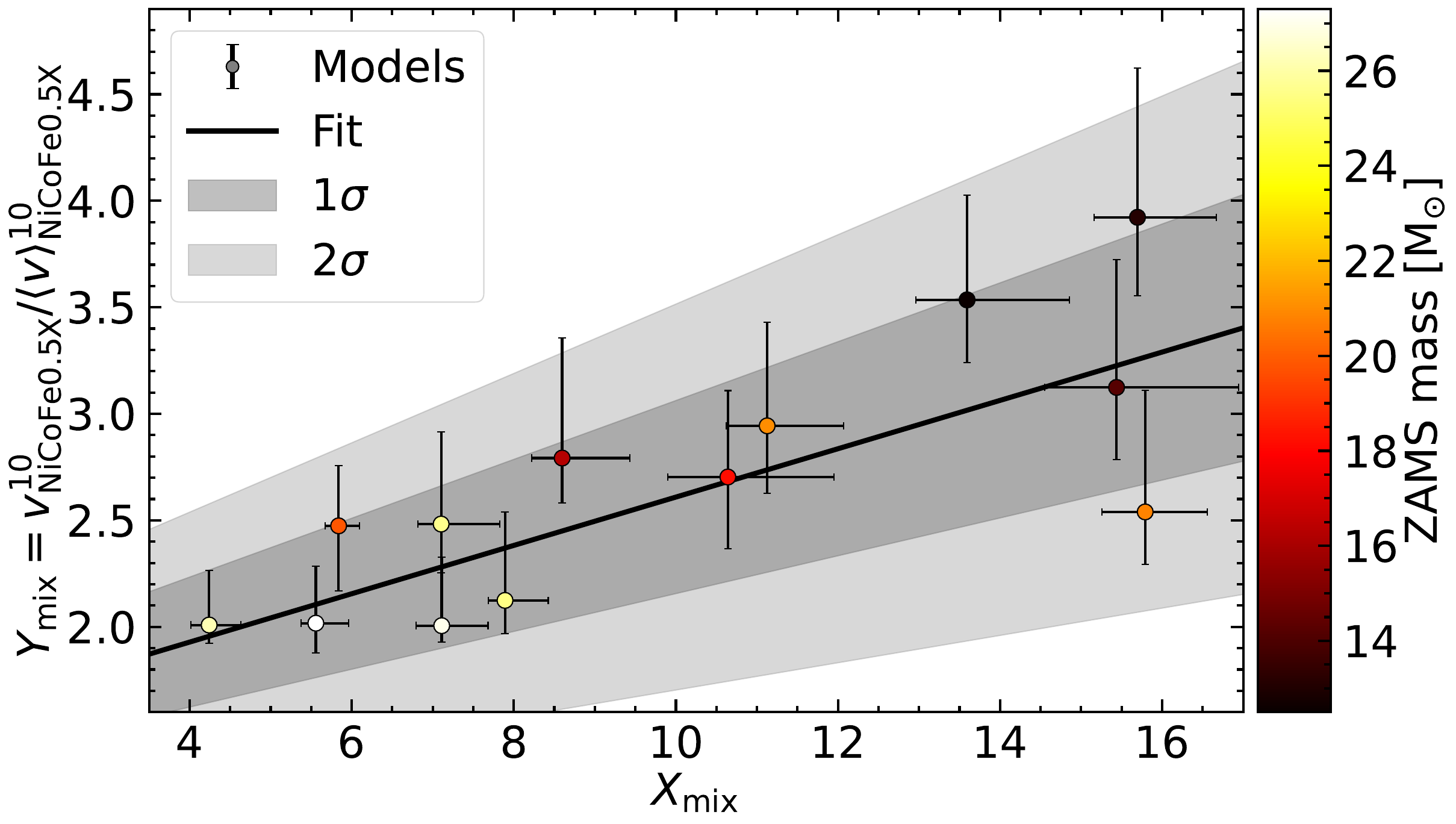}
    \caption{NiCoFe0.5X mixing parameter $Y_\mathrm{mix}$ as defined in equation~\eqref{eq:Y_mix} versus characteristic RT growth parameter $X_\mathrm{mix}$ as defined in equation~\eqref{eq:X_mix} for our set of 3D explosion simulations of \ac*{rsg} progenitors (circles with colour coding corresponding to the \ac*{zams} mass).
    The error bars reflect the uncertainty in velocity, which we estimate by considering two extreme cases for the limiting bulk mass, namely $93\%$ and $99\%$ of the total NiCoFe0.5X mass, respectively. The corresponding $v^{10}_\mathrm{NiCoFe0.5X}$ are then calculated as the mass-weighted averages of the fastest 7\% and 1\% of the total NiCoFe0.5X mass.
    The black line denotes the least-square fit of the data points, and the shaded areas depict the 1$\sigma$ and 2$\sigma$ deviations.}
    \label{fig:correlation}
\end{figure}

We are now equipped with the relations that are suitable to quantify the extent of NiCoFe0.5X mixing in terms of the dependence of $Y_\mathrm{mix}$ on $X_\mathrm{mix}$. Values of all quantities needed for this analysis can be found in Table~\ref{tab:mixing}.
The results are shown in Fig. \ref{fig:correlation} for our set of \ac*{rsg} progenitors. 
We witness a linear correlation between the extent of NiCoFe0.5X mixing in the explosion ejecta and the characteristic RT growth parameter $X_\mathrm{mix}$:
\begin{equation}
    Y_\mathrm{mix} = 0.11 \times X_\mathrm{mix}\,.
    \label{eq:linear_RTI}
\end{equation}

Fig.~\ref{fig:correlation} reveals a clear separation between the different categories defined for our models, which justifies the the chosen grouping. The LM models possess the largest $Y_\mathrm{mix}\gtrsim 3.0$ and $X_\mathrm{mix}\gtrsim 13.5$,
the HM-LE models have $2.4\lesssim  Y_\mathrm{mix} \lesssim 3.0$ and a wide spread of $8.6\lesssim X_\mathrm{mix}\lesssim 15.8$, and the HM-HE are limited to $Y_\mathrm{mix}\lesssim 2.5$ and $X_\mathrm{mix}\lesssim 8.0$.
The separation in $Y_\mathrm{mix}$ works perfectly well. However, $X_\mathrm{mix}$ of SW20.8 lies within the corresponding values of the LM group. These deviations already indicate that the chosen parameters do not describe the properties of the different models perfectly well. The considerable deviation in $X_\mathrm{mix}$ of this particular model comes from the much larger $\Delta t^\mathrm{He}_\mathrm{RS}$ of this model compared to the other models in the HM-LE group. This is probably a consequence of the much lower explosion energy, causing a lower shock speed and thus a later formation time of the reverse shock. Nevertheless, despite its outlier role w.r.t.\ the value of $X_\mathrm{mix}$, also the data point of model SW20.8 is located close to the correlation line defined by equation~(\ref{eq:linear_RTI}). 

\begin{figure}
    \centering
    \includegraphics[width=\linewidth]{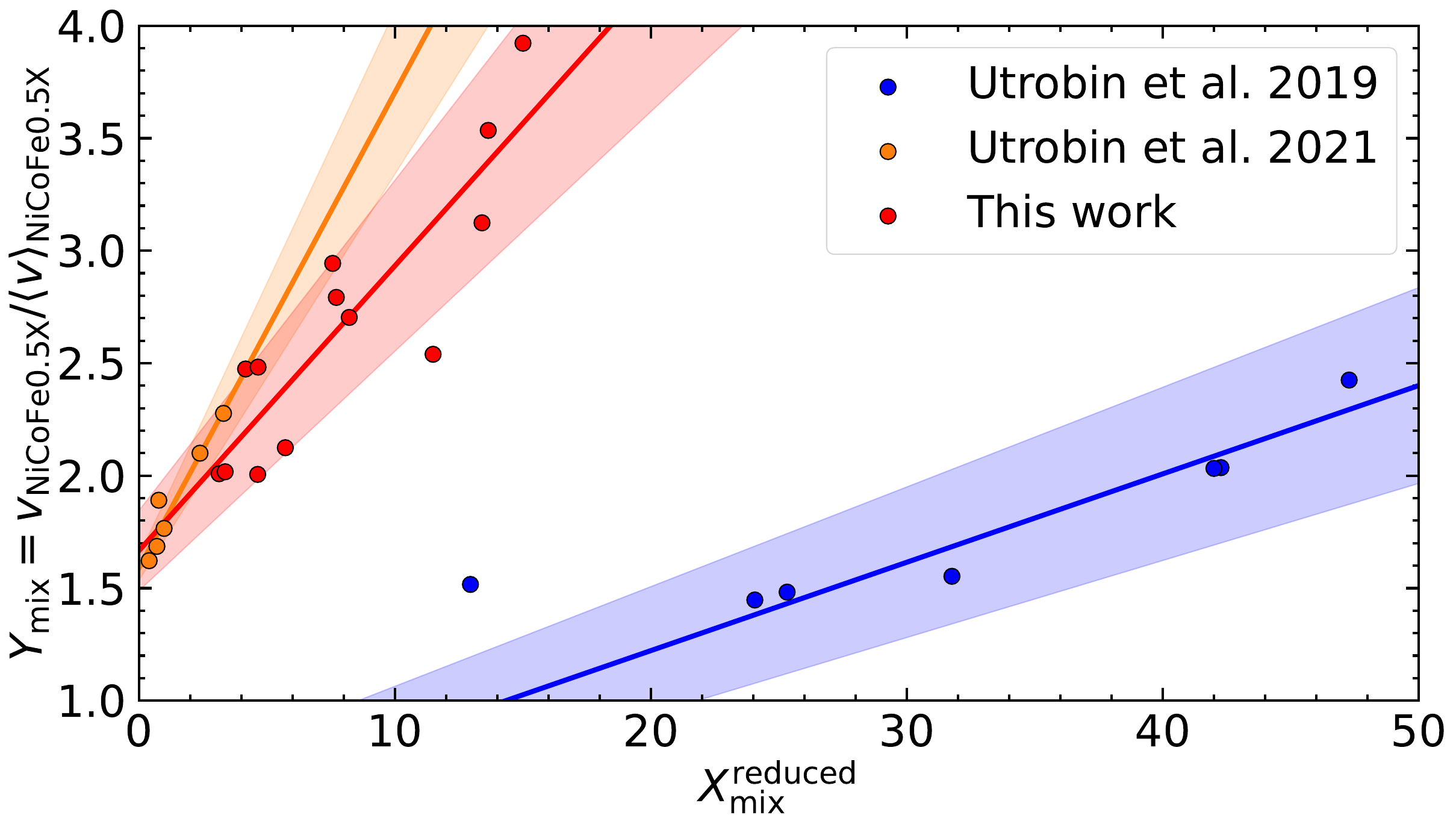}
    \caption{NiCoFe0.5X mixing parameter, defined as the ratio $v_\mathrm{NiCoFe0.5X}/\langle v \rangle_\mathrm{NiCoFe0.5X}$ (measured at 10\,d for our \ac*{rsg} explosions and at 150\,d for the explosions of single-star \ac*{bsg} progenitors of \citealt{utrobin2019} and of binary-merger \ac*{bsg} progenitors of \citealt{utrobin2021}), versus characteristic strength of RT mixing $X_\mathrm{mix}^\mathrm{reduced}$ as defined in equation~\eqref{eq:X_mix_Utrobin}. The shaded bands indicate the 1$\sigma$ deviations from the least-square fits represented by the coloured solid lines.}
    \label{fig:utrobin}
\end{figure}

In Fig.~\ref{fig:utrobin}, we compare our results with those of \citet{utrobin2021}. Since not all information for their models is available to us to add them directly into our Fig.~\ref{fig:correlation}, we redefine the quantity plotted on the abscissa by introducing a simplified version of the characteristic RT growth parameter:
\begin{equation}
    \label{eq:X_mix_Utrobin}
X_\mathrm{mix}^\mathrm{reduced} \equiv  \frac{v_\mathrm{NiCoFe0.5X}^\mathrm{CO} \Delta t^\mathrm{CO}_\mathrm{RS}} {R_\mathrm{RS}} \times \left[\log(\xi / \xi_0)^\mathrm{CO} +\log(\xi / \xi_0)^\mathrm{He} \right]\,.
\end{equation}
The values of the quantities needed to evaluate this expression could be obtained from Table~3 of \citet{utrobin2021} and we re-evaluated our models accordingly. The form of equation~\eqref{eq:X_mix_Utrobin} still differs from the product of RT growth factors considered by \citet{utrobin2021} as explained in the discussion around equation~\eqref{eq:X_mix}. Moreover, we calculated our NiCoFe0.5X mixing parameter $Y_\mathrm{mix}$ at $\unit[10]{d}$, whereas \citet{utrobin2021} used velocities at $\unit[150]{d}$, which introduces another (minor) uncertainty in the comparison.

Our models cover a wider range of values, $2.0\lesssim Y_\mathrm{mix}\lesssim 4.0$, than the \ac*{bsg} explosions of \citet{utrobin2019, utrobin2021}, which yield values up to $Y_\mathrm{mix}\lesssim 2.5$ only. The high values of $Y_\mathrm{mix}$ indicate very strong mixing in particular in our LM and HM-LE models.
The 3D single-star explosions studied by \citet{utrobin2019} (blue bullets in Fig.~\ref{fig:utrobin}) and the 3D explosions of binary-merger progenitors investigated by \citet{utrobin2021} (orange bullets) were based on proposed \ac*{bsg} stellar models for SN~1987A, and therefore quite massive. For progenitor masses $\gtrsim \unit[15]{\Msun}$, we obtain similarly reduced mixing in our \ac*{rsg} explosions, though there may be a tendency that NiCoFe0.5X mixing in explosions of \acp*{rsg} is slightly more efficient than that occurring in SNe of \acp*{bsg} of comparable masses. But also all the generally weakly mixing \ac*{bsg} progenitors exceed a minimum value of $Y_\mathrm{mix}\approx 1.45$.

The characteristic mixing parameter $X_\mathrm{mix}^\mathrm{reduced}$ reveals systematic differences between explosions of the different types of progenitors. While $X_\mathrm{mix}^\mathrm{reduced}\lesssim 3.2$ for binary-merger progenitors (b\acp*{bsg}), it adopts values in the interval $3.1\lesssim X_\mathrm{mix}^\mathrm{reduced}\lesssim 15$ for the \acp*{rsg}, and $X_\mathrm{mix}^\mathrm{reduced}\gtrsim 12.9$ for the single-star \ac*{bsg} progenitors (s\acp*{bsg}).\footnote{Note that in the set of s\ac*{bsg} models the case with the lowest value of $X_\mathrm{mix}^\mathrm{reduced}$ is model N20 from \citet{shigeyama1990}. This model did not result from a self-consistent stellar evolution computation but was artificially created by combining a \ac*{presn} helium core of $\unit[6]{\Msun}$ \citep{nomoto1988} with a hydrogen envelope computed independently by \citet{saio1988}. Both the He core and the H envelope were constructed to satisfy observational constraints from Sanduleak $-$69$^{\circ}$202. For these reasons we did not take it into account in our linear fit.} The basically non-overlapping ranges of values demonstrate that the structures of the different models are intrinsically different between the different types of \ac*{ccsn} progenitors.

The linear correlations indicated in Fig.~\ref{fig:utrobin},
\begin{equation}
Y_\mathrm{mix} = a \times X_\mathrm{mix}^\mathrm{reduced} \,,
\end{equation}
have slope parameters with values $a_\mathrm{RSG}=0.13$, $a_\mathrm{sBSG}=0.04$, and $a_\mathrm{bBSG}=0.21$, respectively. These differences in the correlations for the different types of progenitors are suggestive and may be linked to fundamental differences in the pre-collapse structures of the stars. However, it is presently unclear how much they depend on our simplistic way to characterize the extent of radial NiCoFe0.5X mixing in the 3D explosions due to the influence of the growth factors of \acp*{rti} at the composition-shell interfaces. It will have to be seen whether the differences between different types of progenitors will survive when a more detailed theory becomes available that is able to capture the connection between relevant structural properties and hydrodynamic effects in a more quantitative way \citep[see also][]{utrobin2021}. 

Note that the exact value of $a_\mathrm{RSG}$ obtained in our study may be affected by the chosen prescription of the dilute medium embedding the progenitor and the surrounding stellar wind that we had to assume for our 3D simulations with an Eulerian grid (see Section\,\ref{subsec:physics}; in contrast to the 1D long-term models of \citealt{utrobin2019,utrobin2021}, which were computed with a Lagrangian radiation-hydrodynamics code). For more (less) massive winds, the deceleration of the ejecta in the exterior will be stronger (weaker), and in particular the fastest ejecta will become considerably slower (faster). This will reduce (increase) $Y_\mathrm{mix}$. However, this effect will be systematic for all models, for which reason our conclusions and the linear correlations obtained here should not be affected.

\section{Dependence on the progenitor}\label{sec:dep_prog}

In the previous section, we have shown how the \ac*{rti} growth factors calculated from the 1D explosion simulations as well as NiCoFe0.5X propagation velocities and the time of reverse-shock formation obtained in our 3D models can be used to characterize the mixing behaviour of $\Ni$ in the 3D \ac*{sn} simulations. 
In the following, we aim to go one step further and connect the efficiency of $\Ni$ mixing directly to the structure of the progenitor instead of considering conditions that are determined only in the post-explosion phase. 

In this context it is our primary goal to identify how our characteristic mixing parameter $Y_\mathrm{mix}$, which plays a role for SN light-curve and spectral features, correlates or anti-correlates with those structural properties of the SN progenitors that are directly linked to the physical processes that facilitate the radial element mixing. From the discussion in Section~\ref{sec:long_times} we know that the outward mixing of NiCoFe0.5X from its region of formation close to the PNS is quantified by $Y_\mathrm{mix}$ and is accomplished by effects taking place in the (C+O) and He layers of the exploding RSGs. Therefore, our focus will be on the He-core mass and the density profile between the bottom and the top of the He shell. Other stellar parameters such as the ZAMS mass of the investigated RSGs and their pre-collapse radius and (hydrogen or total) mass also (anti-)correlate with our mixing parameters $X_\mathrm{mix}$ and $Y_\mathrm{mix}$, because they are connected to the stellar properties of our main interest through stellar evolution. However, these stellar parameters are not directly responsible for more or less efficient heavy-element mixing.\footnote{Similarly, there is no causal connection behind the anti-correlations between the 3D explosion energy and our parameters $X_\mathrm{mix}$ and $Y_\mathrm{mix}$ (Tables~\ref{tab:3Dexplosion} and~\ref{tab:mixing}).} Moreover, the values of the pre-collapse radius and mass also depend on the assumed mass-loss history of the progenitors. Therefore we will not much elaborate on various kinds of correlations with such parameters that are not connected to the interior physics that governs the chemical mixing during the explosion.

\begin{figure}
    \centering
    \includegraphics[width=\linewidth]{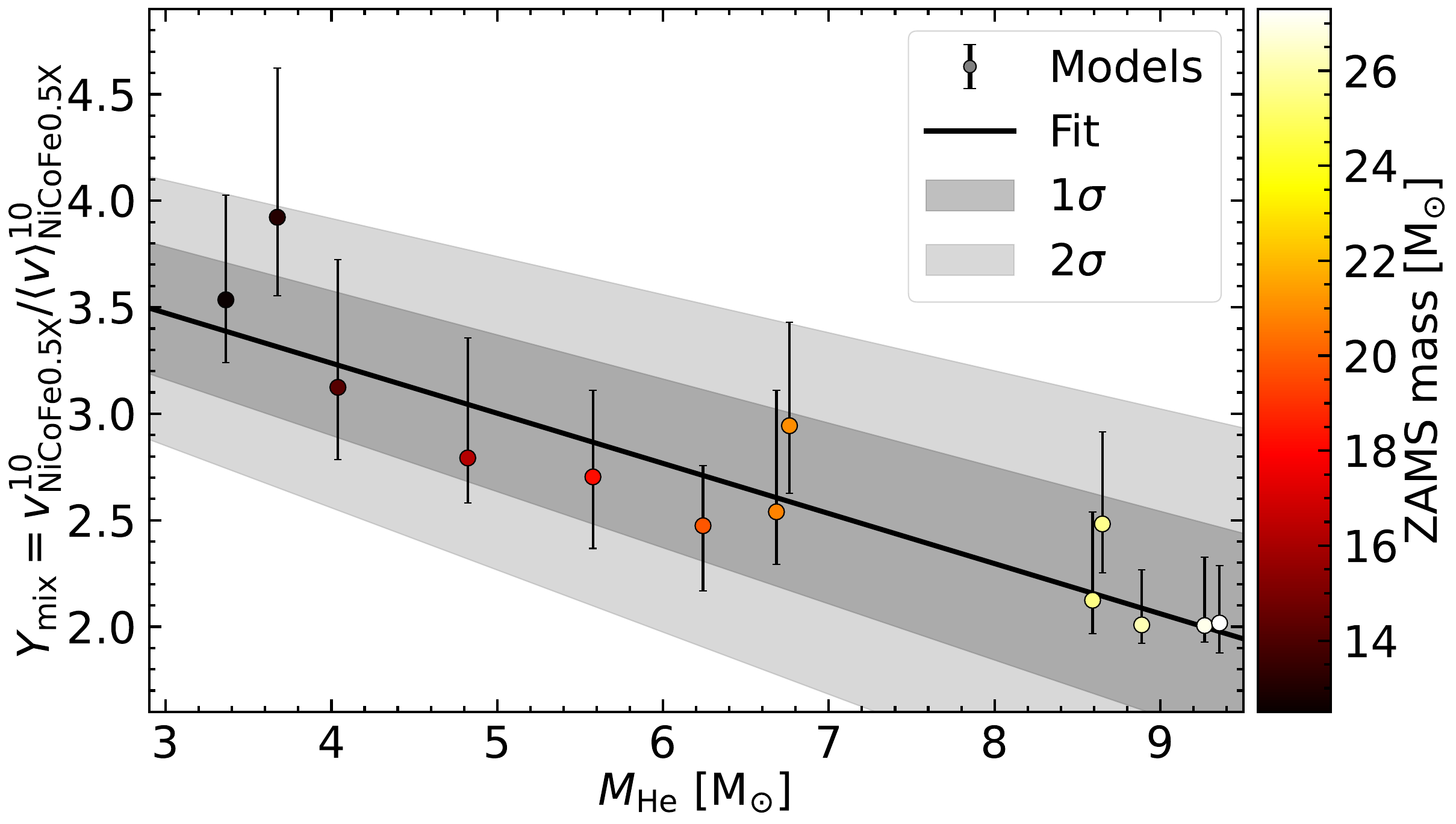}
    \caption{Dependence of the NiCoFe0.5X mixing in terms of $Y_\mathrm{mix}$ (equation~\ref{eq:Y_mix}) on the He-core mass of the progenitor. The error bars are estimated by computing the velocities in $Y_\mathrm{mix}$ with the assumption that the bulk of the NiCoFe0.5X ejecta contains 93\% and 99\% of the total NiCoFe0.5X mass. Colours of the circles indicate \ac*{zams} masses according to the colour bar. The shaded areas depict the 1$\sigma$ and 2$\sigma$ deviations.}
    \label{fig:correlation_masses}
\end{figure}

\subsection{Dependence on the He-core mass}\label{sec:dep_prog_mass}

\citet{utrobin2019} suggested that the extent of $\Ni$ mixing in their 3D \ac*{bsg} explosion simulations depends on the He-core mass. This also holds true for the \ac*{rsg} explosions discussed here, because the He-core mass determines the (C+O)-core mass and correlates tightly with it (Table~\ref{tab:presn}), whereas the pre-SN mass and radius depend on the mass-loss evolution of the progenitors. As discussed in Sections~\ref{sec:RTIs_1d}--\ref{sec:morph} and Section~\ref{sec:linking}, the (C+O)-core mass as well as the stellar structure between the (C+O)/He and the He/H composition interfaces play a crucial role for the NiCoFe0.5X mixing.

Figure~\ref{fig:correlation_masses} shows that the mixing parameter $Y_\mathrm{mix}$ of our models (equation~\ref{eq:Y_mix}) is inversely proportional to $M_\mathrm{He}^\mathrm{core}$.
From the linear fit we performed, we obtain a good anti-correlation between the extent of mixing and the He-core mass with a coefficient of determination $R_\mathrm{fit}^2 = 0.90$.
All models except one lie within a 1$\sigma$ band around the linear fit, as shown by the dark-grey-shaded area in Fig.~\ref{fig:correlation_masses}.
Without providing the plots, we also confirm previous findings that the He-core mass is better to predict the NiCoFe0.5X mixing efficiency than other progenitor properties such as the \ac*{presn} mass $\Mpro$; the \ac*{zams} mass shows a similarly tight anti-correlation with $Y_\mathrm{mix}$ because of its close correlation with the He-core mass in our set of \acp*{rsg} (Table~\ref{tab:presn}). The models of our three classes, LM, HM-LE, and HM-HE, are well separated except model SW19.8, which we count to the HM-HE class despite its average He-core mass of $\unit[6.12]{M_\odot}$ (Table~\ref{tab:presn}).

\begin{figure}
    \centering
    \includegraphics[width=\linewidth]{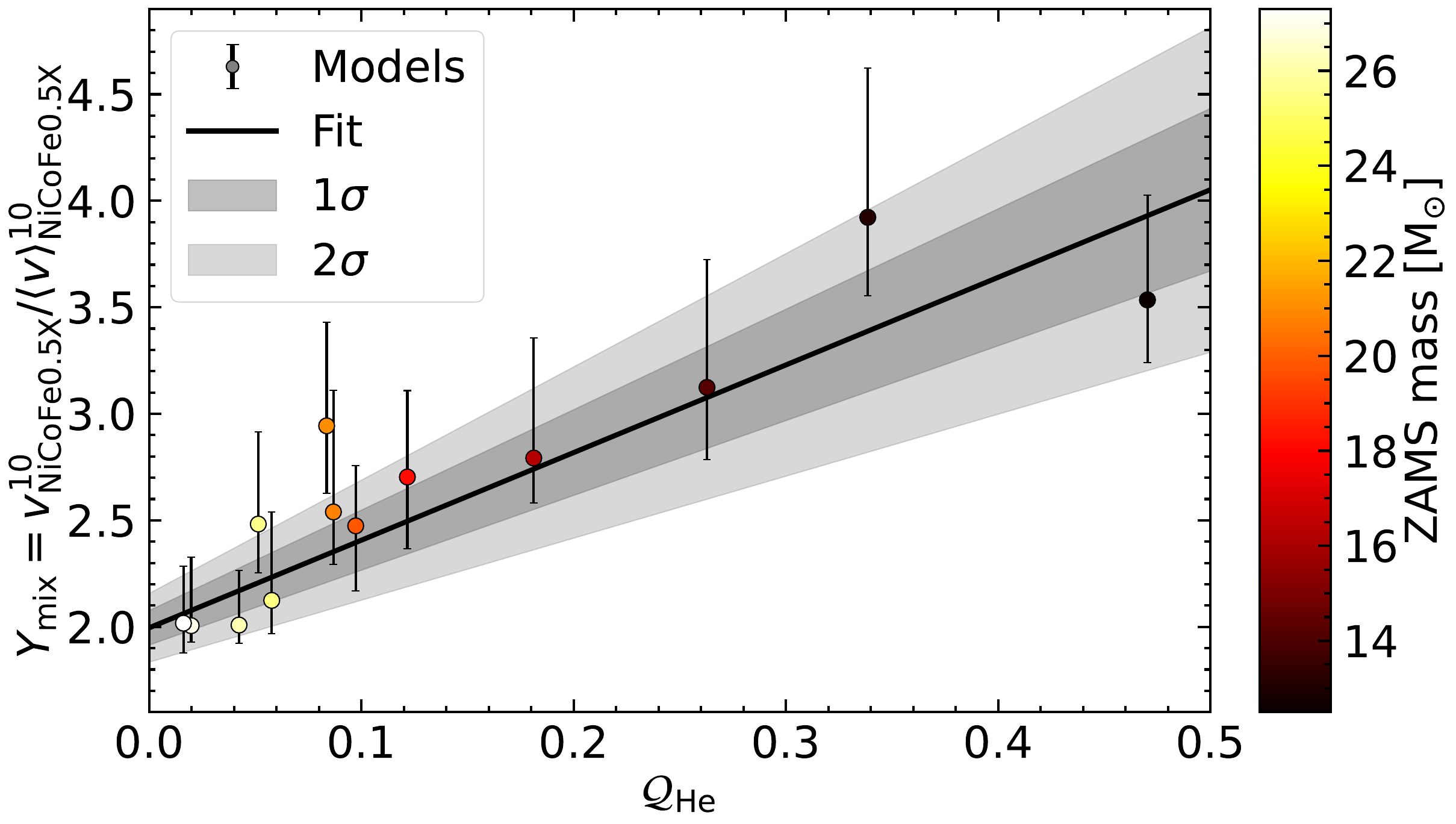}
    \includegraphics[width=\linewidth]{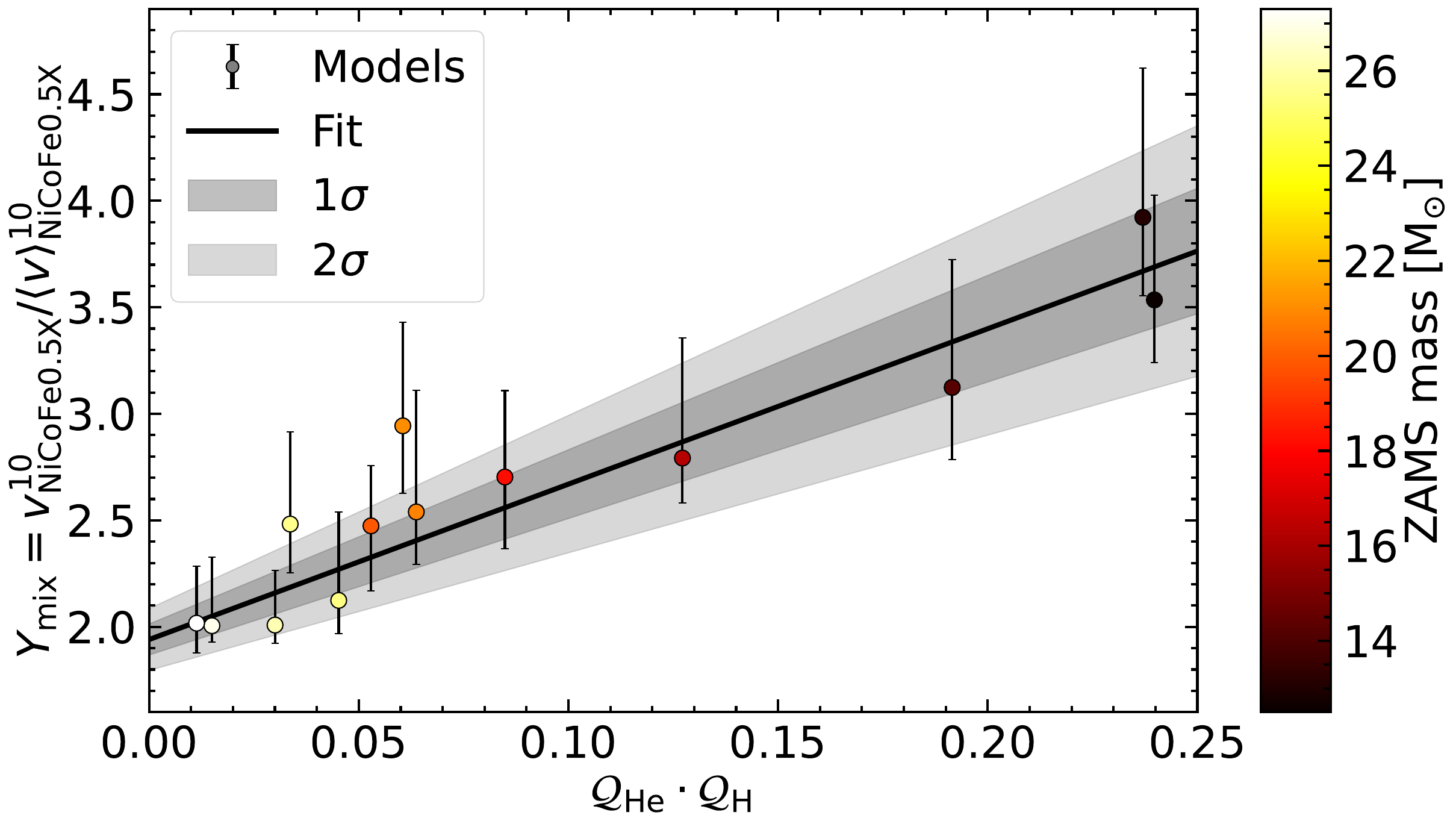}
    \caption{Dependence of the NiCoFe0.5X mixing measured by $Y_\mathrm{mix}$ (equation~\ref{eq:Y_mix}) on the structural parameters $\mathcal{Q}_\mathrm{He}$ (top panel) and $\mathcal{Q}_\mathrm{He}\cdot\mathcal{Q}_\mathrm{H}$ (bottom panel) of the \ac*{rsg} progenitors of our 3D \ac*{sn} models. $\mathcal{Q}_\mathrm{He}$ and $\mathcal{Q}_\mathrm{H}$ are computed according to equations~\eqref{eq:QHe} and~\eqref{eq:QH}, respectively. The error bars are estimated by varying $Y_\mathrm{mix}$ under the assumption that the bulk of the NiCoFe0.5X ejecta contains 93\% and 99\% of the total NiCoFe0.5X mass. Colours of the circles indicate \ac*{zams} masses as given by the colour bar. The shaded areas depict the 1$\sigma$ and 2$\sigma$ deviations.}
    \label{fig:correlation_progenitors}
\end{figure}

\subsection{Dependence on density structure}

As we have seen, the He-core mass performs quite well as an indicator for the efficiency of the NiCoFe0.5X mixing in our \ac*{rsg} explosions. Here, we attempt to advance our understanding further by relating the mixing to progenitor properties that possess some physical relevance for the processes that determine the mixing.

In Section~\ref{sec:RTIs_1d}, we discussed the dependence of the \ac*{rti} growth factors $\xi$ on the function $\rho r^3$, which determines the propagation of the SN shock in the He shell and H envelope. 
In Fig.~\ref{fig:RTgrowth}, we see that model SW13.1 is the only of the considered three representative cases that has a significant growth of \ac*{rti} at the (C+O)/He-interface. It is also the only model with a pronounced maximum in the profile of $\rho r^3$ in the He-shell. Actually, all LM models, showing the strongest mixing of all models, have such a maximum. To obtain this extremum in $\rho r^3$, the latter has to possess a positive density gradient before reaching the maximum. The positive gradient of $\rho r^3$ is directly related to the deceleration of the shock wave before the maximum \citep{sedov1959}. Deceleration leads to a pile-up of material and, hence, to a positive pressure gradient within the exploding ejecta, while the density gradient stays negative in the same region. These are the discussed favourable conditions for the growth of \acp*{rti}.

In order to quantify how strong the gradient or how pronounced the maximum is, we define $\mathcal{Q}_\mathrm{He}$ to be the integral of $\rho r^3$ in the He-shell of the pre-collapse star, normalized by the product of the thickness of the He-shell, $d_\mathrm{He}=R_\mathrm{He/H} - R_\mathrm{CO/He}$, and the value of $\rho r^3$ at the (C+O)/He-shell interface, $\left.(\rho r^3)\right|_\mathrm{CO/He}$:
\begin{equation}
    \mathcal{Q}_\mathrm{He} \equiv \frac{\int_{R_\mathrm{CO/He}}^{R_\mathrm{He/H}} \rho r^3 dr}{d_\mathrm{He} \times \left.(\rho r^3)\right|_\mathrm{CO/He}}\,,
    \label{eq:QHe}
\end{equation}
where $R_\mathrm{CO/He}$ and $R_\mathrm{He/H}$ are the radii of the (C+O)/He and He/H composition interfaces, respectively. The dimensionless ratio $\mathcal{Q}_\mathrm{He}$ quantifies the deviation of the function $\rho r^3$ from a constant value of $\left.(\rho r^3)\right|_\mathrm{CO/He}$, normalizing the area under the curve over the extent of the He shell. The larger $\mathcal{Q}_\mathrm{He}$ is, the higher is the maximum of $\rho r^3$, and the stronger the shock deceleration is expected to be. We also could have chosen properties such as the mean or the maximum gradient of $\rho r^3$ as alternative measures. However, $\mathcal{Q}_\mathrm{He}$ has the advantage of being an integral quantity that captures the behaviour not only at a specific point, e.g., at a composition interface. 

In the upper panel of Fig. \ref{fig:correlation_progenitors}, the extent of NiCoFe0.5X mixing in terms of $Y_\mathrm{mix}$ is plotted as a function of $\mathcal{Q}_\mathrm{He}$. There is a clear correlation between the degree of NiCoFe0.5X mixing and the structure of the progenitor star in the He shell. The linear fit has a slope of $4.15$ and the values of all models, except that of the HM-LE model SW21.0, deviate from the correlation line by less than $2\sigma$. 

So far we have only considered the He-shell structure. Analogue to $\mathcal{Q}_\mathrm{He}$, we define $\mathcal{Q}_\mathrm{H}$ in the H envelope as
\begin{equation}
    \mathcal{Q}_\mathrm{H} \equiv \frac{\int_{R_\mathrm{He/H}}^{2R_\mathrm{He/H}} \rho r^3 dr}{d_\mathrm{H} \times \left.(\rho r^3)\right|_{2R_\mathrm{He/H}}}\,,
    \label{eq:QH}
\end{equation}
where we choose $d_\mathrm{H} = 2 R_\mathrm{He/H} - R_\mathrm{He/H}= R_\mathrm{He/H}$.
Since the instability mainly depends on the change of the gradient close to the He/H interface, we do not integrate over the entire H envelope but set the upper integration boundary to $r=2R_\mathrm{He/H}$. Our results do not sensitively depend on the exact upper bound of the integral. In the denominator of equation~(\ref{eq:QH}) we take the value of $\rho r^3$ also at $r=2R_\mathrm{He/H}$, in contrast to $\mathcal{Q}_\mathrm{He}$, where we take it at the corresponding shell interface. With this definition both values for $\mathcal{Q}_\mathrm{He}$ and $\mathcal{Q}_\mathrm{H}$ are smaller than one. Otherwise both factors could (partially) cancel each other. The opposite behaviour is related to the different slope of $\rho r^3$, which, overall, is decreasing in the He shell and increasing in the H shell (see Fig.~\ref{fig:rhor3}).

Using the product of $\mathcal{Q}_\mathrm{He}$ and $\mathcal{Q}_\mathrm{H}$ to characterize the pre-collapse stellar structure, we obtain the correlation with $Y_\mathrm{mix}$ plotted in the lower panel of Fig.~\ref{fig:correlation_progenitors}. The slope of the linear fit in this case is $7.28$. When comparing the value of $R_\mathrm{fit}^2 = 0.87$ for the coefficient of determination in the case of $\mathcal{Q}_\mathrm{He}$ to $R_\mathrm{fit}^2 = 0.91$ when $\mathcal{Q}_\mathrm{He}\cdot \mathcal{Q}_\mathrm{H}$ is used, we see that the inclusion of $\mathcal{Q}_\mathrm{H}$ provides similarly good results and the main trend is already captured by $\mathcal{Q}_\mathrm{He}$.

In summary, the degree of NiCoFe0.5X mixing measured by the velocity ratio $Y_\mathrm{mix}$ of equation~\eqref{eq:Y_mix} exhibits a linear, negative correlation with the He-core mass (Fig.~\ref{fig:correlation_masses}). In contrast, we find positive, linear correlations of $Y_\mathrm{mix}$ with $\mathcal{Q}_\mathrm{He}$ as well as $\mathcal{Q}_\mathrm{He}\cdot \mathcal{Q}_\mathrm{H}$ (Fig.~\ref{fig:correlation_progenitors}). These results fortify our discussion of Section~\ref{sec:RTIs_1d} that despite their lower explosion energies, the LM progenitors develop more intense NiCoFe0.5X mixing in 3D explosions than the investigated HM-LE and HM-HE explosions of \ac*{rsg} progenitors. There is a clear inverse ordering of the strength of NiCoFe0.5X mixing with the He-core mass, which increases with the ZAMS mass. The same is true for the explosion energy of the different progenitors, which also anti-correlates with the extent of NiCoFe0.5X mixing expressed by our parameter $Y_\mathrm{mix}$, although one would expect stronger mixing for higher explosion energies because of larger velocities of the fastest $^{56}$Ni plumes shortly after the $^{56}$Ni formation during the first second of the explosion. However, this reasoning holds only for a fixed progenitor; in our HM models compared to the LM cases, the influence of the higher explosion energy is overridden by the less mixing-supportive structure of the more massive progenitors. The stronger NiCoFe0.5X mixing in the LM explosion models is explained by a less massive C+O shell and larger growth factors of \ac*{rti} at the (C+O)/He interface due to a more pronounced local maximum of $\rho r^3$ in the He shell of the LM progenitors (Figs.~\ref{fig:RTgrowth} and~\ref{fig:rhor3}). The corresponding growth of \ac*{rti} at the (C+O)/He interface fosters the transport of fast-moving NiCoFe0.5X from the initial asymmetries, which were created by the neutrino-heating mechanism in the first seconds, through the He shell into the ubiquitous RT unstable region near the He/H interface and thus into the H envelopes of the LM \ac*{rsg} explosions. In contrast, this NiCoFe0.5X mixing into the H envelope remains weak in our HM-HE explosion simulations for the highest progenitor masses. The convective plumes carrying iron-group elements are more strongly decelerated in the very massive C+O shells of the HM-HE progenitors and the ejecta of these models undergo only one phase of RTI at the He/H interface. As a consequence, the reverse shock does not allow the NiCoFe0.5X to reach the unstable region in the HM-HE explosions. HM-LE models are intermediate. They have only one dominant RT phase but the reverse shock is not efficiently hindering the NiCoFe0.5X to participate in the RTI.

\section{Summary and Conclusions}\label{sec:conclusion}

We performed long-term 3D simulations of \ac*{ccsn} explosions of 13 \acp*{rsg} from core bounce until 10 days. The progenitors with \ac*{zams} masses between $12.5$ and $\unit[27.3]{\Msun}$ are a subset of the models presented by \citet{woosley2015} and \citet{sukhbold2014}, which have been exploded in 1D by \citet{sukhbold2016}. Our 3D explosions were initiated by neutrino heating using a parametric approach with neutrino luminosities and mean energies imposed at the inner grid boundary such that we roughly reproduced the explosion energies of the 1D SN simulations of \citet{sukhbold2016}. Since this inner grid boundary was located at an optical depth of typically $\gtrsim$\,10 for $\nu_e$ and $\bar\nu_e$ and we applied a grey neutrino transport approximation in the near-surface layers of the PNS exterior to the boundary (a 3D version of the ray-by-ray scheme described in \citealt{scheck2006} implemented by \citealt{wongwathanarat2010,wongwathanarat2013}), the neutrino emission as well as the neutrino heating in our models are not spherically symmetric.

After $\unit[2.5]{s}$, the explosion energies had effectively reached their terminal values in the range between $\unit[0.69]{B}$ and $\unit[1.73]{B}$, and the corresponding $\Ni$ yields were between $\unit[0.059]{\Msun}$ and $\unit[0.110]{\Msun}$. Note that these $\Ni$ yields are our best estimates that include 50\% of the mass of a so-called tracer nucleus produced in neutron-rich ($Y_e < 0.49$) ejecta. Thus we account for uncertainties in the iron-group nucleosynthesis connected to our simplified neutrino transport, which tends to underestimate $Y_e$ in the neutrino-heated ejecta. All values of the explosion energies, estimated $\Ni$ masses, and of the NS properties (masses, kick velocities, and angular momenta) at 2.5\,s are listed in Table~\ref{tab:3Dexplosion}. At 2.5\,s we mapped our 3D explosion models onto a larger radial grid (continuing with the same angular grid) in order to carry the 3D simulations to much later times. Thus we could study the growth the \acp*{rti} at the composition-shell interfaces after the passage of the outward propagating SN shock. These hydrodynamic instabilities at the (C+O)/He and He/H shell boundaries are seeded by the initial explosion asymmetries connected to the neutrino-driven mechanism. The latter creates favourable conditions for the growth of non-radial hydrodynamic instabilities due to SASI activity and convective overturn in the neutrino-heated volume behind the stalled core-bounce shock. This growth is a process that is stochastic and is triggered by initial small-amplitude perturbations in the stellar plasma. The thus developing initial explosion asymmetries with the corresponding high-entropy $^{56}$Ni plumes imprint the final morphology of the heavy-element distribution in SN explosions and their gaseous remnants and determine the radial mixing of heavy elements from the innermost regions into the He shell and H envelope.

The mixing behaviour in 3D simulations is closely linked to RT growth factors as evaluated on the basis of 1D SN simulations with similar explosion energies. These growth factors depend on the radial profile of $\rho r^3$ and typically become large at or near composition interfaces, where the radial derivative of $\rho r^3$ changes its sign. In composition layers where $\rho r^3$ has a positive gradient, the SN shock experiences a deceleration and favourable conditions for the growth of \ac*{rti} develop in the post-shock region. While large growth factors at the He/H interface are ubiquitous, because $\rho r^3$ generally increases in extended regions of the H envelopes of \acp*{rsg}, the strength of \ac*{rti} growth at the (C+O)/He interface depends on a local maximum of $\rho r^3$ in the He layer. Qualitatively, \citet{vartanyan2025b} connect the degree of the mixing in RSGs to the conditions at the He/H interface. However, our analysis is more quantitative and highlights that the explosion properties as well as the progenitor structure in the C+O shell and in the He layer play an important role. Crucial effects that facilitate the outward mixing of heavy elements from the deep core potentially even far into the H envelope, are in fact determined by the stellar conditions in the C+O and He layers, which differ between different types of RSG progenitors. 

In our study we could identify three classes of \ac*{rsg} explosions, which revealed a systematic trend in the strength of $\Ni$ mixing closely correlated with the progenitor mass: (a) Our lowest-mass (LM) progenitors with $12\,M_\odot \lesssim M_\mathrm{ZAMS} \lesssim 15\,M_\odot$ and small C+O cores possess a pronounced maximum of $\rho r^3$ in the He shell. Therefore they show large RT growth factors and strong RT mixing at both the (C+O)/He and He/H interfaces in our 3D explosion simulations. The fastest 4\% of the $\Ni$ are mixed deep into the H envelope in long-stretched RT plumes that are ejected with average velocities between 4000\,km\,s$^{-1}$ and 5000\,km\,s$^{-1}$. (b) Our second group (HM-LE) of models is formed by progenitors with $15\,M_\odot \lesssim M_\mathrm{ZAMS} \lesssim 21\,M_\odot$ and more massive C+O cores. In the explosions of these stars the growth of \ac*{rti} at the (C+O)/He interface is very weak due to only a flat maximum of $\rho r^3$ and little shock deceleration in the He layer. But because of a wide region of high RT growth factors around the He/H interface the mixing of $\Ni$ into the H envelope is still very efficient. The fastest 4\% of the $\Ni$ is also ejected at the tips of very elongated RT fingers with mean velocities between roughly 3000\,km\,s$^{-1}$ and 4000\,km\,s$^{-1}$. In both model classes, LM and HM-LE models, the reverse shock does not efficiently prohibit the inclusion of NiCoFe0.5X-rich ejecta in the RT growth.
(c) Our third class (HM-HE) of models with $M_\mathrm{ZAMS} \gtrsim 21\,M_\odot$ and the most massive C+O cores does not possess any local maximum of $\rho r^3$ in the He shell. It therefore lacks the growth of \ac*{rti} at the (C+O)/He interface. As as consequence, the outermost $\Ni$ resides far behind the reverse shock when the latter forms as the SN shock decelerates in the H envelope. This $\Ni$ is further slowed down when it ultimately collides with the reverse shock and then gets stuck in the base of a dense He region that has built up below the H envelope. Since this He wall had time to accumulate a large mass before it was reached by the fastest $\Ni$, hardly any $\Ni$ can be encompassed by the \ac*{rti} at the He/H interface. The $\Ni$ mixing into the H envelope remains very weak and the average velocities of the fastest 4\% of the $\Ni$ stay below $\sim$3000\,km\,s$^{-1}$ (Figs~\ref{fig:mixing_10days} and~\ref{fig:mix_evol}). Extended $\Ni$-rich fingers are therefore absent in such HM-HE explosions, and due to the strong compression by the reverse shock the overall morphology of the final $\Ni$ distribution develops a less extreme asymmetry than in the LM and HM-LE explosions.

We conclude that the specific progenitor properties of the HM-HE models prevent a temporary increase of the $^{56}$Ni velocities in the He layer and in the H envelope, as is evident from the blue lines in Fig.~\ref{fig:vshock_Ni}, where only the LM and HM-LE models SW13.1 and SW18.2 show significant acceleration around the He/H-shell interface. The HM-HE models are characterized by a very massive C+O core, which decelerates the initially very fast $^{56}$Ni when the high kinetic energy of the innermost ejecta including nickel is shared with the large C+O mass (see Fig.~\ref{fig:vshock_Ni}, bottom panel). Moreover, in the HM-HE models a local maximum of $\rho r^3$ in the He shell is absent (Figs.~\ref{fig:RTgrowth} and~\ref{fig:rhor3}), and therefore there is no shock deceleration in the He layer that could trigger \ac*{rti} growth at the (C+O)/He interface. As a consequence, in these models $^{56}$Ni gets dramatically slowed down in the massive C+O core without being re-accelerated later by \ac*{rti} at the (C+O)/He interface (see Fig.~\ref{fig:vshock_Ni} and compare $v_\mathrm{XX}^\mathrm{CO}$ and $v_\mathrm{XX}^\mathrm{He}$ in Table~\ref{tab:mixing}). Despite their higher explosion energies, these models therefore exhibit much less efficient mixing of $\Ni$ and other heavy elements from the metal core into the outer layers than the other models, in particular also compared to the LM class with its lowest explosion energies. \citet{vartanyan2025b} came to similar conclusions. It implies that the impact of a higher explosion energy in the HM-HE cases is overridden by the influence of the stellar structure of the corresponding \ac*{rsg} progenitors, although for a given star a higher explosion energy is expected to cause stronger outward NiCoFe0.5X mixing, because the velocities of the fastest $^{56}$Ni plumes are bigger shortly after nickel was formed in the first second of the explosion. 

In all of these models, however, independent of the class they belong to and despite the massive hydrogen envelope of the pre-collapse stars, the late-time morphology of the $\Ni$ distribution reflects the spatial asymmetry of the $\Ni$ that was imprinted by the neutrino-heating mechanism and the associated production of $\Ni$ in neutrino-heated and shock-heated ejecta during the very first seconds of the explosion. The growth of hydrodynamic instabilities (SASI and convective overturn) in the neutrino-heated layer behind the stalled SN shock is a chaotic and turbulent process with a high degree of stochasticity, but the big $^{56}$Ni plumes emerging from these processes at the onset of the explosion are not destroyed by their subsequent interaction with the \acp*{rti} growing at the composition-shell interfaces. These secondary instabilities as well as radioactive $^{56}$Ni decay-heating taking place even later \citep{gabler2021} preserve the initially fastest and biggest $^{56}$Ni structures and only lead to substructure due to fragmentation and merging of features on smaller scales. Thus the overall asymmetry and geometry of the initial $^{56}$Ni distribution can still be recognized at later times, because the biggest NiCoFe0.5X fingers correspond to the largest $^{56}$Ni plumes that were present in the first seconds (as can be seen by comparing early and late structures during the time evolution of three representative models shown in Fig.~\ref{fig:ni_iso} and by the rough correlation between the early and late dipole parameters $A_1$ and $A_\mathrm{shock,1}$ in Tables\,\ref{tab:3Dexplosion} and \ref{tab:sb_times}).

As described above, $\Ni$ mixing through the He and H layers depends on the growth factors of \acp*{rti} at the (C+O)/He and He/H composition-shell interfaces, the C+O core mass, and the reverse shock that typically forms below the He/H interface in the expanding ejecta due to the SN-shock deceleration in the H envelope. This leads to some fairly tight correlations between these explosion and progenitor properties on the one hand, and the extent of the radial $\Ni$ mixing measured by $Y_\mathrm{mix}$ of equation~\eqref{eq:Y_mix} (following \citealt{utrobin2019} and \citealt{utrobin2021}, who used it for \ac*{bsg} explosions) on the other hand. $Y_\mathrm{mix}$ describes the spread of the velocity distribution of $\Ni$ by means of the ratio of the average velocities of the fastest 4\% and of the remaining 96\% of the $\Ni$ ejecta at the end of our 3D simulations (at 10 days). For example, we found that $Y_\mathrm{mix}$ increases roughly linearly with the characteristic RT growth parameter $X_\mathrm{mix}$ of equation~\eqref{eq:X_mix}, which accounts for the enhanced mixing due to large \ac*{rti} growth factors as well as the obstructive impact of the interaction of fast $\Ni$ with the reverse shock (Fig.~\ref{fig:correlation}). We thus generalized similar relations obtained by \citet{utrobin2019} and \citet{utrobin2021} for 3D explosions of single-star and binary-merger \ac*{bsg} progenitors of SN~1987A, to our set of \ac*{rsg} explosions (Fig.~\ref{fig:utrobin}). Moreover, we found an inverse (linear) correlation of the $\Ni$ mixing ($Y_\mathrm{mix}$) with the mass of the He core in our \acp*{rsg} (Fig.~\ref{fig:correlation_masses}); such a relation was also reported by \citet{utrobin2019} for their \ac*{bsg} explosions. This inverse correlation is not astonishing in view of the fact that $\Ni$ mixing is reduced in our models with higher C+O core masses and therefore higher He core masses (both core masses are closely correlated, see Table~\ref{tab:presn} and \citealt{ertl2020}). Finally, we quantified the well-known shock-decelerating and thus mixing-conducive influence of pronounced local maxima of $\rho r^3$ in the He layer and H envelope of the pre-collapse \acp*{rsg} by introducing dimensionless quantities $\mathcal{Q}_\mathrm{He}$ and $\mathcal{Q}_\mathrm{H}$ (equations~\ref{eq:QHe} and~\ref{eq:QH}, respectively) and verified positive linear correlations of the extent of the $\Ni$ mixing ($Y_\mathrm{mix}$) with $\mathcal{Q}_\mathrm{He}$ and $\mathcal{Q}_\mathrm{He}\cdot\mathcal{Q}_\mathrm{H}$ (Fig.~\ref{fig:correlation_progenitors}). For all of these correlation relations of $Y_\mathrm{mix}$ with explosion and progenitor properties, we witnessed a systematic ordering and to a large extent a clear separation of our three classes of lowest-mass, intermediate-mass, and highest-mass progenitors. 

We note in passing that besides the mentioned inverse correlations with the He-core and (C+O)-core masses of our \ac*{rsg} progenitors, $Y_\mathrm{mix}$ also tightly anti-correlates with other progenitor properties such as the ZAMS mass and the mass and radius at the pre-collapse stage. However, these correlations do not have any direct causal relations to the processes and the interior physics that are responsible for the radial mixing. Instead, they arise from the fact that different stellar properties are connected through the evolution of massive stars, but in particular the pre-collapse mass and radius also depend sensitively on the assumed mass-loss history of the stars.

Because of the initiation of asymmetric explosions by the neutrino-heating mechanism and the preservation of significant shock asphericity during the expansion of the SN ejecta until shock breakout, we find that the breakout times of the SN shock in different directions can vary in individual \ac*{rsg} progenitors by up to nearly one day and by 0.1 days at least. These results are compatible with recent 3D simulations performed by \citet{vartanyan2025b}. The shortest breakout times (measured by the moment when the maximum shock radius reaches the stellar surface) are around 1.1\,d and the longest breakout times are around 3\,d, roughly correlated with the \ac*{zams} masses and the radii of the progenitors. In contrast, however, we do not observe any systematic variation of the time spans $\Delta t_\mathrm{sb} = t_\mathrm{sb,2}-t_\mathrm{sb,1}$ between the final and initial shock-breakout times with our three classes of explosion models, despite their substantially different mixing behaviour. This can be understood by two effects acting in opposite directions: On the one hand, the degree of shock deformation tends to be higher in models with strong $\Ni$ mixing and with big RT plumes growing in the H envelopes, which means that the SN shocks in the LM models tend to be the most deformed ones. On the other hand, the greater radii of the HM-LE and HM-HE cases shift the shock breakout to later times and thus also stretch the intervals between initial and final shock-breakout times. However, despite similar absolute time spans $\Delta t_\mathrm{sb}$, the relative lengths of the delay, $\Delta t_\mathrm{sb}\,t_\mathrm{sb,1}^{-1}$, exhibit a trend of being largest in the more asymmetric models in terms of the values of $R_\mathrm{shock}^\mathrm{max}/R_\mathrm{shock}^\mathrm{min}$ and of the shock-dipole amplitude $A_\mathrm{shock,1}$ at times $t_\mathrm{sb,1}$ (see Table~\ref{tab:sb_times} and Fig~\ref{fig:shock_ratio}). Moreover, since the outward propagating SN shock decouples from the NiCoFe0.5X distribution in most HM-LE and HM-HE models and its geometry is strongly influenced by the spherical shape of the progenitors, we find only a weak correlation between the PNS kick velocity and $\Delta t_\mathrm{sb}$, and an even weaker correlation of $v_\mathrm{PNS}$ with the other measures of the shock deformation at shock breakout, $\Delta t_\mathrm{sb}\,t_\mathrm{sb,1}^{-1}$, $R_\mathrm{shock}^\mathrm{max}/R_\mathrm{shock}^\mathrm{min}$, and $A_\mathrm{shock,1}$.

Our results are of relevance for the description and interpretation of the shock breakout in \acp*{rsg} and for corresponding conclusions on the progenitor radii drawn from measured shock-breakout bursts. They may also guide artificial prescriptions of $\Ni$ mixing in 1D explosion simulations to improve the modelling of light curves, spectra and spectral-line profiles, and of the gamma-ray and X-ray emission, all of which are very sensitive to the distribution of radioactive material and to its decay-heating, which affects the rise time, peak brightness, peak duration and shape, and the tail of the \ac*{sn} light curves. Observations can therefore provide important information on the explosion asymmetries and the efficiency of mixing of heavy elements during the explosion. Our theoretical framework of connecting the extent of the metal mixing to structural properties (density profiles and masses) of the C+O cores and He cores of the progenitors could help understanding systematic trends in the characteristics of Type-IIP \acp*{sn}. It might thus have a bearing on the use of Type-IIP \acp*{sn} for determining the Hubble constant \citep{vogl2024}. Moreover, our framework might offer new diagnostic possibilities to infer intrinsic progenitor properties at the pre-collapse stage from \ac*{sn} observations. Thus it might open new pathways to assess the ability of late-phase stellar evolution calculations to produce the conditions needed to explain the mixing of heavy elements detected in the \ac*{sn} ejecta.

In a follow-up paper, we will discuss the evolution of our models to even later times, when the $\beta$-decay heating by radioactive isotopes has a significant impact on the geometry of the expanding ejecta and their approach to a homologous state.

\section*{Acknowledgements}
BG and MG acknowledge the support through the Generalitat Valenciana via the grant CIDEGENT/2019/031, 
the grants PID2021-127495NB-I00 funded by MCIN/AEI/10.13039/501100011033 and by the European Union and PID2025-171322NB-C22 funded by MICIU/AEI/10.13039/501100011033, and the Astrophysics and High Energy Physics programme of the Generalitat Valenciana ASFAE/2022/026 funded by MCIN and the European Union NextGenerationEU (PRTR-C17.I1).
BG acknowledges support through the Generalitat Valenciana and the Fundo Social Europeo (FSE) via the grant BEFPI 2023 (CIBEFP/2022).
HTJ is grateful for support by the German Research Foundation (DFG) through the Collaborative Research Centre ``Neutrinos and Dark Matter in Astro- and Particle Physics (NDM),'' Grant No.\ SFB-1258-283604770, and under Germany's Excellence Strategy through the Cluster of Excellence ORIGINS EXC-2094-390783311. The computations were carried out on LluisVives v2 cluster at the University of València.

\section*{Data availability}
The data underlying this paper will be shared upon reasonable request to the corresponding author via the Garching Core-Collapse Supernova Archive,\\ {\tt https://wwwmpa.mpa-garching.mpg.de/ccsnarchive}. See also https://beagiudici.github.io/ for 3D visualization of the data.

\bibliographystyle{mnras}
\bibliography{biblio}





\bsp	
\label{lastpage}

\end{document}